\documentclass[11pt,a4paper]{article}
\usepackage[hmargin=1.1in,vmargin={1.1in,1.1in}]{geometry}
\PassOptionsToPackage{hyphens}{url}\usepackage[colorlinks]{hyperref}
\usepackage{amsmath}
\usepackage{multirow}
\usepackage{mathtools}
\usepackage{amssymb}
\usepackage{bbm}
\usepackage{xurl}
\usepackage{float}
\usepackage{tabularx}
\usepackage{natbib}
\usepackage[font=footnotesize]{caption,subcaption} 
\usepackage{graphicx}
\usepackage[affil-it]{authblk}
\usepackage{array} 
\usepackage{pdflscape}
\usepackage{lscape}
\usepackage{verbatim}
\usepackage{placeins} 
\usepackage[dvipsnames,table,xcdraw,svgnames]{xcolor}
\usepackage{pdfpages}
\usepackage{lineno} 
\usepackage[nottoc,numbib]{tocbibind} 
\usepackage[percent]{overpic} 
\usepackage{chngcntr} 
\usepackage{fullpage}

\newcolumntype{L}[1]{>{\let\newline\\\arraybackslash\hspace{0pt}}m{#1}}

\AtBeginDocument{%
 \hypersetup{
  citecolor=Bittersweet,
    linkcolor=Blue,
  urlcolor=Blue}
}

\title{In and out of lockdown: Propagation of supply and demand shocks in a dynamic input-output model
\\
}

\author{Anton Pichler$^{1,2,3}$, Marco Pangallo$^{4}$, R. Maria del Rio-Chanona$^{1,2}$,\hspace{5cm}
Fran\c{c}ois Lafond$^{1,2}$, J. Doyne Farmer$^{1,2,5}$ \\

\vspace{0.25cm}

\footnotesize{$^{1}$ Institute for New Economic Thinking at the Oxford Martin School, University of Oxford, UK} \\
\footnotesize{$^{2}$ Mathematical Institute, University of Oxford, UK}\\
\footnotesize{$^3$ Complexity Science Hub Vienna, Austria} \\
\footnotesize{$^{4}$Institute of Economics and EMbeDS Department, Sant’Anna School of Advanced Studies, Pisa, Italy} \\
\footnotesize{$^{5}$ Santa Fe Institute, US}\\
}

\date{\today}

\begin{document}

\maketitle

\begin{abstract}
\noindent
Economic shocks due to Covid-19 were exceptional in their severity, suddenness and heterogeneity across industries. To study the upstream and downstream propagation of these industry-specific demand and supply shocks, we build a dynamic input-output model inspired by previous work on the economic response to natural disasters. We argue that standard production functions, at least in their most parsimonious parametrizations, are not adequate to model input substitutability in the context of Covid-19 shocks. We use a survey of industry analysts to evaluate, for each industry, which inputs were absolutely necessary for production over a short time period. We calibrate our model on the UK economy and study the economic effects of the lockdown that was imposed at the end of March and gradually released in May. Looking back at predictions that we released in May, we show that the model predicted aggregate dynamics very well, and sectoral dynamics to a large extent. We discuss the relative extent to which the model's dynamics and performance was due to the choice of the production function or the choice of an exogenous shock scenario. To further explore the behavior of the model, we use simpler scenarios with only demand or supply shocks, and find that popular metrics used to predict a priori the impact of shocks, such as output multipliers, are only mildly useful.

\end{abstract}

\vfill

\footnotesize{
\noindent
This is a substantially revised version of \citet{pichler2020production}, which was released early in the pandemic (May 2020) and included a cost-benefit analysis of both the economic and epidemiological consequences of reopening specific industries.  In this new version we focus solely on the economic model.
\\
\noindent
\emph{Acknowledgements:} We would like to thank Eric Beinhocker, David Van Dijcke, John Muellbauer and David Vines for many useful comments and discussions. 
This work was supported by Baillie Gifford, Partners for a New Economy, the UK's Economic and Social Research Council (ESRC) via the Rebuilding Macroeconomics Network (Grant Ref: ES/R00787X/1), the Oxford Martin Programme on the Post-Carbon Transition,  James S. McDonnell Foundation, and the Institute for New Economic Thinking at the Oxford Martin School.
This research is based upon work supported in part
by the Office of the Director of National Intelligence (ODNI), Intelligence Advanced Research Projects Activity (IARPA), via contract no. 2019-1902010003. The views and conclusions contained herein are those of the authors and should not be interpreted as necessarily representing the official policies, either expressed or implied, of ODNI, IARPA, or the US Government. The US Government is authorized to reproduce and distribute reprints for governmental purposes notwithstanding any copyright annotation therein. We appreciate that IHS Markit provided us with a survey on critical vs. non-critical inputs.  (Note that JDF is on their advisory board). We thank Diana Beltekian for excellent research assistance.
\\
\noindent
\emph{Contacts:}
anton.pichler@maths.ox.ac.uk, 
marco.pangallo@santannapisa.it,
rita.delriochanona@maths.ox.ac.uk,
francois.lafond@inet.ox.ac.uk,
doyne.farmer@inet.ox.ac.uk
}
\normalsize

\newpage

\section{Introduction} 
\label{sec:intro}
The social distancing measures imposed to combat the Covid-19 pandemic created severe disruptions to economic output, causing shocks that were highly industry specific.  Some industries were shut down almost entirely by lack of demand, labor shortages restricted others, and many were largely unaffected.  Meanwhile, feedback effects amplified the initial shocks. The lack of demand for final goods such as restaurants or transportation propagated upstream, reducing demand for the intermediate goods that supply these industries. Supply constraints due to a lack of labor under social distancing propagated downstream, creating input scarcity that sometimes limited production even in cases where the availability of labor and demand would not have been an issue. The resulting supply and demand constraints interacted to create bottlenecks in production, which in turn led to unemployment, eventually decreasing consumption and causing additional amplification of shocks that further decreased final demand. 

In this paper we develop a model that respects the three main features that made the Covid-19 episode exceptional: (1) The shocks were highly heterogeneous across industries, making it necessary to model the economy at the sectoral level, taking sectoral inter-dependencies into account; (2) the shocks affected both supply and demand simultaneously, making it necessary to consider both upstream and downstream propagation; (3) the shocks were so strong and were imposed and relaxed so quickly that the economy never had time to converge to a new steady state, making dynamic models better suited than static models.

In the first version of this work, results were released online on 21 May, not long after social distancing measures first began to take effect in March \citep{pichler2020production}.  Based on data from 2019 and predictions of the shocks by \citet{del2020supply}, we predicted a 21.5\% contraction of GDP in the UK economy in the second quarter of 2020 with respect to the last quarter of 2019.  This forecast was remarkably close to the actual contraction of 22.1\% estimated by the UK Office of National Statistics. (The median forecast by several institutions and financial firms was 16.6\% and the forecast by the Bank of England was 30\%).\footnote{\label{footn:pred}
Sources: ONS estimate: \url{https://www.ons.gov.uk/economy/grossdomesticproductgdp/bulletins/gdpfirstquarterlyestimateuk/apriltojune2020}; median forecast of institutions and financial firms (in May): \url{https://www.gov.uk/government/statistics/forecasts-for-the-uk-economy-may-2020}; forecast by the Bank of England (in May): \url{https://www.bankofengland.co.uk/-/media/boe/files/monetary-policy-report/2020/may/monetary-policy-report-may-2020}.
}

In this substantially revised paper we present our model and describe its results, but we also take advantage of the benefits of hindsight to perform a ``post-mortem''.  This allows us to better understand what worked well, what did not work well, and why. To what extent did we succeed by getting things right vs. just getting lucky?  We systematically investigate the sensitivity of the results under different specifications, including variations in the shock scenario, production function, and key parameters of the model.  We examine the performance at both the aggregate and sectoral levels.  We also look at the ability of the model to reproduce the time series patterns of the fall and recovery of sectoral output.  Surprisingly, we find that the original specification used for out-of-sample forecasting performs about as well at the aggregate level as any of those developed with hindsight; with a minor adjustment its results are also among the best at the sectoral level. We show how getting good aggregate results depends on making the right trade-off between the severity of the shocks and the rigidity of the production function, as well as other key factors such as the right level of inventories. This provides valuable lessons about modeling the economic effects of disasters such as the Covid-19 pandemic.

Our model is inspired by previous work on the economic response to natural disasters \citep{hallegatte2008adaptive, henriet2012firm, inoue2019firm}. As in these models, industry demand and production decisions are based on simple rules of thumb, rather than resulting from optimization in a dynamic general equilibrium setup. We think that the Covid-19 shock was so sudden and unexpected that agent expectations had little time to converge to an equilibrium over the short time period that we consider \citep{evans2012learning}.  Our work here is also one of several studies using sectoral models with input-output linkages that appeared in the first few months of the pandemic \citep{barrot2020sectoral,mandel2020economic,fadinger2020effects,bonadio2020global,baqaee2020nonlinear,guan2020global}. Our paper belongs to this effort, but differs in a number of important ways. 

The most important conceptual difference is our treatment of the production function. The most common production functions can be ordered by the degree to which they allow substitutions between inputs. At one extreme, the Leontief production function assumes a fixed recipe for production, allowing no substitutions and restricting production based on the limiting input \citep{inoue2020propagation}.  Under the Leontief production function, if a single input is reduced, overall production will be reduced proportionately, even if that input is ordinarily relatively small. This can lead to unrealistic behaviours. For example, the steel industry has restaurants as an input, presumably because steel companies have a workplace canteen and sometimes entertain their clients and employees.  A literal application of the Leontief production function predicts that a sharp drop in the output of the restaurant industry will dramatically reduce steel output. This is unrealistic, particularly in the short run.

The alternatives used in the literature are the Cobb-Douglas production function \citep{fadinger2020effects}, which has an elasticity of substitution of 1, and the CES production function, where typically calibration for short term analysis uses an elasticity of substitution less than 1 \citep{barrot2020sectoral,mandel2020economic,bonadio2020global}. Some papers \citep{baqaee2020nonlinear} consider a nested CES production function, which can accommodate a wide range of technologies. In principle, this can be used to allow substitution between some inputs and forbid it between others in an industry-specific manner. However, it is hard to calibrate all the elasticities, so that in practice many models end up using only limited nesting structure or assuming uniform substitutability. Consider again our example of the steel industry: Under the calibrations of the CES production function that are typically used, firms can substitute energy or even restaurants for iron, while still producing the same output.  For situations like this, where the production process requires a fixed technological recipe, this is obviously unrealistic. 

To solve this problem we introduce a new production function that distinguishes between critical and non-critical inputs at the level of the 55 industries in the World Input-Output tables. This production function allows firms to keep producing as long as they have the inputs that are absolutely necessary, which we call {\it critical inputs}. The steel industry cannot produce steel without the critical inputs iron and energy, but it can operate for a considerable period of time without non-critical inputs such as restaurants or management consultants.  We apply the Leontief function only to the critical inputs, ignoring the others.   Thus we make the assumption that during the pandemic the steel industry requires iron and energy in the usual fixed proportions, but the output of the restaurant or management consultancy industries is irrelevant.  Of course restaurants and logistics consultants are useful to the steel industry in normal times -- otherwise they wouldn't use them.  But during the short time-scale of the pandemic, we believe that neglecting them provides a better approximation of economic behavior than either a Leontief or a CES production function with uniform elasticity of substitution. In the appendix, we show that our production function is very close to a limiting case of an appropriately constructed nested CES, which we could have used in principle, but is less well-suited to our calibration procedure.

To determine which inputs are critical and which are not, we use a survey performed by IHS Markit at our request.  This survey asked ``Can production continue in industry X if input Y is not available for two months?''. The list of possible industries X and Y was drawn from the 55 industries in the  World  Input-Output  Database.  This question was  presented  to  30  different  industry analysts who were experts of industry X.  Each of them was asked to rate the importance of each of its inputs Y. They assigned a score of 1 if they believed input Y is critical, 0 if it is not critical, and 0.5 if it is in-between, with the possibility of a rating of NA if they could not make a judgement.  We then apply the Leontief function to the list of critical inputs, ignoring non-critical inputs.  We experimented with several possible treatments for industries with ratings of 0.5 and found that we get somewhat better empirical results by treating them as half-critical (though at present we do not have sufficient evidence to resolve this question unambiguously).

Besides the bespoke production function discussed above, we also introduce a Covid-19-specific treatment of consumption. Most models do not incorporate the demand shocks that are caused by changes in consumer preferences in order to minimize risk of infection. The vast majority of the literature has focused on the ability to work from home, and some studies incorporate lists of essential vs. inessential industries, but almost no papers have also explicitly added shocks to consumer preferences. (\cite{baqaee2020nonlinear} is an exception, but the treatment is only theoretical). Here we use the estimates from \citet{del2020supply}, which are taken from a prospective study by the \citet{CBO2006}. These estimates are crude, but we are not aware of estimates that are any better. The currently available data on actual consumption is qualitatively consistent with the shocks predicted by the CBO, with massive shocks to the hospitality industry, travel and recreation, milder (but large) shocks elsewhere \citep{andersen2020consumer,carvalho2020tracking,chen2020impact,surico2020consumption}. The largest mismatch between the CBO estimates and consumption data is in the healthcare sector, whose consumption decreased during the pandemic, in contrast to the estimated increase by the CBO. There was also an increase in some specific retail categories (groceries) which the CBO estimates did not consider. However, overall, as \citet{del2020supply} argue, the estimates remain qualitatively accurate. Besides the initial shock, we also attempt to introduce realistic dynamics for recovery and for savings.  The shocks to on-site consumption industries are more long lasting, and savings from the lack of consumption of specific goods and services during lockdown are only partially reallocated to other expenses.

Our key finding is that there is a trade-off between the severity of the shocks and the rigidity of the production function. At one extreme, severe shocks with a Leontief production function lead to an almost complete collapse of the economy, and at the other, small shocks with a linear production function fail to reproduce the massive recession observed in the data.  There is a sweet spot in the middle, with empirically good results based on various mixes of production function rigidity and shock severity. We also find that our models' performance improved after we used better data for inventories. 

Our selected scenario predicts the aggregate UK recession in Q2 2020 about as well as our initial forecast. It also correctly predicts a stronger reduction in private consumption, investment and profits than in government consumption, inventories and wages. At the sectoral level, the mean absolute error is around 12 percentage points, slightly lower than in our initial forecast (14 p.p.). The correlation between sectoral reductions in the model and in the data is generally high (the Pearson correlation coefficient weighted by industry size is 0.75). However, this masks substantial differences across industries: while our model makes a good job at predicting sectoral outcomes in most industries, it fails at others such as vehicle manufacturing and air transport. We conjecture that this is due to idiosyncratic features of these industries that our model could not capture. Conversely, we provide examples where our model predicts sectoral outcomes correctly when these depend on inter-industry relationships, both statically and dynamically.

We conclude by investigating some theoretical properties of the model using a simpler setting. We ask whether popular metrics of centrality, such as upstreamness and output multipliers, are useful to understand the diffusion of supply and demand shocks in our model. We find that static measures are only partially able to explain modeling results. Nevertheless, an industry's upstreamness is a strong indicator of its potential to amplify shocks - this is true for supply shocks, as expected, and for demand shocks, which is somewhat more surprising. 

The paper is organized as follows. The details of the model are presented in Section~\ref{sec:model}. We discuss the shocks induced by the pandemic which we use to initialize the model in Section~\ref{sec:pandemic_shock}.
We show our model predictions for the UK economy in Section~\ref{sec:econimpact} and discuss production network effects and re-opening single industries in Section~\ref{sec:nw_effects}. We conclude in Section~\ref{sec:discuss}. 
%
%

\section{A dynamic input-output model} 
\label{sec:model}

Our model combines elements of the input-output models developed by \cite{battiston2007credit, hallegatte2008adaptive, henriet2012firm} and \cite{inoue2019firm}, together with new features that make the model more realistic in the context of a pandemic-induced lockdown. 

The main data input to our analysis is the UK input-output network obtained from the latest year (2014) of the World Input-Output Database (WIOD) \citep{timmer2015illustrated}, allowing us to distinguish 55 sectors.

\subsection{Timeline}

A time step $t$ in our economy corresponds to one day. 
There are $N$ industries\footnote{
See Appendix~\ref{apx:notation} for a comprehensive summary of notations used.
}, 
one representative firm for each industry, and one representative household.
The economy initially rests in a steady-state until it experiences exogenous lockdown shocks. These shocks can affect the supply side (labor compensation, productive capacity) and the demand side (preferences, aggregate spending/saving) of the economy. 
Every day:
\begin{enumerate}
    \item Firms hire or fire workers depending on whether their workforce was insufficient or redundant to carry out production in the previous day.
    \item The representative household decides its consumption demand and industries place orders for intermediate goods.
    \item Industries produce as much as they can to satisfy demand, given that they could be limited by lack of critical inputs or lack of workers.
    \item If industries do not produce enough, they distribute their production to final consumers and to other industries on a pro rata basis, that is, proportionally to demand.
    \item Industries update their inventory levels, and labor compensation is distributed to workers.
\end{enumerate}

\subsection{Model description} 
\label{sec:modeldescription} 
 
It will become important to distinguish between demand, that is, orders placed by customers to suppliers, and actual realized transactions, which might be lower.

\subsubsection{Demand}
\label{sec:demand}

\paragraph{Total demand.} 
The total demand faced by industry $i$ at time $t$, $d_{i,t}$, is the sum of the demand from all its customers,
\begin{equation}
    d_{i,t} = \sum_{j=1}^N O_{ij,t} + c^d_{i,t} + f^d_{i,t},
\end{equation}
where $O_{ij,t}$ (for \emph{orders}) denotes the intermediate demand from industry $j$ to industry $i$, $c_{i,t}^d$ represents (final) demand from households and $f_{i,t}^d$ denotes all other final demand (e.g. government or non-domestic customers).

\paragraph{Intermediate demand.} 
Intermediate demand follows a dynamics similar to the one studied in \cite{henriet2012firm}, \cite{hallegatte2014modeling}, and \cite{inoue2019firm}. Specifically, the demand from industry $i$ to industry $j$ is
\begin{equation} \label{eq:order_interm}
    O_{ji,t} = A_{ji} d_{i,t-1} + \frac{1}{\tau} [ n_i Z_{ji,0} - S_{ji,t} ].
\end{equation}
Intermediate demand thus is the sum of two components. 
First, to satisfy incoming demand (from $t-1$), industry $i$ demands an amount $A_{ji} d_{i,t-1}$ from $j$.  
Therefore, industries order intermediate inputs in fixed proportions of total demand, with the proportions encoded in the technical coefficient matrix $A$, i.e. $A_{ji} = Z_{ji,0}/x_{i,0}$ where $Z_{ji,0}$ is realized intermediate consumption and $x_{i,0}$ is total output of industry $i$. Before the shocks, both of these variables are considered to be in the pre-pandemic steady state.
While we will consider several scenarios where industries do not strictly rely on fixed recipes in production, demand always depends on the technical coefficient matrix.

The second term in Eq.~\eqref{eq:order_interm} describes intermediate demand induced by desired reduction of inventory gaps. Due to the dynamic nature of the model, demanded inputs cannot be used immediately for production. 
Instead industries use an inventory of inputs in production. $S_{ji,t}$ denotes the stock of input $j$ held in $i$'s inventory. Each industry $i$ aims to keep a target inventory $n_i Z_{ji,0}$ of every required input $j$ to ensure production for $n_i$ further days\footnote{
Considering an input-specific target inventory would require generalizing $n_i$ to a matrix with elements $n_{ji}$, which is easy in our computational framework but difficult to calibrate empirically.
}. 
The parameter $\tau$ indicates how quickly an industry adjusts its demand due to an inventory gap. Small $\tau$ corresponds to responsive industries that aim to close inventory gaps quickly. In contrast, if $\tau$ is large, intermediate demand adjusts slowly in response to inventory gaps.

\paragraph{Consumption demand.}  We let consumption demand for good $i$ be
\begin{equation}\label{eq:cd}
    c^d_{i,t}= \theta_{i,t} \Tilde{c}^d_t, 
\end{equation}
where $\theta_{i,t}$ is a preference coefficient, giving the share of goods from industry $i$ out of total consumption demand $\Tilde{c}^d_t$. The coefficients $\theta_{i,t}$ evolve exogenously, following assumptions on how consumer preferences change due to exogenous shocks (in our case, differential risk of infection across industries, see Section \ref{sec:pandemic_shock}).

Total consumption demand evolves following an adapted and simplified version of the consumption function in \cite{muellbauer2020}. 
In particular, $\Tilde{c}^d_t$ evolves according to
\begin{equation}\label{eq:consdemand}
     \Tilde{c}^d_t= \left(1-\Tilde{\epsilon}^D_t\right) \exp\left\{ \rho \log \Tilde{c}^d_{t-1} + \frac{1-\rho}{2} \log\left( m  \Tilde{l}_t \right) + \frac{1-\rho}{2} \log \left( m \Tilde{l}_t^p \right) \right\}.
\end{equation}
In the equation above, the factor $\left(1-\Tilde{\epsilon}^D_t\right)$ accounts for direct aggregate shocks and will be explained in Section \ref{sec:pandemic_shock}. The second factor accounts for the endogenous consumption response to the state of the labor market and future income prospects. In particular, $\Tilde{l}_t$ is current labor income, $\Tilde{l}_t^p$ is an estimation of permanent income (see Section \ref{sec:pandemic_shock}), and  $m$ is the share of labor income that is used to consume final domestic goods, i.e. that is neither saved nor used for consumption of imported goods. 
In the pre-pandemic steady state with no aggregate exogenous shocks, $\Tilde{\epsilon}^D_t=0$ and by definition permanent income corresponds to current income, i.e. $\Tilde{l}_t^p=\Tilde{l}_t$. In this case, total consumption demand corresponds to $m \Tilde{l}_t $.\footnote{
To see this, note that in the steady state $\Tilde{c}^d_t=\Tilde{c}^d_{t-1}$. Taking logs on both sides, moving the consumption terms on the left hand side and dividing by $1-\rho$ throughout yields $\log \Tilde{c}^d_t=\log\left( m  \Tilde{l}_t \right)$.
}

\paragraph{Other components of final demand.} In addition, an industry $i$ also faces demand $f_{i,t}^d$ from sources that we do not model as endogenous variables in our framework, such as government or industries in foreign countries. We discuss the composition and calibration of $f_{i,t}^d$ in detail in Section \ref{sec:pandemic_shock}.

\subsubsection{Supply}
\label{sec:supply}
Every industry aims to satisfy incoming demand by producing the required amount of output. Production is subject to the following two economic constraints:

\paragraph{Productive capacity.}
First, an industry has finite production capacity $x_{i,t}^\text{cap}$, which depends on the amount of available labor input.
Initially every industry employs $l_{i,0}$ of labor and produces at full capacity 
$x_{i,0}^{\text{cap}} =  x_{i,0}$.
We assume that productive capacity depends linearly on labor inputs,
\begin{equation} \label{eq:xcap}
x_{i,t}^{\text{cap}} =  \frac{l_{i,t}}{l_{i,0}}x_{i,0}^{\text{cap}}.
\end{equation}

\paragraph{Input bottlenecks.}
Second, the production of an industry might be constrained due to an insufficient supply of critical inputs.  This can be caused by production network disruptions. 
Intermediate input-based production capacities depend on the availability of inputs in an industry's inventory and its production technology, i.e.
\begin{equation} \label{eq:xinp_general}
x_{i,t}^{\text{inp}} = \text{function}_i( S_{ji,t}, A_{ji} ).
\end{equation}
We consider five different specifications for how input shortages impact production, ranging from a Leontief form, where inputs need to be used in fixed proportions, to a Linear form, where inputs can be substituted arbitrarily. As intermediate cases we consider specifications with industry-specific dependencies of inputs. For this purpose, IHS Markit analysts rated at our request whether a given input is \emph{critical}, \emph{important} or \emph{non-critical} for the production of a given industry (see Appendix \ref{apx:ihs} for details). We then make different assumptions on how the criticality ratings of inputs affect the production of an industry. We now introduce the five specifications in order of stringency with respect to inputs.

\noindent 
\emph{(1) Leontief:} 
As first case we consider the Leontief production function, in which every positive entry in the technical coefficient matrix $A$ is a binding input to an industry.
This is the most rigid case we are considering,
leading to the functional form
\begin{equation} \label{eq:xinp_leo}
x_{i,t}^{\text{inp}} = 
\min_{ \{ j: \; A_{ji} > 0 \} } 
\left \{ \;  \frac{ S_{ji,t} }{ A_{ji} } \right \}.
\end{equation}
In this case, an industry would halt production immediately if inventories of any input are run down, even for small and potentially negligible inputs.

\noindent 
\emph{(2) IHS1:}
As most stringent case based on the IHS Markit ratings, we assume that production is constrained by critical and important inputs, which need to use in fixed proportions. In contrast to the Leontief case, however, production is not constrained by the lack of non-critical inputs.
Thus, an industry's production capacity with respect to inputs is
\begin{equation} \label{eq:xinp_ihs1}
x_{i,t}^{\text{inp}} = 
\min_{j \in \{ \mathcal{V}_i \; \cup \; \mathcal{U}_i \} } 
\left \{ \;  \frac{ S_{ji,t} }{ A_{ji} } \right \},
\end{equation}
where $\mathcal{V}_i$ is the set of \textit{critical} inputs and 
$\mathcal{U}_i$ is the set of \textit{important} inputs to industry $i$.

\noindent 
\emph{(3) IHS2:}
As second case using the input ratings, we leave the assumptions regarding \emph{critical} and \emph{non-critical} inputs unchanged but assume that the lack of an \emph{important} input reduces an industry's production by a half. We implement this production scenario as
\begin{equation} \label{eq:xinp_ihs2}
    x_{i,t}^{\text{inp}} =
    \min_{ 
    \{ 
     j \in \mathcal{V}_i, \; k \in \mathcal{U}_i
    \} 
} 
    \left \{ \;  \frac{ S_{ji,t} }{ A_{ji} },  
   \frac{1}{2} \left(\frac{ S_{ki,t} }{ A_{ki} } +  x_{i,0}^\text{cap}\right)
     \right \}.     
\end{equation}
This means that if an \textit{important} input goes down by 50\% compared to initial levels, production of the industry would decrease by 25\%. When the stock of this input is fully depleted, production drops to 50\% of initial levels.

\noindent 
\emph{(4) IHS3:}
Next we treat all \emph{important} inputs as \emph{non-critical}, such that only \emph{critical} suppliers can create input bottlenecks. This reduces the input bottleneck equation, Eq.~\eqref{eq:xinp_general}, to
\begin{equation} \label{eq:xinp_ihs3}
x_{i,t}^{\text{inp}} = 
\min_{j \in \mathcal{V}_i } 
\left \{ \;  \frac{ S_{ji,t} }{ A_{ji} } \right \}.
\end{equation}

\noindent 
\emph{(5) Linear:}
Finally, we also implement a linear production function for which all inputs are perfectly substitutable. Here, production in an industry can continue even when inputs cannot be provided, as long as there is sufficient supply of alternative inputs. In this case we have 
\begin{equation} \label{eq:xinp_linear}
    x_{i,t}^{\text{inp}} =   \frac{ \sum_j S_{ji,t} }{ \sum_j A_{ji} } .     
\end{equation}
Note that while production is linear with respect to intermediate inputs, the lack of labor supply cannot be compensated by other inputs. 

We assume for any production function that imports never cause bottlenecks. 
Thus, imports are treated as non-critical inputs or, equivalently, there is no shortages in foreign intermediate goods.

Input bottlenecks are most likely to arise under the Leontief assumption, and least likely under the linear production function. The IHS production functions assume intermediate levels of input specificity. 
In Appendix \ref{apx:ces_prodfun} we show that the IHS production functions are almost equivalent to (suitably parameterised) CES production functions and that results are even identical using these CES specifications instead of the IHS production functions.

\paragraph{Output level choice and input usage.}
Since an industry aims to satisfy incoming demand within its production constraints, realized production at time step $t$ is
\begin{equation} \label{eq:x_act}
    x_{i,t} = \min \{ x_{i,t}^{\text{cap}}, x_{i,t}^{\text{inp}}, d_{i,t}  \}.
\end{equation}
Thus, the
output level of an industry is constrained by the smallest of three values: labor-constrained production capacity $x_{i,t}^{\text{cap}}$, intermediate input-constrained production capacity $x_{i,t}^{\text{inp}}$, or total demand $d_{i,t}$.

The output level $x_{i,t}$ determines the quantity used of each input according to the production recipe. 
Industry $i$ uses an amount $A_{ji}x_{i,t}$ of input $j$, unless $j$ is not critical and the amount of $j$ in $i$'s inventory is less than $A_{ji}x_{i,t}$. In this case, the quantity consumed of input $j$ by industry $i$ is equal to the remaining inventory stock of $j$-inputs $S_{ji,t} < A_{ji}x_{i,t}$.

\paragraph{Rationing.}
Without any adverse shocks, industries can always meet total demand, i.e. $x_{i,t} = d_{i,t}$. However, in the presence of production capacity and/or input bottlenecks, industries' output may be smaller than total demand (i.e., $x_{i,t} < d_{i,t}$) in which case industries ration their output across customers. We assume simple proportional rationing, although alternative rationing mechanisms could be considered \citep{pichler2021modeling}. The final delivery from industry $j$ to industry $i$ is a share of orders received
\begin{equation}
    Z_{ji,t} = O_{ji,t} \frac{x_{j,t}}{d_{j,t}}.
\end{equation}
Households receive a share of their demand
\begin{equation}
    c_{i,t} = c_{i,t}^d \frac{x_{i,t}}{d_{i,t}},
\end{equation}
and the realized final consumption of agents with exogenous final demand is 
\begin{equation}
    f_{i,t} = f_{i,t}^d \frac{x_{i,t}}{d_{i,t}}.
\end{equation}

\paragraph{Inventory updating.} The inventory of $i$ for every input $j$ is updated according to
\begin{equation}
    S_{ji,t+1} = \min \left\{ S_{ji,t} + Z_{ji,t} - A_{ji} x_{i,t}, 0 \right\}.
\end{equation}
In a Leontief production function, where every input is critical, the minimum operator would not be needed since production could never continue once inventories are run down. It is necessary here, 
since industries can produce even after inventories of non-critical input $j$ are depleted and inventories cannot turn negative.

\paragraph{Hiring and separations.} 
Firms adjust their labor force depending on which production constraints in Eq.~\eqref{eq:x_act} are binding. If the capacity constraint $x_{i,t}^{\text{cap}}$ is binding,  industry $i$ decides to hire as many workers as necessary to make the capacity constraint no longer binding. Conversely, if either input constraints $x_{i,t}^{\text{inp}}$ or demand constraints $d_{i,t}$ are binding, industry $i$ lays off workers until capacity constraints become binding.  
More formally, at time $t$ labor demand by industry $i$ is given by $l^d_{i,t}=l_{i,t-1}+\Delta l_{i,t}$, with 
\begin{equation}
    \Delta l_{i,t} = \frac{l_{i,0}}{x_{i,0}}\left[ \min\{x_{i,t}^{\text{inp}},d_{i,t}\} - x_{i,t}^{\text{cap}}\right].
\end{equation} 
The term $l_{i,0}/x_{i,0}$ reflects the assumption that the labor share in production is constant over the considered period. We assume that it takes time for firms to adjust their labor inputs.
Specifically, we assume that industries can increase their labor force only by a fraction $\gamma_{\text{H}}$ in direction of their target. Similarly, industries can decrease their labor force only by a fraction $\gamma_{\text{F}}$ in the direction of their target. In the absence of strong labor market regulations, we usually have $\gamma_{\text{F}}>\gamma_{\text{H}}$, indicating that it is easier for firms to lay off employed workers than to hire new workers.
Industry-specific employment evolves then according to
\begin{equation} \label{eq:labor_evolution}
    l_{i,t} = 
    \begin{cases} 
    l_{i,t-1} + \gamma_{\text{H}} \Delta l_{i,t} &\mbox{if } \; \Delta l_{i,t} \ge 0, \\
    l_{i,t-1} + \; \gamma_{\text{F}} \Delta l_{i,t} &\mbox{if }  \Delta l_{i,t} < 0.
    \end{cases}
\end{equation}
The parameters $\gamma_{\text{H}}$ and $\gamma_{\text{F}}$ can be interpreted as policy variables. 
For example, the implementation of a furloughing scheme makes re-hiring of employees easier, corresponding to an increase in $\gamma_{\text{H}}$.

\section{Pandemic shock} \label{sec:pandemic_shock}

Simulations of the model described in Section \ref{sec:modeldescription} start in the pre-pandemic steady state. While there is evidence that consumption started to decline prior to the lockdown \citep{surico2020consumption}, for simplicity we apply the pandemic shock all at once, at the date of the start of the lockdown (March $23^{rd}$ in the UK). 

The pandemic shock is a combination of supply and demand shocks that propagate downstream and upstream and get amplified through the supply chain. During the lockdown, workers who cannot work on-site and are unable to work from home become unproductive, resulting in lowered productive capacities of industries. At the same time demand-side shocks hit as consumers adjust their consumption preferences to avoid getting infected, and reduce overall consumption out of precautionary motives due to the depressed state of the economy. 

For diagnostic purposes, in addition to the shocks that we used for our original predictions, we consider a few additional supply shock scenarios that are grounded on what happened in the UK specifically\footnote{
As described below, here we use estimates of the shocks that we develop a priori, based on empirically motivated assumptions on labor supply constraints and changes in preferences. We could have, in principle, used a traditional macro approach (see \citet{brinca2020measuring} in the case of Covid-19) where one can infer the shocks based on data. We did not pursue this here for several reasons.  Our model does not feature price changes (which are crucial to separately identify supply and demand shocks), and we would have had to develop a method to infer the shocks based on model's fit to the data, a problem that is unlikely to have a unique solution.  Most importantly, inferring shocks this way is only useful in hindsight, whereas our goal was to make predictions.
}. We then compare the outcomes under each scenario based on its aggregate and sectoral forecasts (see Section \ref{sec:scenarioselection}). In the following, we describe all the scenarios for supply and demand shocks that we consider. Further details and industry-level shock statistics are shown in Appendix \ref{apx:shocks}.

\subsection{Supply shock scenarios}
\label{sec:supplyshockscenarios}

At every time step during the lockdown an industry $i$ experiences an (exogenous) first-order labor supply shock $\epsilon^S_{i,t} \in [0,1]$ that quantifies reductions in labor  availability. Letting $l_{i,0}$ be the initial labor supply before the lockdown, the maximum amount of labor available to industry $i$ at time $t$ is given as
\begin{equation}
    l_{i,t}^\text{max} = (1- \epsilon^S_{i,t}) l_{i,0}.
\end{equation}
If $\epsilon^S_{i,t} > 0$, the productive capacity of industry $i$ is smaller than in the initial state of the economy. 
We assume that the reduction of total output is proportional to the loss of labor. In that case the productive capacity of industry $i$ at time $t$ is
\begin{equation}
x_{i,t}^{\text{cap}} = \frac{l_{i,t}}{l_{i,0}} x_{i,0}^{\text{cap}} \le (1-\epsilon^S_{i,t}) x_{i,0}.
\end{equation}
Recall from Section \ref{sec:supply} that firms can hire and fire to adjust their productive capacity to demand and supply constraints. Thus, productive capacity can be lower than the initial supply shock, in case industry $i$ has some idle workers that are not prevented to go to work by lockdown measures. In any case, during lockdown firms can never hire more than $l_{i,t}^\text{max}$ workers. When the lockdown is lifted for a specific industry $i$, first-order supply shocks are removed, i.e., we set $\epsilon^S_{i,t} = 0$, for $t\geq t_\text{end\_lockdown}$.

For diagnostic purposes we consider six different scenarios for the supply shocks $\epsilon^S_{i,t}$, $\text{S}_\text{1}$ to $\text{S}_\text{6}$, ordered from lowest to the highest severity of lockdown restrictions. Scenario $\text{S}_\text{5}$ is the one that was used to produce out of sample forecasts in our original paper.  We follow \cite{del2020supply} and estimate the supply shocks in each scenario by calculating for each industry the number of workers who can work remotely, and considering the government regulations that dictate whether an industry is essential i.e., whether an industry can operate during a lockdown even if working from home is not possible. For instance, if an industry is non-essential, and none of its employees can work from home, it faces a labor supply reduction of 100\% during lockdown i.e., $\epsilon^S_{i,t}=1, \ \forall t \in [t_\text{start\_lockdown}, t_\text{end\_lockdown}$. Instead, if an industry is classified as fully essential, it faces no labor supply shock and $\epsilon^S_{i,t}=0 \ \forall t$. In some scenarios, we further refine the supply shocks estimates by taking into account the difficulty of adjusting to social distancing measures. To do this refinement we use the Physical Proximity work context provided by O*NET, as others have done \citep{mongey2020workers,koren2020business}.

To estimate the shocks we define a \emph{Remote Labor Index}, an \emph{Essential Score} and a \emph{Physical Proximity Index} at the WIOD industry level. We interpret these indices as follows. The Remote Labor Index of industry $i$ is the probability that a worker from industry $i$ can work from home. The Essential Score is the probability that a worker from industry $i$ has an essential job. The Physical Proximity Index is the probability that an essential worker that cannot work from home cannot go to work due to social distancing measures. In Appendix \ref{apx:shocks} we explain how, similarly to \cite{del2020supply,dingel2020,gottlieb2020working,koren2020business}, we use O*NET data to estimate our Remote Labor Index and Physical Proximity Index. Below we explain each scenario in more detail and how we determine the essential score in each scenario. Figure \ref{fig:supplyscenarios} in Appendix \ref{apx:shocks} shows time series of supply shocks $\epsilon^S_{i,t}$ for all scenarios throughout our simulations, while Table \ref{tab:FO_shocks_supply} shows the cross section of shocks.

\paragraph{$\text{S}_\text{1}$: UK policy.} 
In contrast to some European countries such as Italy and Spain, in the UK shutdown orders were only issued for a few industries\footnote{\label{footn:ukleg}
https://www.legislation.gov.uk/uksi/2020/350/pdfs/uksi\_20200350\_en.pdf
}. Although social distancing guidelines were imposed for all industries (see $\text{S}_\text{2}-\text{S}_\text{4}$ below), strictly speaking only non-essential retail, personal and recreational services, and the restaurant and hospitality industries were mandated to shut down. While some European countries shut down some manufacturing sectors and construction, the UK did not explicitly forbid these sectors from operating. 

Based on the UK regulations, we assume that all WIOD industries have an essential score of one with the exception of industries G45 and G47 (vehicle and general retail), I (hotels and restaurants) and R\_S (recreational and personal services). We break down these industries into smaller subcategories that we can directly match to shutdown orders in the UK, and compute an essential score from a weighted average of these subcategories, where weights correspond to output shares (see Appendix \ref{apx:shocks} for more details). The resulting essential scores are 0.64 for G45, 0.71 for G47, 0.05 for I and 0.07 for R\_S.

Similarly to \cite{del2020supply}, we assume an industry's supply shock is given by the fraction of workers that cannot work. If we interpret the Remote Labor Index and the essential score as independent probabilities, the expected value of the fraction of workers of industry $i$ that cannot work is 
$$\epsilon^S_{i,t} = (1 - \text{RLI}_i)(1 - \text{ESS}_i) \quad \forall t \in [t_\text{start\_lockdown}, t_\text{end\_lockdown}),$$

where $\text{RLI}_i$ and $\text{ESS}_i$ are the Remote Labor Index and Essential Score of industry $i$ respectively. 
We lift labor supply shocks to trade industries on June $15^\text{th}$ and shocks to other sectors in July, as per official guidance. That is, $ \epsilon^S_{i,t} = 0, \quad \forall t \in [t_\text{end\_lockdown}, \infty).$

\paragraph{$\text{S}_\text{2}$, $\text{S}_\text{3}$, $\text{S}_\text{4}$ : UK policy + difficulty to adapt to social distancing.} In monthly communications on the impact of Covid-19 on the UK economy\footnote{
\label{ft1}See, e.g., \url{https://www.ons.gov.uk/economy/grossdomesticproductgdp/articles/coronavirusandtheimpactonoutputintheukeconomy/may2020}
}, the ONS reported that several manufacturing industries and the construction sector were struggling to comply with the requirements on social distancing imposed by the government. Thus, further to the legal constraints considered in scenario $\text{S}_\text{1}$, we also consider the practical constraints of operating under the new guidance. For all industries that were not explicitly forbidden to operate, we consider a supply shock due to the difficulties of adapting to social distancing measures. 

We assume that industries with a higher index of physical proximity have more difficulty to adhere to social distancing, so that only a portion of workers that cannot work from home can actually work at the workplace. In particular, in-person work can only be performed by a fraction of workers that is proportional to the Physical Proximity Index of industry $i$ (see Appendix \ref{apx:shocks} for details). We rescale this index so that it varies in an interval between zero and $\iota$, and consider three values $\iota=0.1,0.4,0.7$. These three values distinguish between scenarios $\text{S}_\text{2}$, $\text{S}_\text{3}$ and $\text{S}_\text{4}$, which are reported in order of severity. At the time when lockdown starts, the supply shocks are given by  
$$\epsilon^S_{i,t_\text{start\_lockdown}} = \left(1 - \text{RLI}_i\right) \left(1 - \text{ESS}_i\left(1 - \iota \frac{\text{PPI}_{i}}{\max_j(\text{PPI}_j)}\right) \right),$$
where $\text{PPI}_{i}$ is the Physical Proximity Index. 

Differently from the scenarios above, we do not assume that shocks due to the difficulty to adapt to social distancing are constant during lockdown. In line with ONS reports, firms are able to bring a larger fraction of their workforce back to work as they adapt to the new guidelines. For simplicity, we assume that these shocks vanish when lockdown is lifted (May $13^\text{th}$), and interpolate linearly between their maximum value (which is when lockdown is imposed, March $23^\text{rd}$) and the time when lockdown is lifted. This leads to the following supply shocks
$$\epsilon^S_{i,t} = \left(1 - \text{RLI}_i\right) \left(1 - \text{ESS}_i\left(1 - \iota \frac{\text{PPI}_{i,t}}{\max_j(\text{PPI}_j)}\right) \right),$$
where 
$$
\text{PPI}_{i,t} = \text{PPI}_{i} \left(1 - \frac{t - t_\text{start\_lockdown}}{t_\text{end\_lockdown}- t_\text{start\_lockdown}} \right).
$$

Note that lockdown did not have a clear end date in the UK. However, we conventionally take May $13^\text{th}$ as $t_\text{end\_lockdown}$. This is the day when the UK government asked all workers to go back to work, and we assume that at this time most firms had had sufficient time to comply to social distancing guidelines.

\paragraph{$\text{S}_\text{5}$: Original shocks.}

This baseline scenario corresponds to the original shocks $\epsilon^S_{i,t}$ we used in our previous work \citep{pichler2020production} and was based on the estimates by \cite{del2020supply}.  The supply shocks were estimated by quantifying which work activities of different occupations can be performed from home based on the Remote Labor Index and using the occupational compositions of industries. The predictions also considered whether an industry was essential in the sense that on-site work was allowed during the lockdown. We then compiled a list of essential industries within the NAICS classification system, based on the list of essential industries provided by the Italian government using the NACE classification system. Then, we computed an essential score $\text{ESS}_i$ for each industry $i$ in their sample of NAICS industries and calculated the supply shock using the following equation
\begin{equation}
   \epsilon^S_{i} = (1 - \text{RLI}_i)(1 - \text{ESS}_i).
    \label{eq:supply_shock_original}
\end{equation}

Following our original work \citep{pichler2020production}, we map these supply shocks from the NAICS 4-digit industry classification system to the WIOD classification using a NAICS-WIOD crosswalk. To deal with one-to-many and many-to-one maps in these crosswalks, we split each NAICS industry's contributions using employment data (see Appendix \ref{apx:shocks} for details).  We follow this approach to weigh each worker in the NAICS industries' sample from \cite{del2020supply} equally. In  Appendix \ref{apx:shocks} we discuss the implications of this crosswalk methodology in more detail. 
Finally, for the Real Estate sector, we assume that the supply shock does not apply to imputed rents (which represent about 2/3 of gross output).

\paragraph{$\text{S}_\text{6}$: European list of essential industries.} For comparison, we also consider the list of essential industries produced by \cite{fana2020covid}. This list was compiled independently from \cite{del2020supply}, and listed which industries were considered essential by governments of Spain, Germany and Italy. Using this list, we compute the essential score for each industry by taking the mean across essential scores for the three countries considered. We keep labor supply shocks constant during lockdown, and lift them all when lockdown ends. This scenario produces the largest supply shocks of all. For comparison, the average supply shock for $\text{S}_\text{1}$ is 3\%, while the average shock for $\text{S}_\text{6}$ is 28\%.

\subsection{Demand shocks}
\label{sec:consumptiondemandshockscenarios}

The Covid-19 pandemic caused strong shocks to all components of demand. We consider shocks to private consumption demand, which we further distinguish into shocks due to fear of infection and due to fear of unemployment, and shocks to other components of final demand, such as investment, government consumption and exports. 
We outline the basic assumptions on demand shocks below and show in in Appendix \ref{apx:shocks}
the detailed cross-sectional and temporal shock profiles. We further demonstrate in Appendix \ref{apx:sensitivity} that alternative plausible demand shock assumptions only mildly influence model results.

\paragraph{Demand shocks due to fear of infection.} During a pandemic, consumption/saving decisions and consumer preferences over the consumption basket are changing, leading to first-order demand shocks \citep{CBO2006, del2020supply}. For example, consumers are likely to demand less services from the hospitality industry, even when the hospitality industry is open. Transport is very likely to face substantial demand reductions, despite being classified as an essential industry in many countries. A key question is whether reductions in demand for ``risky'' goods and services is compensated by an increase in demand for other goods and services, or if lower demand for risky goods translates into higher savings.

We consider a demand shock vector ${\epsilon}^D_t$, whose components $\epsilon_{i,t}^D$ are the relative changes in demand for goods of industry $i$ at time $t$. Recall from Eq.~\eqref{eq:cd}, $c^d_{i,t}= \theta_{i,t} \Tilde{c}^d_t$, that consumption demand is the product of the total consumption scalar $\Tilde{c}^d_t$ and the preference vector $\theta_t$, whose components $\theta_{i,t}$ represent the share of total demand for good $i$. 
We initialize the preference vector by considering the initial consumption shares, that is $\theta_{i,0}=c_{i,0}/\sum_j c_{j,0}$. 
By definition, the initial preference vector $\theta_{0}$ sums to one, and we keep this normalization at all following time steps. To do so, we consider an auxiliary preference vector $\bar{\theta}_t$, whose components $\bar{\theta}_{i,t}$ are obtained by applying the shock vector $\epsilon^D_{i,t}$. That is, we define $\bar{\theta}_{i,t}=\theta_{i,0} (1-\epsilon^D_{i,t})$ and define $\theta_{i,t}$ as
\begin{equation} \label{eq:theta}
    \theta_{i,t}= 
    \frac{\bar{\theta}_{i,t}}{\sum_j \bar{\theta}_{j,t}} = 
    \frac{ (1-\epsilon^D_{i,t}) \theta_{i,0} }{\sum_j  (1-\epsilon^D_{j,t}) \theta_{j,0} } .
\end{equation} 

The difference $1-\sum_i \bar{\theta}_{i,t}$ is the aggregate reduction in consumption demand due to the demand shock, which would lead to an equivalent increase in the saving rate. However, households may not want to save all the money that they are not spending. For example, they most likely want to spend on food the money that they are saving on restaurants. Therefore, we define the aggregate demand shock $\Tilde{\epsilon}^D_t$ in Eq.~\eqref{eq:consdemand} as 
\begin{equation} \label{eq:epsilon}
\Tilde{\epsilon}_t^D=\Delta s \left(1- \sum_{i=1}^N \bar{\theta}_{i,t} \right) ,
\end{equation}
where $\Delta s$ is the change in the savings rate. When $\Delta s=1$, households save all the money that they are not planning to spend on industries affected by demand shocks; when $\Delta s=0$, they spend all that money on goods and services from industries that are affected less.

To parameterize $\epsilon^D_{i,t}$, we adapt consumption shock estimates by the \cite{CBO2006} and \cite{del2020supply}. 
Roughly speaking, these shocks are massive for restaurants and transport, mild for manufacturing and null for utilities.
We make two modifications to these estimates. First, we remove the positive shock to the health care sector, as in the UK the cancellation of non-urgent treatment for other diseases than Covid-19 far exceeded the additional demand for health due to Covid-19.\footnote{
\url{https://www.ons.gov.uk/economy/grossdomesticproductgdp/bulletins/gdpfirstquarterlyestimateuk/apriltojune2020}
} 
Second, we apportion to manufacturing sectors the reduced demand due to the closure of non-essential retail. For example, retail shops selling garments and shoes were mandated to shut down, and so we apply a consumption demand shock to the manufacturing sector producing these goods.\footnote{
To be fully consistent with the definition of demand shock, we should model non-essential retail closures as supply shocks, and propagate the shocks to manufacturing through reduced intermediate good demand. However, there are two practical problems that prevent us to do so: (i) the sectoral aggregation in the WIOD is too coarse, comprising only one aggregate retail sector; (ii) input-output tables only report margins of trade, i.e. they do not model explicitly the flow of goods from manufacturing to retail trade and then from retail trade to final consumption. Given these limitations, we conventionally interpret non-essential retail closures as demand shocks.
} 

We keep the intensity of demand shocks constant during lockdown. We then reduce demand shocks when lockdown is lifted according to the situation of the Covid-19 pandemic in the UK. In particular, we assume that consumers look at the daily number of Covid-19 deaths to assess whether the pandemic is coming to an end, and that they identify the end of the pandemic as the day in which the death rate drops below 1\% of the death rate at the peak. Given official data\footnote{\url{https://coronavirus.data.gov.uk/}}, this happens on August $11^\text{th}$. Thus, we reduce $\epsilon^D_{i,t}$ from the time lockdown is lifted (May $13^\text{th}$) by linearly interpolating between the value of $\epsilon^D_{i,t}$ during lockdown and $\epsilon^D_{i,t}=0$ on August $11^\text{th}$. The choice of modeling behavioral change in response to a pandemic by the death rate has a long history in epidemiology \citep{funk2010modelling}.

\paragraph{Demand shocks due to fear of unemployment.}

A second shock to consumption demand occurs through reductions in current income and expectations for permanent income. 

Reductions in current income are due to firing/furloughing, due to both direct shocks and subsequent upstream or downstream propagation, resulting in lower labor compensation, i.e. $\tilde{l}_t < \tilde{l}_0$, for $t\geq t_\text{start\_lockdown}$. To support the economy, the government pays out social benefits to workers to compensate income losses. In this case, the total income $\tilde{l}_t$ that enters Eq.~\eqref{eq:consdemand} is replaced by an effective income $\tilde{l}^\star_t=b \tilde{l}_0 + (1-b) \tilde{l}_t $, where $b$ is the fraction of pre-pandemic labor income that workers who are fired or furloughed are able to retain.

A second channel for shocks to consumption demand due to labor market effects occurs through expectations for permanent income. These expectations depend on whether households expect a V-shaped vs. L-shaped recovery, that is, whether they expect that the economy will quickly bounce back to normal or there will be a prolonged recession. Let expectations for permanent income $\Tilde{l}_t^p$ be specified by
\begin{equation} \label{eq:perm_income}
\Tilde{l}_t^p=\xi_t  \Tilde{l}_0    
\end{equation}
In this equation, the parameter $\xi_t$ captures the fraction of pre-pandemic labor income $\Tilde{l}_0$ that households expect to retain in the long run. We first give a formula for $\xi_t$ and then explain the various cases.
\begin{equation}{\label{eq:xit}}
\xi_t =
\begin{cases}
    1, & t<t_\text{start\_lockdown}, \\
    \xi^L = 1-\frac{1}{2}\frac{\tilde{l}_0-\tilde{l}_{t_\text{start\_lockdown}}}{\tilde{l}_0}, & t \in [t_\text{start\_lockdown}, t_\text{end\_lockdown} ],\\
    1-\rho + \rho\xi_{t-1} + \nu_{t-1}, & t > t_\text{end\_lockdown}.
\end{cases}
\end{equation}
Before lockdown, we let $\xi_t \equiv 1$, i.e. permanent income expectations are equal to current income. 
During lockdown, following \cite{muellbauer2020} we assume that $\xi_t$ is equal to one minus half the relative reduction in labor income that households experience due to the direct labor supply shock, and denote that value by $\xi^L$. (For example, given a relative reduction in labor income of 16\%, $\xi^L=0.92$.)\footnote{During lockdown, labor income may be further reduced due to firing. For simplicity, we choose not to model the effect of these further firings on permanent income.} After lockdown, we assume that 50\% of households believe in a V-shaped recovery, while 50\% believe in an L-shaped recovery. We model these expectations by letting $\xi_t$ evolve according to an autoregressive process of order one, where the shock term $\nu_t$ is a permanent shock that reflects beliefs in an L-shaped recovery. With 50\% of households believing in such a recovery pattern, it is $\nu_t\equiv-(1-\rho)(1-\xi^L)/2$.\footnote{The specification in Eq.~\eqref{eq:xit} reflects the following assumptions: (i) time to adjustment is the same as for consumption demand, Eq.~\eqref{eq:consdemand}; (ii) absent permanent shocks, $\nu_t=0$ after some $t$, $\xi_t$ returns to one, i.e. permanent income matches current income; (iii) with 50\% households believing in an L-shaped recovery, $\xi_t$ reaches a steady state given by $1-(1-\xi^L)/2$: with $\xi^L=0.92$ as in the example above, $\xi_t$ reaches a steady state at 0.96, so that permanent income remains stuck four percentage points below pre-lockdown income.}

\paragraph{Other final demand shock scenarios}

Note that WIOD distinguishes five types of final demand: 
(I) \textit{Final consumption expenditure by households},	(II) \textit{Final consumption expenditure by non-profit organisations serving households}, 
(III) \textit{Final consumption expenditure by government}	(IV) \textit{Gross fixed capital formation} and
(V) \textit{Changes in inventories and valuables}.
Additionally, all final demand variables are available for every country, meaning that it is possible to calculate imports and exports for all categories of final demand. The endogenous consumption variable $c_{i,t}$ corresponds to (I), but only for domestic consumption. All other final demand categories, including all types of exports, are absorbed into the variable $f_{i,t}$.

We apply different shocks to $f_{i,t}$. We do not apply any exogenous shocks to categories (III) \textit{Final consumption expenditure by government}	and (V) \textit{Changes in inventories and valuables}, while we apply the same demand shocks to category (II) as we do for category (I). To determine shocks to investment (IV) and exports we start by noticing that, before the Covid-19 pandemic, the volatility of these variables has generally been three times the volatility of consumption.\footnote{
This is computed by calculating the standard deviation of consumption, investment and export growth over all quarters from 1970Q1 to 2019Q4. These are 1.03\%, 2.87\% and 3.24\% respectively. Source: \url{https://www.ons.gov.uk/file?uri=\%2feconomy\%2fgrossdomesticproductgdp\%2fdatasets\%2frealtimedatabaseforukgdpcomponentsfortheexpenditureapproachtothemeasureofgdp\%2fquarter2aprtojune2020firstestimate/gdpexpenditurecomponentsrealtimedatabase.xls}
} 
The overall consumption demand shock is around 5\% so, as a baseline, we assume shocks to investment and exports to be 15\%. In Appendix \ref{apx:sensitivity} we show that the model results are fairly robust with respect to alternative choices.


\section{Economic impact of Covid-19 on the UK economy}
\label{sec:econimpact}

As already mentioned, in the first version of this work we released results in May predicting a 21.5\% contraction of GDP in the UK economy in the second quarter of 2020 with respect to the last quarter of 2019\footnote{%
Due to an error in how we dealt with Real Estate shocks, our prediction was slightly worse than it would have been without the error, see Appendix \ref{apx:shocks}
}. This is in comparison to the contraction of 22.1\% that was actually observed.  In this section we do a post-mortem to understand the factors that influenced the quality of the forecasts. To do this, we compare results under different scenarios defined by different shocks and production functions. Our analysis includes a sectoral breakdown of the forecasts and a comparison of the time series of observed vs. predicted behavior.

\subsection{Definition of scenarios and calibration}

The model and shock scenarios that we described in the previous sections has several degrees of freedom that can be tuned when exploring the model. These are either model assumptions such as the production function, shock scenarios, or parameters (see Table \ref{tab:econmodelpars}). To understand the factors that influenced the quality of the forecasts, we focus on the assumptions and parameters that cannot easily be calibrated from data and that have a strong effect on the results.

As the sensitivity analysis in Appendix \ref{apx:sensitivity} shows, the model is most sensitive to two assumptions: the supply shock scenario and the production function. So, we consider all combinations of the six scenarios for supply shocks, $\text{S}_\text{1}$ to $\text{S}_\text{6}$ (Section \ref{sec:supplyshockscenarios}), and of the five production functions mentioned in Section \ref{sec:supply}. These are: standard Leontief, three versions of the IHS-Markit-modified Leontief that treat important inputs as critical (IHS1), half-critical (IHS2), and non-critical (IHS3) and the linear production function. Combining these two assumptions, we get $6\times5=30$ scenarios that we compare to the data.

\begin{table}[htbp]
	\centering
	\caption{Assumptions and parameters of the model that: (top) are varied across scenarios; (middle) are fixed due to little effect on results; (bottom) are fixed due to direct data calibration. }
		\begin{tabular}{|l|c|c|}
			\hline
			 Name & Symbol & Value \\
			 \hline
			 Supply shocks & S &  $\text{S}_\text{1}$, $\text{S}_\text{2}$, $\text{S}_\text{3}$, $\text{S}_\text{4}$, $\text{S}_\text{5}$, $\text{S}_\text{6}$ \\
			 Production function & & Leontief, IHS1, IHS2, IHS3, linear \\
			 \hline
			 Final demand shocks &  & Appendix \ref{apx:shocks} \\
			 Inventory adjustment & $\tau$ &10 \\
			 Upward labor adjustment & $\gamma_H$ & 1/30 \\
			 Downward labor adjustment & $\gamma_F$ & 1/15 \\
			 Change in savings rate & $\Delta s$ & 0.50 \\
			 Consumption adjustment & $\rho$ & 0.99 \\
             \hline
			 Inventory targets & $n_i$ & Appendix \ref{apx:inventory} \\
			 Propensity to consume & $m$ & 0.82 \\
			 Government benefits & $b$ & 0.80 \\
			\hline						
		\end{tabular}
	\label{tab:econmodelpars}
\end{table}

Other parameters have less effect on the model (Appendix \ref{apx:sensitivity}) and so we fix them to reasonable values. These are:
\begin{itemize}
    \item The parameter $\tau$, capturing responsiveness to inventory gaps. We fix $\tau=10$ days, which indicates that firms aim at filling most of their inventory gaps within two weeks. This lies in the range of values used by related studies (e.g. $\tau=6$ in \cite{inoue2019firm}, $\tau=30$ in \cite{hallegatte2014modeling}). 
    \item The hiring and firing parameters $\gamma_{\text{H}}$ and $\gamma_{\text{F}}$. We choose $\gamma_{\text{H}}=1/30$ and $\gamma_{\text{F}} = 2 \gamma_{\text{H}}$. Given our daily time scale, this is a rather rapid adjustment of the labor force, with firing happening faster than hiring.
    \item The parameter  $\rho$, indicating sluggish adjustment to new consumption levels. We select the value assumed by \cite{muellbauer2020}, adjusted for our daily timescale\footnote{ 
Assuming that a time step corresponds to a quarter, \cite{muellbauer2020} takes $\rho=0.6$, implying that more than 70\% of adjustment to new consumption levels occurs within two and a half quarters. 
We modify $\rho$ to account for our daily timescale: By letting $\bar{\rho}=0.6$, we take $\rho=1-(1-\bar{\rho})/90$ to obtain the same time adjustment as in \cite{muellbauer2020}. Indeed, in an autoregressive process like the one in Eq.~\eqref{eq:consdemand}, about 70\% of adjustment to new levels occurs in a time $\iota$ related inversely to the persistency parameter $\rho$. Letting $Q$ denote the quarterly timescale considered by \cite{muellbauer2020}, time to adjustment $\iota^Q$ is given by $\iota^Q=1/(1-\bar{\rho})$. Since we want to keep approximately the same time to adjustment considering a daily time scale, we fix $\iota^D=90\iota^Q$. We then obtain the parameter $\rho$ in the daily timescale such that it yields $\iota^D$ as time to adjustment, namely $1/(1-\rho)=\iota^D=90\iota^Q=90/(1-\bar{\rho})$. Rearranging gives the formula that relates $\rho$ and $\bar{\rho}$.
}. 
\item The savings parameter $\Delta s$. We take $\Delta s = 0.5$, meaning that households save half the money they are not spending in goods and services due to fear of infection, and direct half of that money to spending for other ``safer'' goods and services.
\end{itemize}

Finally, we are able to directly calibrate some parameters against the data. For example, we calibrate the inventory target parameters $n_i$ using ONS data for the usual stock of inventories that different industries typically have (see details in Appendix \ref{apx:inventory}). These parameters are highly heterogeneous across industries; typically manufacturing and trade have much higher inventory targets than services. Another parameter which we can directly calibrate from data is the propensity to consume $m$ (see Eq.~\ref{eq:consdemand}). Directly reading the share of labor income that is used to buy final domestic goods from the input-output tables, we find $m=0.82$. Finally, we calibrate benefits $b$ based on official UK policy, $b=0.8$.

\subsection{Comparing our initial forecast with alternative scenarios}
\label{sec:scenarioselection}

We have chosen not to search for a calibration of the model that best fits the data. We do this because we only have one real world example and this would lead to overfitting. Instead we start from the out-of-sample forecast we made in May and try to understand how performance would have changed if we had made different choices.  This helps us understand what is important in determining the accuracy and gives some insight into how the model works, and where care needs to be taken in making forecasts of this type.

Since our initial forecast, we have made small changes to the model (in particular the consumption function), and obtained better data to calibrate inventories (using ONS rather than US BEA data). Our initial forecast featured a supply shock scenario identical to S5 except for Real Estate (Section \ref{sec:supplyshockscenarios} and Appendix \ref{apx:shocks}), and an IHS3 production function.  In this section, we compare our original forecast to the forecasts made using the 30 possible scenarios discussed above. To do so, we evaluate the forecast errors for each scenario at both the sectoral and aggregate levels, allowing us to better understand the trade-off involved in selecting amongst various assumptions and calibrations. 

For all 30 scenarios, we simulate the model for six months, from January $1^{st}$ to June $30^{th}$, 2020. We start lockdown on March $23^{rd}$, at which point we apply the supply and demand shocks described in Section \ref{sec:pandemic_shock}.\footnote{
We do not run the model further in the future, both because we focus on the first UK lockdown and the immediate aftermath, and because our assumptions on non-critical inputs are only valid for a limited time span.
}

We then compare the monthly sectoral output of each scenario against empirical data from the indexes of agriculture, production, construction and services, all provided by the ONS (see Appendix \ref{apx:validation_data}). Specifically, we compute the sector-level Absolute Forecast Errors (AFE) for April, May, and June,\footnote{
We exclude January, February and March from our comparison as we do not model the reaction of the economy before the lockdown, when e.g. international supply chains started to be disrupted.
} 
\[
\mathcal{E}_{i,t}^z=|y_{i,t}-\hat{y}_{i,t}^z|,
\]
where $y_{i,t} = x_{i,t}/x_{i,0}$ is the output of sector $i$ during month $t$ expressed as a percentage of the output of sector $i$ during February, $x_{i,0}$. Here, $\hat{y}_{i,t}^z$ is the equivalent quantity in simulations of scenario $z$.  We then obtain a scenario-specific average sectoral AFE by taking a weighted mean of $\mathcal{E}_{i,t}^z$ across all sectors $i$ and months $t$, where weights correspond to output shares in the steady state (forecast errors for important sectors are more relevant than forecast errors for small sectors). This quantity is defined as 
\begin{equation}
    \text{AFE}_\text{sec}^z= \frac{1}{3N} \sum_t \sum_i \frac{x_{i,0}}{\sum_j x_{j,0}} \mathcal{E}_{i,t}^z.
\end{equation}

We also compare aggregate output reductions in all different scenarios in April, May and June against empirical data. We compute a scenario-specific average aggregate AFE by averaging aggregate forecast errors over the three months we are considering.  This is 
\begin{equation}
    \text{AFE}_\text{agg}^z= \frac{1}{3} \sum_t \left(\sum_i x_{i,t}-\sum_i \hat{x}_{i,t}^z\right),
\end{equation}
where $\hat{x}_{i,t}^z$ is output of sector $i$ at time $t$ for scenario $z$.  Note that, unlike the measure of sectoral error $\text{AFE}_\text{sec}^z$,  the aggregate measure $\text{AFE}_\text{agg}^z$ does not include an absolute value -- it is positive when true production is greater than predicted, and negative when it is smaller than predicted.

\begin{figure}[hbt]
    \centering
\includegraphics[width = 1.0\textwidth]{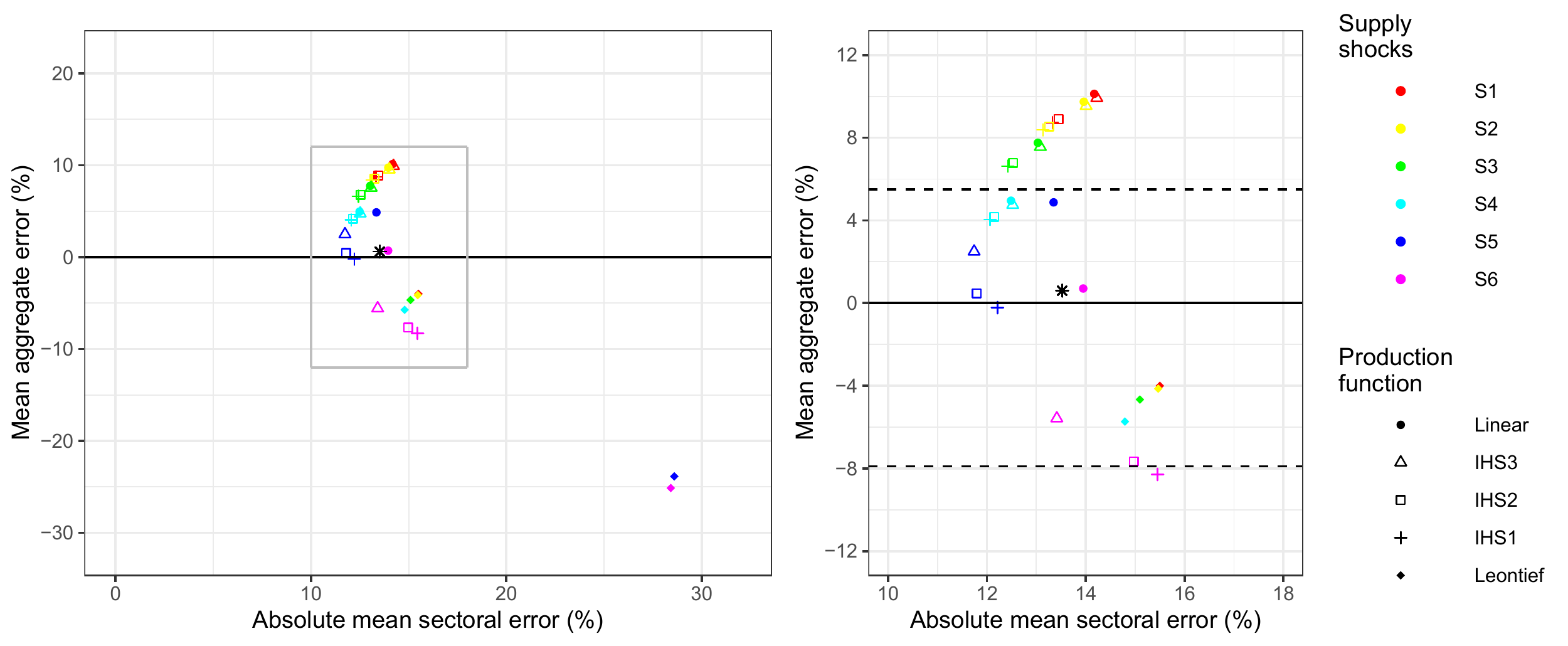}
    \caption{\textbf{Sectoral and aggregate errors across scenarios.} We plot aggregate Average Forecast Error $\text{AFE}^\text{agg}_z$ vs. sectoral Average Forecast Error $\text{AFE}^\text{sec}_z$ for 30 different scenarios $z$, corresponding to all of the six possible shock scenarios (indicated by color) and the five possible production functions (indicated by symbol), as well as the original model forecast, indicated by a black asterisk. The other parameter values are indicated in Table \ref{tab:econmodelpars}. Both sectoral and aggregate AFE are multiplied by 100 to be interpreted as percentages. The right panel zooms on the region with lowest sectoral and aggregate errors. The dashed lines refer to an average forecast by institutions and financial firms (above zero) and to a forecast by the Bank of England (below zero), see footnote \ref{footn:pred}. }
        \label{fig:scenarios_comparison_main}
\end{figure}

Figure \ref{fig:scenarios_comparison_main} plots sectoral and aggregate AFE for the 30 scenarios, plus the original out-of-sample forecast. The sectoral and aggregate errors vary considerably across scenarios. The left panel, which includes all 31 scenarios, contains two outliers where the sectoral error reaches almost 30\% and the aggregate error is roughly -30\%, meaning that the model predicts a downturn 30 percentage points larger than that actually observed (i.e., about a 50\% downturn).  This occurs when the Leontief production function is combined with the two most severe supply shock scenarios $S_5$ and $S_6$. The right panel blows up the region containing the other 28 scenarios.  The sectoral errors range between roughly 10\% and 20\%, while  the aggregate forecast errors range from roughly -10\% to 10\%.   

A close examination of the figure makes it clear that, as expected, the predicted downturn generally gets stronger as the severity of the shock and the rigidity of the production function increases.  With one exception, the choice of scenario has a bigger effect on the error than the production function.  This is evident from the fact that there are clusters of points with the same color associated with each scenario.  The clusters are particularly tight for the less severe scenarios.  The exception is the Leontief production function:  The two most severe shock scenarios produce outliers with downturns that are much too strong and the remainder all produce results clustered together, with aggregate errors in the range of $-4\%$ to $-6\%$.  This is true even for the weakest shock scenario $S_1$; by comparison,  every other production function predicts a downturn that is too weak under scenario $S_1$, in the range of $8\% - 12\%$.  

The original model forecast, shown as a black asterisk, has a aggregate error of 0.6\% and a sectoral error of 13.5\%. It is useful to compare this to its counterpart in the retrospective analysis, which uses the $\text{S}_\text{5}$ scenario in combination with the IHS3 production function. This forecast performs better at the sectoral level, but worse at the aggregate level (the error is 2.5\%). Using an IHS2 production function with the $S_5$ shock scenario seems to provide the best combination of sectoral and aggregate errors.

To put these results in perspective, it is worth comparing to the other out-of-sample forecasts that were made around the same time.  We do not know their sectoral errors but we can compare the aggregate errors.  The Bank of England forecast, for example, predicted a downturn of about 30\%, which corresponds to an aggregate error of about -8\%, comparable to the scenarios with $S_6$ supply shocks and with Leontief production function and supply shocks $S_1$ to $S_4$. Conversely, the average forecast by institutions and financial firms for the UK economy in Q2 2020 was -16.6\%, which is 5.5\% less than what was observed in reality. This forecast is in line with supply shock scenarios $S_3$ to $S_4$, as well as with $S_5$ combined with a linear production function.

Therefore, it seems that the more accurate predictions of our model are obtained by combining shock scenario $S_5$ -- as in our original prediction -- with one of the IHS production functions. Why does this scenario, which was not designed to capture specific features of the UK economy, work so well? The evidence suggests that this is because there were many voluntary firm shutdowns, such as for the car manufacturing industry in the UK. The $S_5$ shock scenario does a better job of identifying industries that we not mandated to shut down, but did so in practice, and seems to better capture the behavioral response to the pandemic than a literal reading of UK regulations.

\subsection{Analysis of the selected scenario}
\label{sec:selectedscenarioanalysis}

Given that it minimizes a combination of aggregate and sectoral error, and given that it is close to the original scenario used for our out-of-sample predictions, we focus on the scenario that combines $S_5$ and the IHS2 production function to illustrate the outcomes of our model in more detail. We start by showing the dynamics produced by the model, and then we evaluate how well the selected scenario explains some aspects of the economic effects of Covid-19 on the UK economy that we did not consider so far.

\begin{figure}[hbt]
    \centering
\includegraphics[width = 1.0\textwidth]{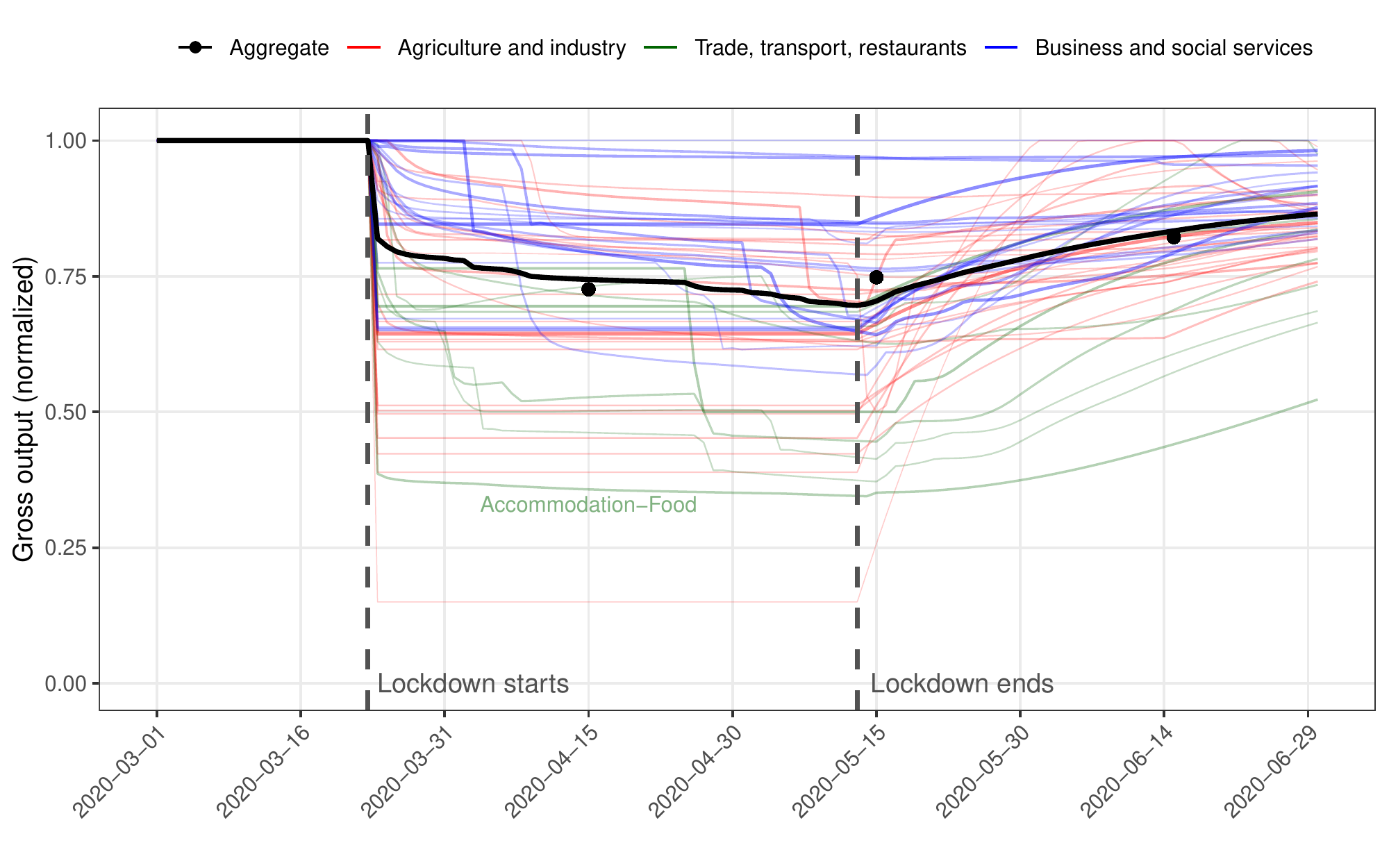}
    \caption{\textbf{Economic production for the chosen scenario as a function of time.} We plot production (gross output) as a function of time for each of the 55 industries.  Aggregate production is a thick black line and each sector is colored. Agricultural and industrial sectors are colored red; trade, transport, and restaurants are colored green; service sectors are colored blue. All sectoral productions are normalized to their pre-lockdown levels, and each line size is proportional to the steady-state gross output of the corresponding sector. For comparison, we also plot empirical gross output as normalized with respect to March 2020.  }
        \label{fig:aggregatevssectoral}
\end{figure}

\begin{figure}[tb]
    \centering
\includegraphics[width = 1.0\textwidth]{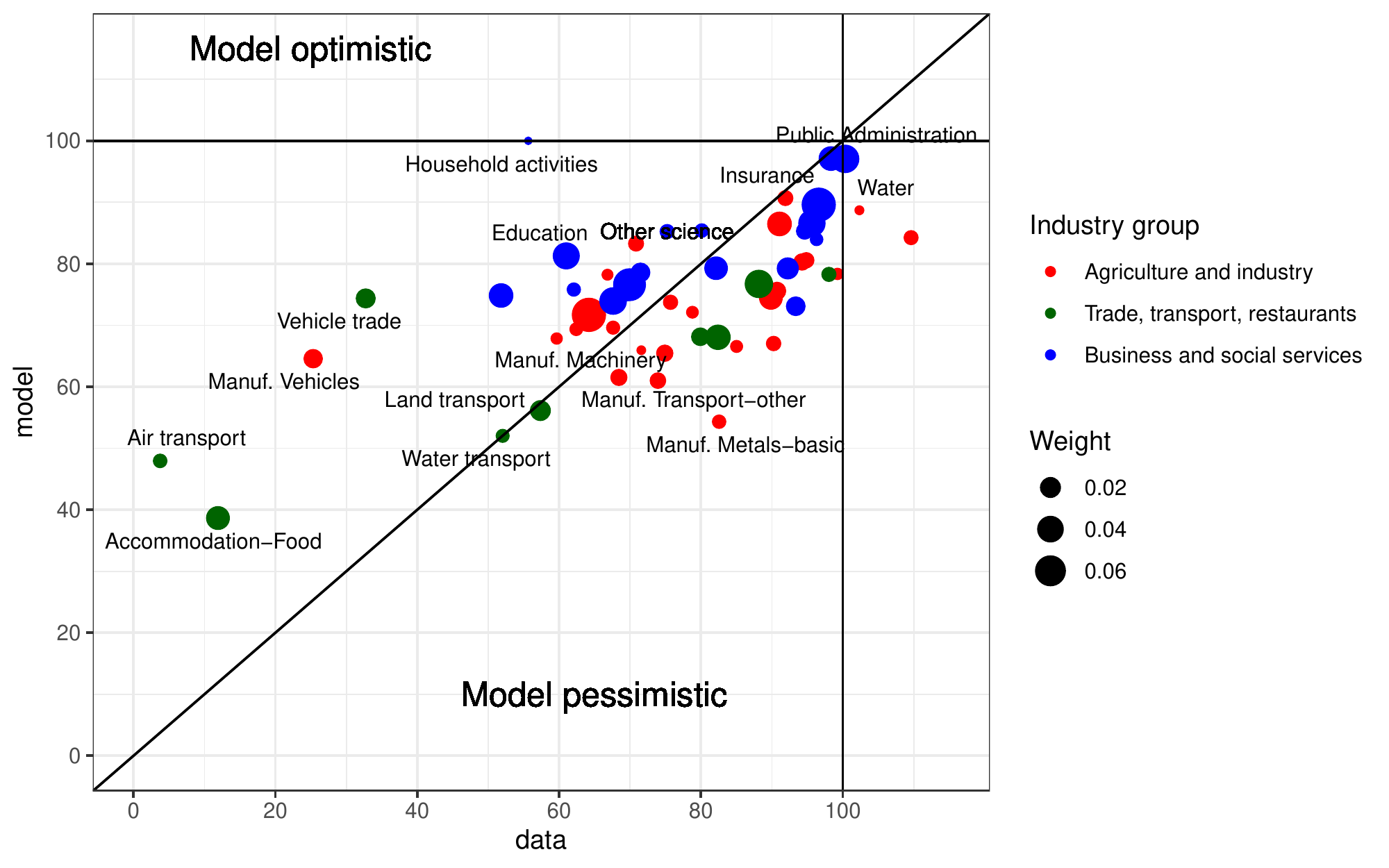}
    \caption{\textbf{Comparison between model predictions and empirical data.} We plot predicted production (gross output) for each of 53 industries against the actual values from the ONS data, averaged across April, May and June. Values are relative to pre-lockdown levels and dot size is proportional to the steady-state gross output of the corresponding sector. Agricultural and industrial sectors are colored red; trade, transport, and restaurants are colored green; service sectors are colored blue.   Dots above the identity line means that the predicted recession is less severe than in the data, while the reverse is true for dots below the identity line.
    }
        \label{fig:model_data_comparison}
\end{figure}

Figure \ref{fig:aggregatevssectoral} shows model results for the selected scenario for production (gross output); results for other important variables, such as profits, consumption and labor compensation (net of government benefits) are similar. When the lockdown starts, there is a sudden drop in economic activity, shown by a sharp decrease in production. Some other industries further decrease production over time as they run out of critical inputs.\footnote{The reduction in production due to input bottlenecks is somewhat limited in this scenario as compared to the outliers in Figure \ref{fig:model_data_comparison}. With supply shocks $S_5$ or $S_6$ and a Leontief production function, the economy collapses by 50\% due to the strong input bottlenecks created in a substantially labor-constrained economy in which all inputs are critical for production.} Throughout the simulation, service sectors tend to perform better than manufacturing, trade, transport and accommodation sectors. The main reason for that is most service sectors face both lower supply and demand shocks, as a high share of workers can effectively work from home, and business and professional services depend less on consumption demand.  

In the UK, there was not a clear-cut lifting of the lockdown but, under scenario $S_5$, we take May 13th as a conventional date in which lockdown measures are lifted (see Figure \ref{fig:supplyscenarios} in Appendix \ref{apx:shocks} for shock dynamics). By the end of June, the economy is still far from recovering.  In part, this is due to the fact that the aggregate level of consumption does not return to pre-lockdown levels, due to a reduction in expectations of permanent income associated with beliefs in an L-shaped recovery (Section \ref{sec:demand}), and due to the fact that we do not remove shocks to investment and exports (see Section \ref{sec:pandemic_shock}).

\begin{table}[htbp]
	\centering
		\begin{tabular}{|l|c|c|}
			\hline
			Variable (compared to Q4-2019) & Data & Model \\
			\hline
			Gross output April & -27.4\% & -25.3\% \\
			Gross output May & -25.2\% & -26.9\% \\
			Gross output June & -17.8\% & -16.8\% \\
			Value added Q2 & -21.5\% & -22.1\% \\
			 \hline
			 Private consumption Q2  & -25.3\% & -21.3\% \\
			 Investment Q2  & -26.3\% & -29.7\% \\
			 Government consumption Q2  & -17.5\% & -14.2\%\\
			 Inventories Q2  & -2.2\% & -0.5\% \\
			 Exports Q2  & -23.3\% & -27.8\% \\
			 Imports Q2  & -30.6\% & -23.9\% \\
			 \hline
			 Wages and Salaries Q2  & -1.1\% & -4.3\% \\
			 Profits Q2  & -26.7\% & -22.3\%\\
			\hline						
		\end{tabular}
		\caption{Comparison between data and predictions of the selected scenario for the main aggregate variables. All percentage changes refer to the last quarter of 2019, which we take to represent the pre-pandemic economic situation.}
	\label{tab:aggregate-data-model}
\end{table}

We now turn to evaluating how well the selected scenario describes the economic effects of Covid-19 on the UK economy. In terms of gross output, one can see in Table \ref{tab:aggregate-data-model} that the model slightly underestimates the recession in April and slightly overestimates it in May, while it correctly estimates a strong recovery in June.\footnote{
We aggregate empirical output from sectoral indexes using our steady-state output shares as weights.
}  Additionally,  aggregate value added in the second quarter of 2020 is very close to the data.

Our model, however, considers other macroeconomic variables than gross output and value added, so we also compare these other variables to data (Table \ref{tab:aggregate-data-model}). From national accounts, we collect data on private and government consumption, investment, change in inventories, exports and imports (expenditure approach to GDP); wages and salaries and profits (income approach to GDP). Looking at the expenditure approach to GDP, some variables have a worse reduction in the model, such as investment and exports, while other variables have a worse reduction in the data, such as private and government consumption, inventories and imports.\footnote{If one considers Exports-Imports as a component, the model predicts a current account deficit, while in reality there was a current account surplus. We should however note that we do not model international trade, we treat exports as exogenous and imports like locally produced goods and services.}  However, the model predicts the relative reductions fairly well, as we find a stronger collapse in private consumption and investment than in government consumption or inventories, as in the data. Finally, considering the income approach to GDP, we overestimate the reduction in wages and salaries and underestimate the reduction in profits.\footnote{
Note that these categories are not jointly exhaustive, as the ONS also considers mixed income and taxes less subsidies, which are difficult to compare to variables in our model.
} Nonetheless, we correctly predict that the absolute reduction in wages and salaries is much smaller than in profits (due to government subsidies).

Turning to the performance of our model at the disaggregate level of industries, Figure \ref{fig:model_data_comparison} shows gross output as predicted by the model and in the data. Here, gross output is averaged over the values it takes in March, April and June, both in the model and in the data, and compared to the value it had in Q4 2019. To interpret this figure, note that for all points on the left of the identity line, model predictions are lower than in the data, i.e. the model is pessimistic. Conversely, on the right of the identity line model predictions are higher than in the data, i.e. the model is optimistic. Note that model predictions and empirical data are correlated, although not perfectly: the Pearson correlation coefficient weighted by industry size is 0.75.  The majority of sectors decreased production up to 60\% of initial levels, both in the model and in the data, but a few sectors were forced to decrease production much more. 

While the predictions are generally very good, there are also some dramatic failures.  We conjecture that this is due to idiosyncratic features of these industries that we could not take into account without overly complicating our model. For example, one sector for which the predictions of our model are completely off is C29 - Manufacturing of vehicles. Almost all car manufacturing plants were completely closed in the UK in April and May, and so production was essentially zero (7\% of the pre-pandemic level in April and 14\% in May). While they reopened in June, production in Q2 is slightly above 20\% of the pre-pandemic level. However, our model predicts a level of production around 63\% of the pre-pandemic level. It is difficult to account for the complete shutdown of car manufacturing plants in our model, as our selected scenario for supply shocks does not require manufacturing plants to completely close during lockdown. We think that this discrepancy between model predictions and data can be explained by two factors. First, car manufacturing is highly integrated internationally, and, in a period where most developed countries were implementing lockdown measures, international supply chains were highly disrupted. For simplicity, however, in our model we did not model input bottlenecks due to lack of imported goods. Second, it is possible that firms producing non-essential goods voluntarily decided to stop production to protect the health of their workers, even if they were not forced to do so. Another example for which the predictions of our model are off is H51 -  Air transport. Production in the data is 3\% of pre-lockdown levels, while our model predicts around 50\%. In the model, most activity of the air transport industry during lockdown is due to business travel, which is a non-critical input to many industries that we do not exogenously shut down (recall that industries aim at using non-critical inputs if they are available).

In some other cases, however, our model gave accurate predictions, even when the answer was far from obvious. Compare, for instance, industries M74\_M75 (Other Science) and O84 (Public Administration). Both received a very weak supply shock (3\% for Other Science and 1.1\% for Public Administration, see Table \ref{tab:FO_shocks_supply} in Appendix \ref{apx:shocks}), and no private consumption demand shocks. Yet, Public Administration had almost no reduction in production, while Other Science reduced its production to about 75\% of its pre-pandemic level. The ONS report in May (see footnote \ref{ft1}) quotes reduced intermediate demand as the reason why Other Science reduced its activity. Conversely, Public Administration's output is almost exclusively sold to the government, which did not reduce its consumption. Because of its ability to take into account supply chain effects and the resulting reductions in intermediate demand, our model is able to endogenously capture the difference between these two sectors, even though their shocks are small in both cases. 

\begin{figure}[tb]
    \centering
\includegraphics[width = 1.0\textwidth]{fig/new_model_april_june_all.pdf}
    \caption{\textbf{Comparison between model predictions and empirical data.} We plot predicted production (gross output) vs. observed values from ONS data.  Production is relative to pre-lockdown levels and dot size is proportional to the steady-state gross output of the corresponding sector.  Yellow is April and red is June. Black lines connect the same industry from April to June. Only a few industries are highlighted (see main text).
    }
        \label{fig:new_model_april_may_june_all}
\end{figure}

Figure \ref{fig:new_model_april_may_june_all} shows the ability of the model to predict sectoral dynamics.  It is similar to Figure \ref{fig:model_data_comparison} but shows output in both April and June.  The dots that represent the same industry in April and June are connected by a black line. We focus on a few industries that we discuss in this section, making all other points light grey (Figure \ref{fig:new_model_april_may_june_facets} in Appendix \ref{apx:selectedscenario} shows labels for all industries).  To interpret changes from April to June, note that a line close to vertical implies that a given industry had a much stronger recovery in the data than in the model, while a horizontal line implies the opposite. A line parallel to the identity line indicates that the recovery was as strong in the data as in the model. Almost all sectors experience a substantial recovery from April to June, both in the model and in the data.

 An example in which our model correctly predicts dynamic supply chain effects is the recovery experienced by C23 (Manufacture of other non-metallic mineral products) as a consequence of the recovery by F (Construction). According to ONS reports, increased activity in construction in June is explained by the lifting of the lockdown and by adaptation to social distancing guidelines by construction firms. At the same, industry C23 recovers due to the production of cement, lime, plaster, etc to satisfy intermediate demand by construction (construction is by far the main customer of C23, buying almost 50\% of its output). This pattern is faithfully reproduced by the endogenous dynamics in our model.

\section{Propagation of supply and demand shocks}
\label{sec:nw_effects}

In the standard Cobb-Douglas Equilibrium IO model, productivity shocks propagate downstream and demand shocks propagate upstream \citep{carvalho2019production}. Thus, the elasticity of aggregate output to a shock to one sector depends on the type of shock, and on the position of an industry in the input-output network. What can we say about these questions in our model? Are there properties of an industry that could be computed ex-ante to know how systemic it is? Is this different for supply and demand shocks?

To answer these questions, we run the model with a single shock -- either supply or demand -- to a single industry. 
All the other industries do not experience any shocks but have initial productive capacities and face initial levels of final demand.
We then let the economy evolve under this specific setting for one month\footnote{
We also did the analysis with model simulations up to two months after the initial shock is applied. Since results are similar for the two cases, we only report results for the one month simulations.
}. 
We repeat this procedure for every industry, every production function specification and different shock magnitudes. We then investigate if the decline in total output can be explained by simple measures such as shock magnitude, output multipliers or upstreamness levels which we formally define below. We first explain the supply and demand shock based scenarios in somewhat more detail.

\paragraph{Supply shock scenarios.}
When considering supply shocks only, we completely switch off any adverse demand effects, i.e. $\epsilon_{i,t}^D = 0$ (which implies $\theta_{i,t} = \theta_{i,0}$ and $\tilde \epsilon_t^D = 0$), $\xi_t = 1$ and $f^d_{i,t} = f^d_{i,0}$ for all $i$ and $t$. 
We also set all supply shocks equal to zero $\epsilon_{i,t}^S = 0$, except for a single industry $j$ which experiences a supply shock from the set $\epsilon_{j,t}^S \in \{0.1,0.2,...,1\}$. We then loop over every possible $j$. We do this for each of the different production function assumptions.

\paragraph{Demand shock scenarios.}
In our demand shock scenarios there are no supply shocks ($\epsilon_{i,t}^S = 0$ $\forall \; i,t$) and similarly there is no demand shocks for all but one industry $j$ ($\epsilon_{i,t}^D = 0$, $f^d_{i,t} = f^d_{i,0}$ $ \forall i \ne j, \forall t$).
For the single industry $j$ we again let shocks vary between 10 and 100\%; $\epsilon_{j,t}^D \in \{0.1,0.2,...,1\}$. For simplicity we assume uniform shocks across all final demand categories of the given industry, resulting in $f^d_{j,t} = (1-\epsilon_{j,t}^D) f^d_{j,0}$. To keep things as simple as possible, we further assume that there is no fear of unemployment $\xi_t = 1$ and that final consumers do not switch to alternative products at all ($\Delta s = 1$).  
Under these assumptions the values for $\theta_{i,t} = \theta_{i,t}$ and $\tilde \epsilon_t^D$ are then computed as outlined in Section \ref{sec:consumptiondemandshockscenarios}.
\newline

\begin{figure}[H]
    \centering
\includegraphics[width = \textwidth]{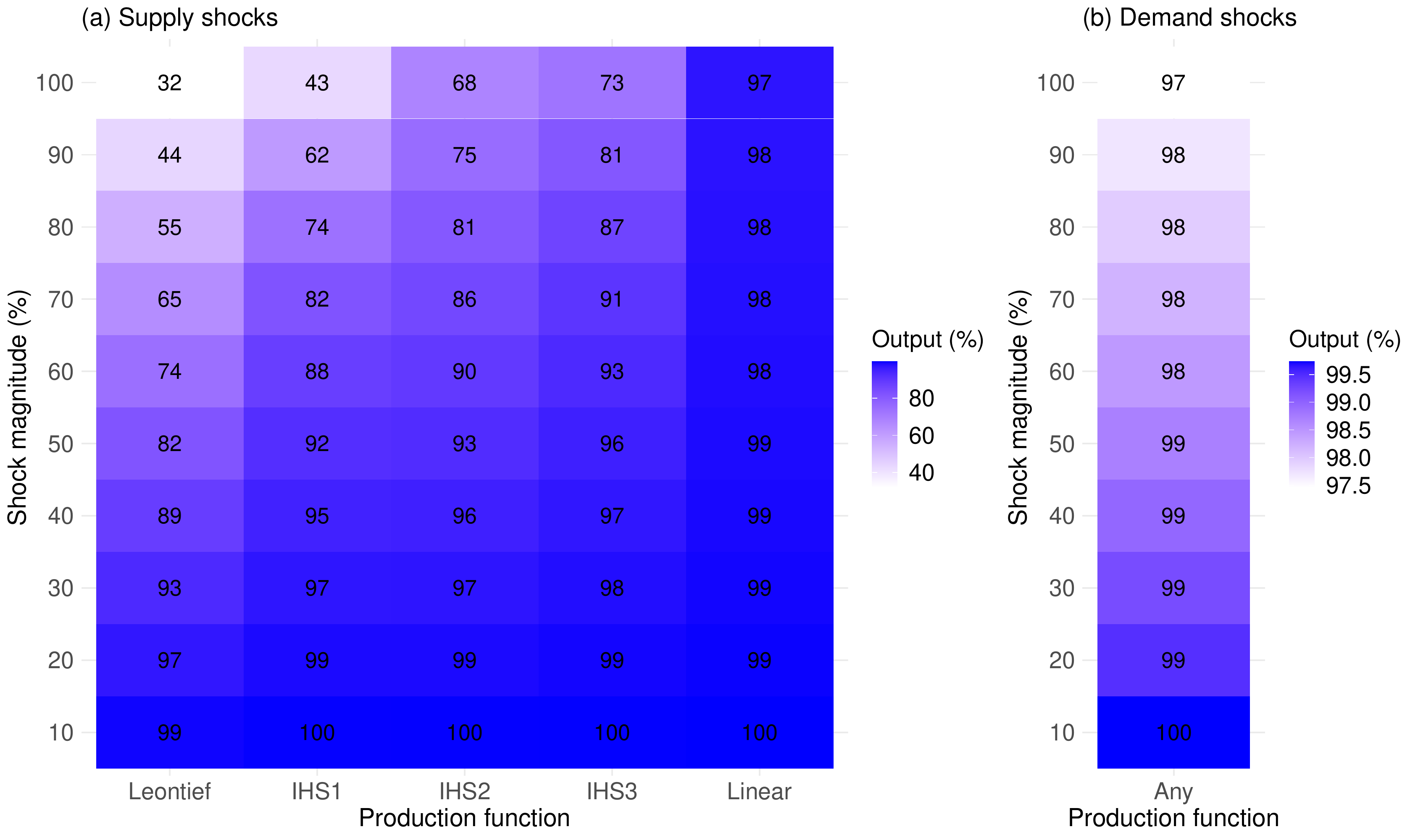}
    \caption{
\textbf{Aggregate gross output as percentage of pre-shock levels after shocking single industries.} 
A column depicts different production functions and rows distinguish the supply (a) and demand (b) shock magnitude which an industry is exposed to. Results in (b) are only shown for one production function, since they are identical across alternative specifications.
The values in the tiles and their coloring denote aggregate output levels as percentage of pre-shock levels one month after the shock hits a given industry. This values are computed as averages from $N$ runs, always shocking only a single industry.
}
        \label{fig:output_table}
\end{figure}

Figure \ref{fig:output_table} shows the simulation results broken down into the various shock magnitudes (vertical axis) and production function categories (horizontal axis).
The coloring and the value of a tile represent the average aggregate output (as a fraction of initial output), where the average is taken over all $N$ runs. 
Results obtained from the demand shock scenarios, Figure \ref{fig:output_table}(b), do not differ across alternative production function specifications and thus we only show one representative case. The reason for this is that demand shocks to an industry are passed on proportionally to all of its suppliers as shown in Eq.~\eqref{eq:order_interm}, regardless of the underlying production mechanisms.
This is in stark contrast to the supply shock scenarios, Figure \ref{fig:output_table}(a), where economic impacts depend strongly on the choice of production function.
While the economic impact of supply constraints is limited in the linear production model and depends fairly smoothly on the shock magnitude, this is not the case in the Leontief model. Here the economy experiences a major truncation if single industries are shut down. The results lie in-between these two extremes when production functions differentiate between critical and non-critical inputs (IHS 1-3). 

Except for the linear production function simulations, we observe for several severe supply shock scenarios fairly wide distributions of economic impacts.
Thus, not only the shock size matters but also \emph{which} industries are affected by these shocks. 
For example, applying an 80\% supply shock shock to industry Repair-Installation (C33) under the IHS2 assumption, collapses the economy by about 50\%, although the industry accounts for less than half a percent of the overall economy. On the other hand, applying the same shock to the comparatively large industry Other Services (R\_S, 3\% of the economy) leads to a mere 6\% reduction of aggregate output.

For a policymaker it is important to know what properties of an industry drive the amplification of shocks, as this could inform both the design of lockdown measures as well as reopening policies. To explore this more systematically, we regress output levels against potential explanatory factors such as upstreamness, output multipliers and industry sizes. An industry's upstreamness in a production network is its average distance to the final consumer \citep{antras2012measuring} and also known as Total Forward Linkages \citep{miller2017output}. 
High upstreamness implies that the output of this industry requires several subsequent production steps before it is purchased by final consumers. Thus, relaxing shocks on industries with high upstreamness could potentially stimulate further economic activity. Since upstreamness boils down to the row sums of the Gosh inverse \citep{miller2017output}, we obtain the $N$-dimensional upstreamness vector as
$ u = (\mathbb{I} - B)^{-1} \mathbf{1}$,
where a matrix element $B_{ij} = Z_{ij,0}/x_{i,0}$ represents ``allocation coefficients''. Upstreamness ranges from 1.004 (Household activities) to 2.742 (Warehousing) in our sample of UK industries.

Output multipliers, or alternatively Total Backward Linkages, are a core metric in many economic studies. In input-output analysis multipliers quantify the impact of a change in final demand in a given sector on the entire economy. Multipliers are related to various network centrality concepts and have been shown to be strongly predictive of long-term growth \citep{mcnerney2018production}.
Since shocking an industry with a high multiplier should lead to larger decreases in intermediate demand, it is plausible that high-multiplier industries tend to amplify shocks more. The output multiplier is computed as the column sum of the Leontief inverse, i.e. $m = (\mathbb{I} - A^\top)^{-1} \mathbf{1}$. 
Multipliers range from 1 (Household activities) to 2.379 (Forestry) in our sample.
Upstreamness and multipliers are different, but are fairly highly correlated, with a Pearson correlation of 0.45.

We then regress log aggregate output one month after the initial shock hits the economy, $\log (\sum_k x_{k,30})$, against the shocked industry's log upstreamness and multiplier levels. 
Naturally, we would expect a larger decline if supply shocks hit an overall large industry and similarly for demand shocks and the size of final consumption. 
Thus we also control for total industry size measured in log gross output, $\log (x_{i,0})$, and industry-specific total demand, $\log ( c^d_{i,0} + f^d_{i,0} )$, in our regressions\footnote{
We do not include gross output and final demand values together as regressors to avoid multicolinearity. Industry gross output and final consumption are highly correlated; $\text{cor} \{ \log (x_{i,0}), \log (c_{i,0}+ f_{i,0}) \} =0.89$ (p-value $< 10^{-16}$).
}.
We then run the regression for every given shock magnitude and production function separately.

Tables \ref{tab:regsuppshock} summarizes the regression results for the supply shock scenarios.
For supply shocks we find that upstreamness is a very good predictor of adverse economic impacts if the economy builds upon Leontief production mechanisms. If industries use linear production technologies, on the other hand, it is rather the size of the industry than its upstreamness level that explains reductions in aggregate output. When using the intermediate assumption of an IHS2 production function, both upstreamness levels and industry size significantly affect aggregate impacts, although the overall model fit ($R^2$) drops substantially. 

Note that supply shocks are to a large extent a policy variable as they are directly coupled to non-pharmaceutical interventions such as industry-specific shutdowns. Our results indicate that upstreamness levels of industries are an important aspect for designing lockdown scenarios. In case of limited production flexibilities,  upstreamness might be even a better indicator of industry closure related aggregate impacts than the actual size of the industry.

Regression results for the demand shock experiments are shown in Table \ref{tab:regdemashock}.
The first four columns show the results from univariate regressions where we include only one of the potential covariates: upstreamness, multipliers, final consumption and gross output.
Somewhat surprisingly, output multipliers, a key metric for quantifying aggregate impacts resulting from demand side perturbations in simpler input-output models, do not exhibit any predictive power in our case. Upstreamness, on the other hand, is positively associated with aggregate output values, indicating that demand shocks to upstream industries have less adverse impacts on the economy than downstream industries.
Better model fits, however, are obtained when regressing aggregate output against indicators of industry size \- gross output or final consumption. 

Interestingly, combining final consumption values with the network-based metrics in a multivariate regression (column five) does not improve explanatory power at all. On the other hand, upstreamness and output size explain complementary parts of aggregate impacts (columns six), resulting in a better model fit than any regression using final consumption values.
Thus, industries' upstreamness and output sizes do not only play a role for the propagation of supply shocks but are also key determinants for demand shock amplification.

Overall, we find that static measures are indicative of our model results but only explain them partially, even in this simplified case where we studied either demand or supply to only one industry at time. In reality, supply and demand shocks are mixed and multiple industries affected at the same time to varying degrees. In that case the propagation of pandemic shocks will be more complex. Nevertheless, our results indicate that upstream industries play an important role in the amplification of exogenous shocks.

\FloatBarrier

\begin{table}[!htbp] \centering 
\footnotesize
\begin{tabular}{@{\extracolsep{5pt}}lcccccc} 
\\[-1.8ex]\hline 
\hline \\[-1.8ex] 
\\[-1.8ex] & \multicolumn{6}{r}{ \emph{Dependent variable:} $\log( \sum_i x_{i,30})$ } \\ 
\cline{2-7} \\[-1.8ex]
 Shock size: & 40\% & 40\% & 40\% & 80\% & 80\% & 80\% 
 \\ 
Production: & Leontief & IHS2 & Linear & Leontief & IHS2 & Linear 
 \\ 
\hline \\[-1.8ex] 
 $\log (u_{i,0})$ & $-$0.233$^{***}$ & $-$0.114$^{***}$ & 0.008 & $-$0.440$^{***}$ & $-$0.383$^{***}$ & 0.016 \\ 
  & (0.018) & (0.019) & (0.003) & (0.023) & (0.071) & (0.007) \\ 
  & & & & & & \\ 
 $\log (m_{i,0})$ & 0.067 & 0.065 & $-$0.013 & 0.037 & 0.306 & $-$0.025 \\ 
  & (0.038) & (0.041) & (0.007) & (0.049) & (0.150) & (0.014) \\ 
  & & & & & & \\ 
 $\log (x_{i,0})$ & $-$0.006 & $-$0.019$^{***}$ & $-$0.008$^{***}$ & $-$0.011 & $-$0.079$^{***}$ & $-$0.016$^{***}$ \\ 
  & (0.004) & (0.004) & (0.001) & (0.005) & (0.016) & (0.001) \\ 
  & & & & & & \\ 
 Constant & 15.512$^{***}$ & 15.673$^{***}$ & 15.560$^{***}$ & 15.224$^{***}$ & 16.175$^{***}$ & 15.641$^{***}$ \\ 
  & (0.050) & (0.053) & (0.009) & (0.065) & (0.197) & (0.018) \\ 
  & & & & & & \\ 
\hline \\[-1.8ex] 
Observations & 55 & 55 & 55 & 55 & 55 & 55 \\ 
Adjusted R$^{2}$ & 0.776 & 0.456 & 0.722 & 0.890 & 0.448 & 0.719 \\ 
\hline 
\hline \\[-1.8ex] 
\textit{Note:}  & \multicolumn{6}{r}{$^{*}$p$<$0.01; $^{**}$p$<$0.001; $^{***}$p$<$0.0001} \\ 
\end{tabular} 
  \caption{
  \textbf{Regression results for supply shock experiments. \\
  }
  } 
  \label{tab:regsuppshock}
\begin{tabular}{@{\extracolsep{5pt}}lcccccc} 
\\[-1.8ex]\hline 
\hline \\[-1.8ex] 
\\[-1.8ex] & \multicolumn{6}{r}{ \emph{Dependent variable:} $\log( \sum_i x_{i,30})$ } \\ 
\cline{2-7} \\[-1.8ex]
& (1) & (2) & (3) & (4) & (5) & (6) 
 \\ 
\hline \\[-1.8ex] 
 $\log (u_{i,0})$ & 0.040$^{**}$ &  &  &  & 0.012 & 0.036$^{***}$ \\ 
  & (0.010) &  &  &  & (0.010) & (0.008) \\ 
  & & & & & & \\ 
 $\log (m_{i,0})$ &  & 0.021 &  &  & $-$0.024 & $-$0.030 \\ 
  &  & (0.024) &  &  & (0.019) & (0.018) \\ 
  & & & & & & \\ 
 $\log (c_{i,0}+f_{i,0})$ &  &  & $-$0.012$^{***}$ &  & $-$0.011$^{***}$ &  \\ 
  &  &  & (0.002) &  & (0.002) &  \\ 
  & & & & & & \\ 
 $\log (x_{i,0})$ &  &  &  & $-$0.014$^{***}$ &  & $-$0.013$^{***}$ \\ 
  &  &  &  & (0.002) &  & (0.002) \\ 
  & & & & & & \\ 
 Constant & 15.443$^{***}$ & 15.453$^{***}$ & 15.588$^{***}$ & 15.618$^{***}$ & 15.586$^{***}$ & 15.599$^{***}$ \\ 
  & (0.006) & (0.013) & (0.016) & (0.023) & (0.025) & (0.023) \\ 
  & & & & & & \\ 
\hline \\[-1.8ex] 
Observations & 55 & 55 & 55 & 55 & 55 & 55 \\ 
Adjusted R$^{2}$ & 0.217 & $-$0.004 & 0.522 & 0.452 & 0.522 & 0.580 \\ 
\hline 
\hline \\[-1.8ex] 
\textit{Note:}  & \multicolumn{6}{r}{$^{*}$p$<$0.01; $^{**}$p$<$0.001; $^{***}$p$<$0.0001} \\ 
\end{tabular} 
  \caption{
  \textbf{Regression results for demand shock experiments.}
  As representative case we show regression results for the scenario of 60\% shocks to final consumption, because results are similar for alternative shock sizes. Note that economic impacts are identical across alternative production functions. 
  } 
\label{tab:regdemashock} 
\end{table}

\FloatBarrier

\section{Discussion} \label{sec:discuss}

In this paper we have investigated how locking down and re-opening the UK economy as a policy response to the Covid-19 pandemic affects economic performance.
We introduced a novel economic model specifically designed to address the unique features of the current pandemic. 
The model is industry-specific, incorporating the production network and inventory dynamics. We use survey results by industry experts to model how critical different inputs are in the production of a specific industry.

We found in simulation experiments where we studied simpler shock scenarios and a simplified model setup that an industry's upstreamness is predictive of shock amplification. However, the relationship is noisy and strongly depends on the underlying production mechanism in case of downstream propagation of supply shocks. These results underline the necessity of more sophisticated macroeconomic models for quantifying the economic impacts resulting from national lockdowns and subsequent re-opening.

Real time GDP predictions for the UK economy made in an early version of this paper turned out to be very accurate \citep{pichler2020production}. But was this because we did things right, or because we just got lucky?  Our analysis here shows that it was a mixture of the two.  By investigating both alternative shocks scenarios, alternative production functions and studying the sensitivity to parameters and initial conditions we are able to see how the quality of the predictions depends on these factors.  We find that the shock scenarios are the most important determinant, but the production function can also be very important, and some of the other parameters can affect the results as well.

To make a real time forecast we had to act quickly. There were no data available about which industry classifications were considered essential in the UK and the few data available on UK jobs that could be performed form home was based on US O*NET  data\footnote{
\url{https://www.ons.gov.uk/employmentandlabourmarket/peopleinwork/employmentandemployeetypes/articles/whichjobscanbedonefromhome/2020-07-21}
}. In the interest of time, we estimated the UK supply shocks using predicted US supply shocks \cite{del2020supply}. These supply shocks were based on a list of essential industries that was considerably less permissive (i.e., less industries were considered essential) than the UK guidelines. This turned out to be lucky: respecting social distancing guidelines caused many industries in the UK to close even though they were not formally and explicitly required to do so. With hindsight, this was fortuitous -- if we had had a list of essential British industries our supply shocks would have been too weak, or we would have had to model social distancing constraints by industry, which is difficult. Even if it missed some of the details, the supply shocks estimated by \cite{del2020supply} provided a reasonable approximation to the truth.

The choice of production function also matters a great deal. Our results suggest that the Leontief production function, which is widely used for understanding the response to disasters, is a poor choice.  This is for an intuitive reason:  Some inputs are not critical, and an industry can operate reasonably well without them, at least for a few months.  Our results here show that production functions that incorporate this fact can do well.  This could be further developed by calibrating CES production functions to correctly incorporate when substitutions are appropriate.  The IHS Markit survey that was performed should eventually be performed with larger samples and tested in detail (but that is beyond the scope of this paper). At the other extreme, our results also suggest that the linear production function is a poor choice.  It comes close to the correct aggregate error only with the strongest shock scenario and never performs well at the sectoral level.  

Our results indicate that dynamic models of the type that we developed here can do a good job of forecasting disruptions in the economy.  We want to emphasize that, while there is wide variation in the results under different scenarios, this variation could be dramatically reduced by collecting better data.  A clear example is the choice of inventory levels.  In our original model we had no data for inventory levels of UK industries, so we used data for the US.  Replacing this by UK data made a substantial improvement in our results.  Similarly, the economic data we had available was from 2019, and the IO data was from 2014 -- better real time measurements about the response of industries to the pandemic would likely have improved our predictions. A more extensive study of the importance of different inputs to production could have reduced ambiguity about the choice of production function.

At the highest level, our model illustrates the value of its key features.  These are:  modeling at the sectoral level, allowing both supply and demand shocks to operate simultaneously, using a realistic production function that properly captures nonlinearities without exaggerating them, and using a dynamic model that incorporates inventory effects.  With better data and better measurement of parameters, our results demonstrate that a model of this type can give useful insight into the economics of a disaster such as the Covid-19 pandemic.

\FloatBarrier

\small
\bibliographystyle{agsm}
\bibliography{tech_ref}
\FloatBarrier
\normalsize

\newpage
\appendix
\section*{Appendix}
\label{sec:Appendix}

\section{First-order supply and demand shocks} 
\label{apx:shocks}

\subsection{Supply shocks}
Due to the Covid-19 pandemic, industries experience supply-side reductions due to the closure of non-essential industries, workers not being able to perform their activities at home, and difficulties adapting to social distancing measures. Many industries also face substantial reductions in demand. \cite{del2020supply} provide quantitative predictions of these first-order supply and demand shocks for the U.S. economy. To calculate supply-side predictions, \cite{del2020supply} constructed a Remote Labor Index, which measures the ability of different occupations to work from home, and scored industries according to their essentialness based on the Italian government regulations. 

We follow a similar approach. We score industries essentialness based on the UK government regulations and use occupational data to estimate the fraction of workers that could work remotely and the difficulties sectors faced in adapting to social distancing measures for on-site work. Several of our estimates are based on indexes and scores available for industries in the NAICS 4-digit classification system. An essential step in our methodology is to map these estimates into the WIOD industry classification system. We explain our mapping methodology below.

\paragraph{NAICS to WIOD mapping} We build a crosswalk from the NAICS 4-digit industry classification to the classification system used in WIOD, which is a mix of ISIC 2-digit and 1-digit codes. We make this crosswalk using the NAICS to ISIC 2-digit crosswalk from the European Commission and then aggregating the 2-digit codes presented as 1-digit in the WIOD classification system. We then do an employment-weighted aggregation of the index or score in consideration from the 277 industries at the NAICS 4-digit classification level to the 55 industries in the WIOD classification. Some of the 4-digit NAICS industries map into more than one WIOD industry classification. When this happens, we assume employment is split uniformly among the WIOD industries the NAICS industry maps into.

\begin{figure}
    \centering
\includegraphics[width = 0.5\textwidth]{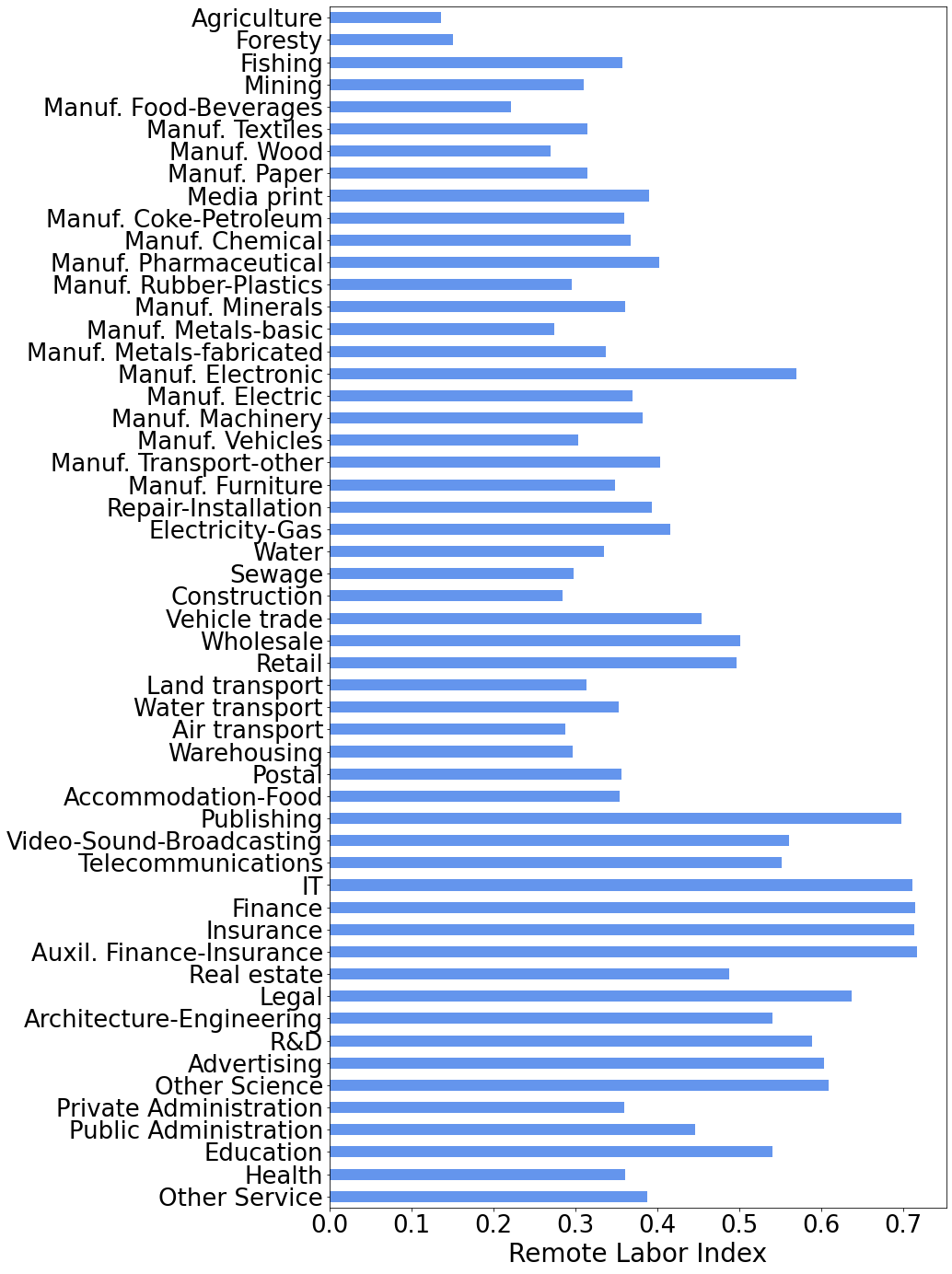}
    \caption{\textbf{Remote Labor Index of industries.} Remote labor index of the WIOD industry classification. See Table \ref{tab:FO_shocks_supply} for code-industry name.}
        \label{fig:RLI}
\end{figure}

\begin{figure}
    \centering
\includegraphics[width = 0.5\textwidth]{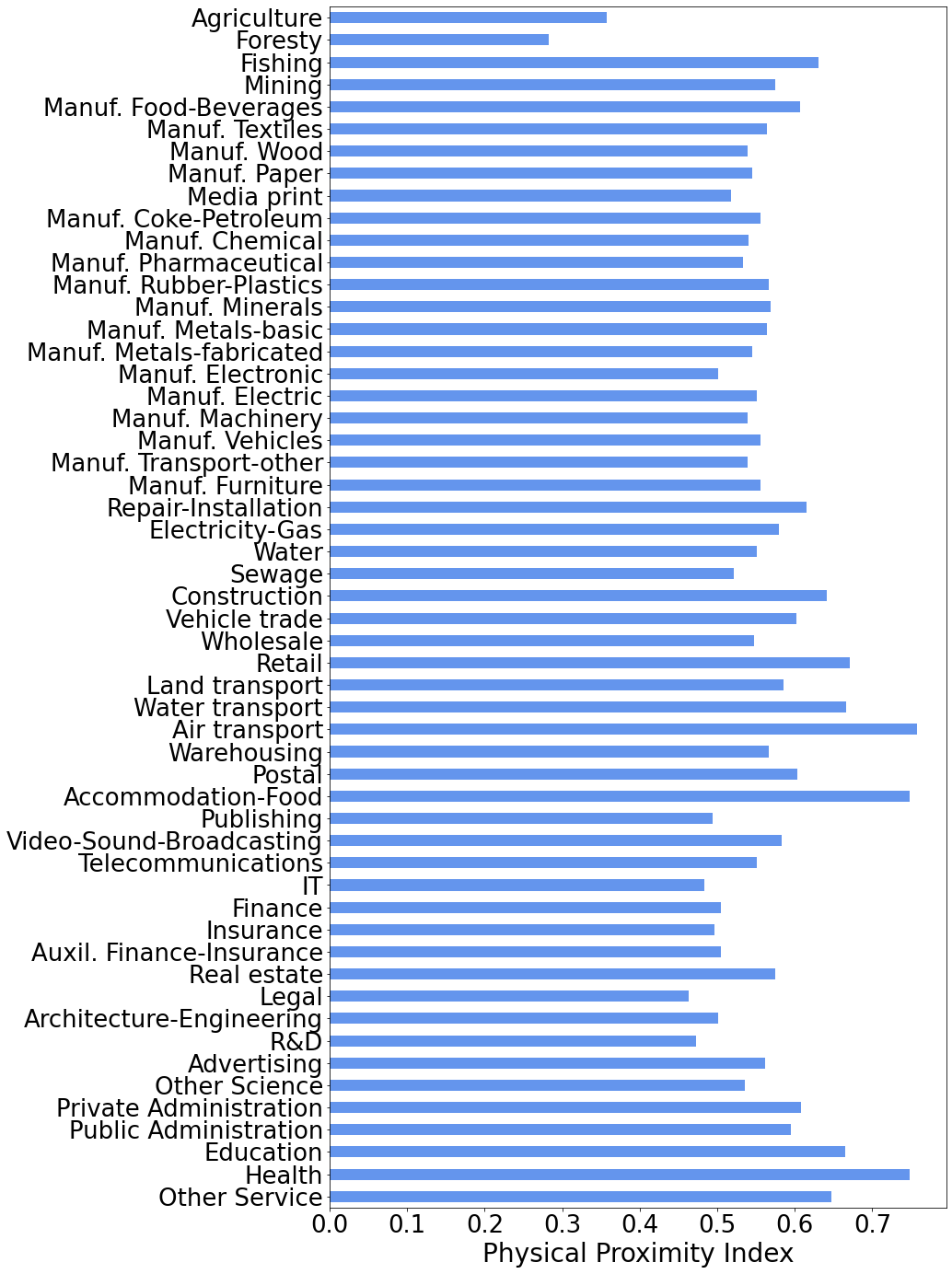}
    \caption{\textbf{Physical Proximity Index of industries.} Physical Proximity Index of the WIOD industry classification. See Table \ref{tab:FO_shocks_supply} for code-industry name.}
        \label{fig:PPI}
\end{figure}

\paragraph{Remote Labor Index and Physical Proximity Index} To estimate the fraction of workers that could work remotely, we use the Remote Labor Index from \cite{del2020supply}. To understand the difficulties different sectors face in adapting to the social distancing measures, we compute an industry-specific Physical Proximity Index. Other works have also used the Physical Proximity of occupations to understand the economic consequences of the lockdown \citep{mongey2020workers,koren2020business}. We map the Physical Proximity Work Context\footnote{
https://www.onetonline.org/find/descriptor/result/4.C.2.a.3
} of occupations provided by O*NET into industries, using the same methodology that \cite{del2020supply} used to map the Remote Labor Index into industries. That is, we use BLS data that indicate the occupational composition of each industry and take the employment-weighted average of the occupation's work context employed in each industry at the NAICS 4-digit classification system. 
Under the assumption that the distribution of occupations across industries and that the percentage of essential workers within an industry is the same for the US and the UK, we can map the Remote Labor Index by \cite{del2020supply} and the Physical Proximity Index into the UK economy following the mapping methodology explained in the previous paragraph. The WIOD industry sector 'T' ("Activities of households as employers; undifferentiated goods- and services-producing activities of households for own use") does only maps into one NAICS code for which we do not have RLI or PPI. Since We consider  sector 'T' to be similar to the 'R\_S' sector ("Other service activities") we use the  RLI and PPI from 'R\_S' sector in the 'T' sector. In Figures~\ref{fig:RLI} and \ref{fig:PPI} we show the Remote Labor Index and the Physical Proximity Index of the WIOD sectors.

\paragraph{Essential score} 
To determine the essential score of industries in scenarios $S_1$ to $S_4$ we follow the UK government guidelines. We break down WIOD sectors, which are aggregates of 2-digit NACE codes, into finer 3- and 4-digit industries. The advantage of having smaller subsectors is that we can associate shutdown orders to the subsectors and then compute an average essential score for the aggregate WIOD sector. In the following, we provide some details on how we calculate essential scores while referring the reader to the online repository to check all our assumptions.
\begin{itemize}
    \item Essential score of G45 (Wholesale and retail trade and repair of motor vehicles and motorcycles). The only shops in this sector that were mandated to close were car showrooms (see footnote \ref{footn:ukleg} for a link to official legislation).  Lacking a disambiguation between retail and wholesale trade of motor vehicles, we assign a 0.5 essential score to subsector 4511, which comprises 72\% of the turnover of G45, and an essential score of 1 to all other subsectors. The average essential score for G45 turns out to be 64\%.
    \item Essential score of G47 (Retail trade, except of motor vehicles and motorcycles). All ``non-essential'' shops were mandated to close, except food and alcohol retailers, pharmacies and chemists, newsagents, homeware stores, petrol stations, bicycle shops and a few others. We assigned an essential score that could be either 0 or 1 to all 37 4-digit NACE subsectors that compose G47. Weighing essential scores by turnover results in an average essential score of 71\%.
    \item Essential score of I (Accommodation and food service activities). Almost all economic activities in this sector were mandated to close, except hotels for essential workers (e.g. those working in transportation) and workplace canteens where there is no practical alternative for staff at that workplace to obtain food. Assigning a 10\% essential score to the main hotels subsector (551) and a 50\% essential score to subsector 5629, including workplace canteens, results in an overall 5\% essential score for sector I.
     \item Essential score of R\_S (Recreational and other services). Almost all activities in this sector were mandated to close, except those related to repair, washing and funerals. Considering all 34 subsectors yields an essential score of 7\%.
\end{itemize}

\paragraph{Real Estate.} This sector includes imputed rents, which account for $69\%$ of the monetary value of the sector,\footnote{
\url{https://www.ons.gov.uk/economy/grossvalueaddedgva/datasets/nominalandrealregionalgrossvalueaddedbalancedbyindustry}, Table 1B.
}. Because we think applying a supply shock to imputed rent does not make sense, we compute that the supply shock derived from the RLI and Essential Score (which is around $50\%$) only affects $31\%$ of the sector, leading to a $15\%$ final supply shock to Real Estate (due to an error, our original work used a value of $4.7\%$).If we had used the correct real estate shock we would have predicted a 22.1\% reduction, exactly like in the data. In the main text, we still describe our prediction as having been a 21.5\% contraction, as explicitly stated in our original paper. Note that these differences are within the range of expected data revisions\footnote{see for instance \url{https://www.ons.gov.uk/economy/grossdomesticproductgdp/datasets/revisionstrianglesforukgdpabmi} for ONS GDP quarterly national accounts revisions, which are of the order of 1 pp.)}

In scenario $S_5$, we mapped the supply shocks estimated by \cite{del2020supply} at the NAICS level into the WIOD classification system as explained previously. The WIOD industry sector 'T' ("Activities of households as employers; undifferentiated goods- and services-producing activities of households for own use")  does only maps into one NAICS code for which we do not have a supply shock. Since this sector is likely to be essential, we assume a zero supply shock. 

The supply shocks at the NAICS level depend on the list of industries' essential score at the NAICS 4-digit level provided by \cite{del2020supply}.  It is important to note that, although the list of industries' essential score provided by \cite{del2020supply} is based on the Italian list of essential industries, these lists are based on different industry classification systems (NAICS and NACE, respectively) and do not have a one-to-one correspondence. To derive the essential score of industries at the NAICS level \cite{del2020supply} followed three steps. (i) the authors considered a 6-digit NAICS industry essential if the industry had correspondence with at least one essential NACE industry. (ii) \cite{del2020supply} aggregated the 6-digit NAICS essential lists into the 4-digit level taking into account the fraction of NAICS 6-digit subcategories that were considered essential. (iii) the authors revised each 4-digit NAICS industry's essential score to check for implausible classifications and reclassified ten industries whose original essential score seemed implausible. Step (i) likely resulted in a larger fraction of industries at the NAICS level to be classified as essential than in the NACE level.  This, in turn, results in supply shocks $S_5$ that are milder than they would have been if the essential score was mapped directly from NACE to WIOD and the supply shocks calculated at the WIOD level.

For scenario $S_6$ we used the list of essential industries compiled by \cite{fana2020covid} for Italy, Germany, and Spain. We make one list by taking the mean over the three countries of the essential score of each industry. This list is at the ISIC 2-digit level, which we aggregate to WIOD classification weighting each sector by its gross output in the UK.

\begin{figure}
    \centering
\includegraphics[width = \textwidth]{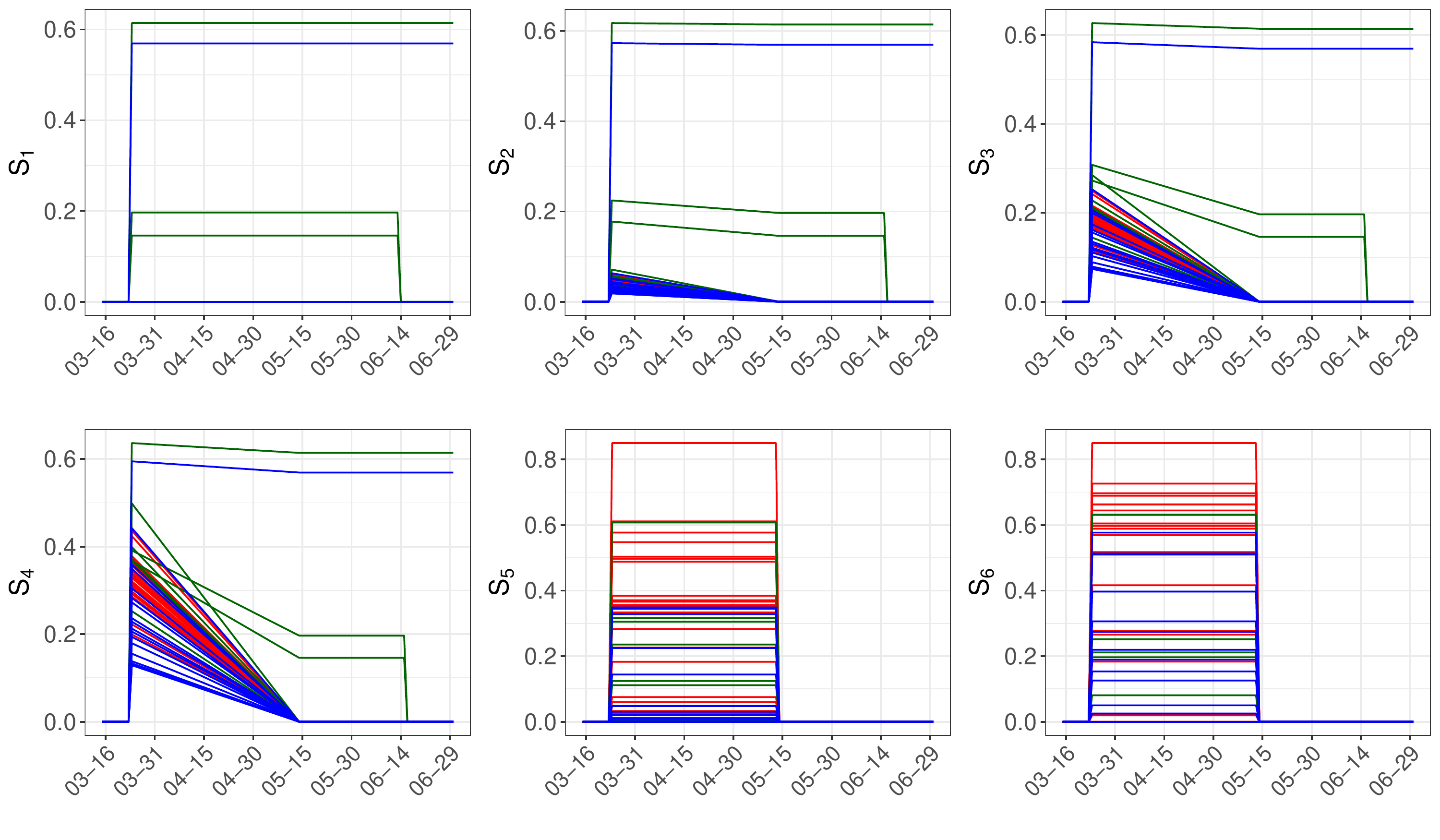}
    \caption{
    \textbf{Comparison of supply shock scenarios over time.} 
    Each panel shows industry specific supply shocks $\epsilon_{i,t}^S$ for a given scenario. The coloring of the lines is based on the same code as in Figure \ref{fig:aggregatevssectoral}. 
    }
        \label{fig:supplyscenarios}
\end{figure}

Table \ref{tab:FO_shocks_supply} gives an overview of first-order supply shocks and Figure \ref{fig:supplyscenarios} show the supply scenarios over time.

\subsection{Demand shocks}
For calibrating consumption demand shocks, we use the same data as \cite{del2020supply} which are based on the \cite{CBO2006} estimates. These estimates are available only at the more aggregate 2-digit NAICS level and map into WIOD ISIC categories without complications. To give a more detailed estimate of consumption demand shocks, we also link manufacturing sectors to the closure of certain non-essential shops as follows. 
\begin{itemize}
     \item Consumption demand shock to C13-C15 (Manufacture of textiles, wearing apparel, leather and other related products). Four subsectors (4751, 4771, 4772, 4782) selling goods produced by this manufacturing sector were mandated to close, while one subsector was permitted to remain open as it sells homeware goods (4753). Lacking more detailed information about the share of C13-C15 products that these subsectors sell, we simply give equal shares to all subsectors, leading to an 80\% consumption demand shock to C13-C15. 
      \item Consumption demand shock to C20 (Manufacture of chemicals and chemical products). Three subsectors (4752, 4773, 4774) selling homeware and medical goods were considered essential, while subsector 4775, selling cosmetic and toilet articles, was mandated to close. Using the same assumptions as above, we get a 25\% consumption demand shock for this sector.
      \end{itemize}
      The same procedure leads to consumption demand shocks for all other manufacturing subsectors.

\begin{figure}
    \centering
\includegraphics[width = \textwidth]{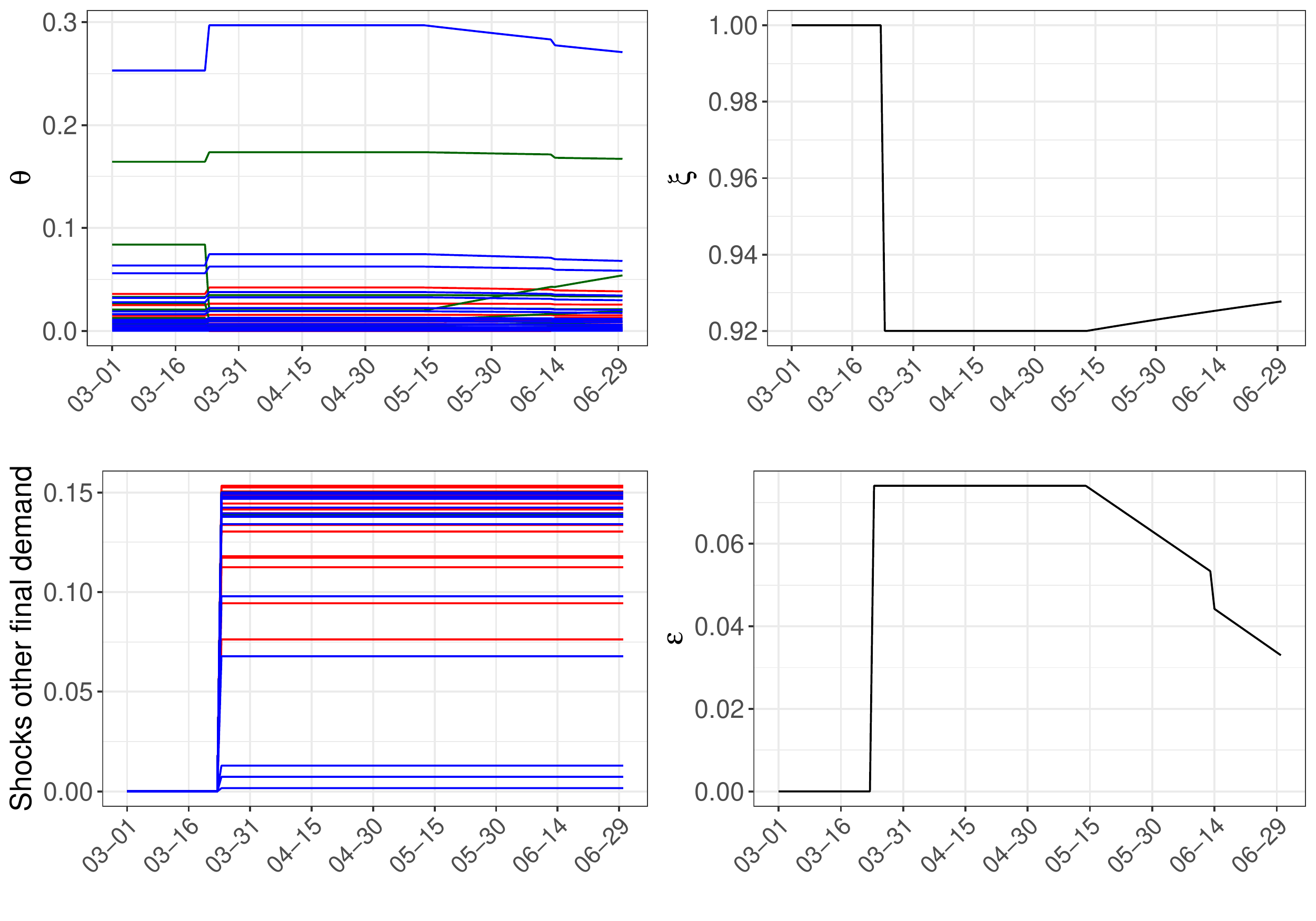}
    \caption{
    \textbf{Industry-specific demand shocks over time.} The upper left panel shows the change in preferences $\theta_{i,t}$. 
    The bottom left panel shows shock magnitudes to investment and export.
    The upper right panel shows demand shocks due to fear of unemployment $\xi_{t}$. 
    The bottom right panel is the aggregate demand shock $\tilde \epsilon_t$ taking the savings rate of 50\% into account. The coloring of the lines for industry-specific results follows the same code as in Figure \ref{fig:aggregatevssectoral}. 
    }
        \label{fig:demand_time}
\end{figure}

Table \ref{tab:FO_shocks_demand} shows the demand shock for each sector and \ref{fig:demand_time} illustrates the demand shock scenarios over time.

\begin{table}[ht]
\scriptsize
\centering
\begin{tabular}{|ll|rrrrrrr|}
  \hline
ISIC & Sector & $x$ &  $\text S_1$ & $\text S_2$ & $\text S_3$ & $\text S_4$ & $\text S_5$ & $\text S_6$  \\ 
  \hline
A01 & Agriculture & 0.8 & 0.0 & 4.1 & 16.3 & 28.6 & 0.0 & 0.0 \\ 
  A02 & Forestry & 0.0 & 0.0 & 3.2 & 12.7 & 22.2 & 85.0 & 85.0 \\ 
  A03 & Fishing & 0.1 & 0.0 & 5.3 & 21.4 & 37.4 & 0.0 & 0.0 \\ 
  B & Mining & 1.3 & 0.0 & 5.2 & 20.9 & 36.6 & 35.3 & 18.4 \\ 
  C10.C12 & Manuf. Food-Beverages & 2.8 & 0.0 & 6.2 & 25.0 & 43.7 & 0.6 & 0.0 \\ 
  C13.C15 & Manuf. Textiles & 0.4 & 0.0 & 5.1 & 20.4 & 35.7 & 37.1 & 64.5 \\ 
  C16 & Manuf. Wood & 0.2 & 0.0 & 5.2 & 20.8 & 36.4 & 61.1 & 68.9 \\ 
  C17 & Manuf. Paper & 0.4 & 0.0 & 4.9 & 19.7 & 34.5 & 7.5 & 27.6 \\ 
  C18 & Media print & 0.3 & 0.0 & 4.2 & 16.7 & 29.2 & 6.0 & 0.0 \\ 
  C19 & Manuf. Coke-Petroleum & 0.9 & 0.0 & 4.7 & 18.8 & 32.9 & 18.3 & 2.1 \\ 
  C20 & Manuf. Chemical & 1.1 & 0.0 & 4.5 & 18.1 & 31.6 & 2.6 & 26.6 \\ 
  C21 & Manuf. Pharmaceutical & 0.7 & 0.0 & 4.2 & 16.8 & 29.4 & 1.1 & 0.0 \\ 
  C22 & Manuf. Rubber-Plastics & 0.7 & 0.0 & 5.3 & 21.1 & 36.9 & 28.3 & 51.0 \\ 
  C23 & Manuf. Minerals & 0.5 & 0.0 & 4.8 & 19.2 & 33.6 & 50.3 & 63.1 \\ 
  C24 & Manuf. Metals-basic & 0.6 & 0.0 & 5.4 & 21.6 & 37.8 & 57.7 & 72.6 \\ 
  C25 & Manuf. Metals-fabricated & 1.1 & 0.0 & 4.8 & 19.1 & 33.4 & 54.8 & 66.3 \\ 
  C26 & Manuf. Electronic & 0.8 & 0.0 & 2.8 & 11.4 & 19.9 & 38.5 & 41.6 \\ 
  C27 & Manuf. Electric & 0.4 & 0.0 & 4.6 & 18.3 & 32.1 & 33.3 & 58.9 \\ 
  C28 & Manuf. Machinery & 1.1 & 0.0 & 4.4 & 17.6 & 30.8 & 49.7 & 56.9 \\ 
  C29 & Manuf. Vehicles & 1.6 & 0.0 & 5.1 & 20.5 & 35.8 & 22.6 & 69.7 \\ 
  C30 & Manuf. Transport-other & 1.0 & 0.0 & 4.2 & 17.0 & 29.7 & 48.8 & 59.7 \\ 
  C31\_C32 & Manuf. Furniture & 0.6 & 0.0 & 4.8 & 19.2 & 33.5 & 36.6 & 60.5 \\ 
  C33 & Repair-Installation & 0.4 & 0.0 & 4.9 & 19.7 & 34.5 & 3.3 & 51.8 \\ 
  D35 & Electricity-Gas & 3.2 & 0.0 & 4.5 & 17.9 & 31.3 & 0.0 & 1.9 \\ 
  E36 & Water & 0.2 & 0.0 & 4.8 & 19.4 & 33.9 & 0.0 & 0.0 \\ 
  E37.E39 & Sewage & 0.8 & 0.0 & 4.8 & 19.3 & 33.8 & 0.0 & 0.0 \\ 
  F & Construction & 7.9 & 0.0 & 6.1 & 24.2 & 42.4 & 35.6 & 66.3 \\ 
  G45 & Vehicle trade & 1.7 & 19.7 & 22.4 & 30.8 & 39.1 & 31.6 & 19.7 \\ 
  G46 & Wholesale & 3.5 & 0.0 & 3.6 & 14.4 & 25.3 & 23.6 & 21.1 \\ 
  G47 & Retail & 4.7 & 14.6 & 17.8 & 27.3 & 36.8 & 30.5 & 25.2 \\ 
  H49 & Land transport & 2.0 & 0.0 & 5.3 & 21.2 & 37.1 & 11.1 & 0.0 \\ 
  H50 & Water transport & 0.6 & 0.0 & 5.7 & 22.8 & 39.9 & 12.4 & 0.0 \\ 
  H51 & Air transport & 0.6 & 0.0 & 7.1 & 28.5 & 49.9 & 0.1 & 8.1 \\ 
  H52 & Warehousing & 1.4 & 0.0 & 5.3 & 21.0 & 36.8 & 0.5 & 0.0 \\ 
  H53 & Postal & 0.7 & 0.0 & 5.1 & 20.5 & 35.9 & 0.0 & 0.0 \\ 
  I & Accommodation-Food & 2.9 & 61.4 & 61.7 & 62.7 & 63.6 & 60.8 & 63.2 \\ 
  J58 & Publishing & 0.6 & 0.0 & 2.0 & 7.9 & 13.8 & 14.4 & 2.5 \\ 
  J59\_J60 & Video-Sound-Broadcasting & 0.9 & 0.0 & 3.4 & 13.5 & 23.7 & 32.8 & 5.0 \\ 
  J61 & Telecommunications & 1.6 & 0.0 & 3.3 & 13.1 & 22.9 & 0.9 & 0.0 \\ 
  J62\_J63 & IT & 2.3 & 0.0 & 1.8 & 7.4 & 12.9 & 0.2 & 12.6 \\ 
  K64 & Finance & 4.3 & 0.0 & 1.9 & 7.6 & 13.3 & 0.0 & 0.0 \\ 
  K65 & Insurance & 3.2 & 0.0 & 1.9 & 7.5 & 13.2 & 0.0 & 0.0 \\ 
  K66 & Auxil. Finance-Insurance & 1.1 & 0.0 & 1.9 & 7.6 & 13.2 & 0.0 & 0.0 \\ 
  L68 & Real estate & 7.8 & 0.0 & 3.9 & 15.6 & 27.3 & 4.8 & 51.3 \\ 
  M69\_M70 & Legal & 2.8 & 0.0 & 2.2 & 8.9 & 15.5 & 2.0 & 18.9 \\ 
  M71 & Architecture-Engineering & 1.7 & 0.0 & 3.0 & 12.2 & 21.3 & 0.0 & 30.6 \\ 
  M72 & R\&D & 0.5 & 0.0 & 2.6 & 10.3 & 17.9 & 0.0 & 27.4 \\ 
  M73 & Advertising & 0.6 & 0.0 & 2.9 & 11.8 & 20.6 & 22.5 & 39.7 \\ 
  M74\_M75 & Other Science & 0.7 & 0.0 & 2.8 & 11.1 & 19.4 & 3.0 & 22.0 \\ 
  N & Private Administration & 4.4 & 0.0 & 5.1 & 20.5 & 36.0 & 34.9 & 51.3 \\ 
  O84 & Public Administration & 4.8 & 0.0 & 4.4 & 17.4 & 30.5 & 1.1 & 0.0 \\ 
  P85 & Education & 4.2 & 0.0 & 4.0 & 16.2 & 28.3 & 0.0 & 15.3 \\ 
  Q & Health & 7.0 & 0.0 & 6.3 & 25.3 & 44.2 & 0.1 & 0.0 \\ 
  R\_S & Other Service & 3.2 & 56.9 & 57.3 & 58.4 & 59.5 & 34.5 & 57.7 \\ 
  T & Household activities & 0.2 & 0.0 & 5.0 & 20.0 & 35.0 & 0.0 & 51.0 \\ 
   \hline
\end{tabular}
\caption{
{\bf Industry-specific supply shock details. }
Column $x$ denotes relative shares of gross output. Supply shocks $\epsilon_{i}^S$ for different scenarios are shown in the columns $\text S_1$ to $\text{S}_6$. 
All values are in \%.
} 
\label{tab:FO_shocks_supply}
\end{table}

\begin{table}[ht]
\scriptsize
\centering
\begin{tabular}{|ll|rrrr|}
  \hline
ISIC & Sector & $c$ & $\epsilon^D_i$ & $f$  & $f$ shock \\ 
  \hline
A01 & Agriculture & 0.9 & 10.0 & 0.3 & 13.8 \\ 
  A02 & Forestry & 0.0 & 10.0 & 0.0 & 11.9 \\ 
  A03 & Fishing & 0.0 & 10.0 & 0.1 & 14.8 \\ 
  B & Mining & 0.1 & 10.0 & 1.5 & 15.3 \\ 
  C10.C12 & Manuf. Food-Beverages & 2.4 & 10.0 & 1.4 & 15.0 \\ 
  C13.C15 & Manuf. Textiles & 0.1 & 80.0 & 0.5 & 13.4 \\ 
  C16 & Manuf. Wood & 0.1 & 10.0 & 0.1 & 11.2 \\ 
  C17 & Manuf. Paper & 0.1 & 10.0 & 0.2 & 14.1 \\ 
  C18 & Media print & 0.1 & 66.0 & 0.1 & 9.4 \\ 
  C19 & Manuf. Coke-Petroleum & 1.4 & 10.0 & 0.7 & 14.8 \\ 
  C20 & Manuf. Chemical & 0.3 & 25.0 & 1.7 & 14.7 \\ 
  C21 & Manuf. Pharmaceutical & 0.3 & 10.0 & 1.2 & 14.9 \\ 
  C22 & Manuf. Rubber-Plastics & 0.1 & 10.0 & 0.6 & 14.0 \\ 
  C23 & Manuf. Minerals & 0.1 & 10.0 & 0.2 & 13.0 \\ 
  C24 & Manuf. Metals-basic & 0.0 & 10.0 & 1.7 & 15.0 \\ 
  C25 & Manuf. Metals-fabricated & 0.1 & 10.0 & 0.8 & 14.4 \\ 
  C26 & Manuf. Electronic & 0.2 & 100.0 & 1.5 & 14.9 \\ 
  C27 & Manuf. Electric & 0.1 & 10.0 & 0.8 & 14.9 \\ 
  C28 & Manuf. Machinery & 0.2 & 10.0 & 2.2 & 15.0 \\ 
  C29 & Manuf. Vehicles & 1.2 & 100.0 & 2.8 & 14.8 \\ 
  C30 & Manuf. Transport-other & 0.1 & 10.0 & 2.6 & 15.1 \\ 
  C31\_C32 & Manuf. Furniture & 0.2 & 40.0 & 0.8 & 14.7 \\ 
  C33 & Repair-Installation & 0.0 & 10.0 & 0.0 & 11.8 \\ 
  D35 & Electricity-Gas & 3.4 & 0.0 & 0.1 & 14.8 \\ 
  E36 & Water & 0.5 & 0.0 & 0.0 & 14.8 \\ 
  E37.E39 & Sewage & 0.5 & 0.0 & 1.1 & 7.6 \\ 
  F & Construction & 0.3 & 10.0 & 12.1 & 15.2 \\ 
  G45 & Vehicle trade & 1.9 & 10.0 & 0.6 & 15.0 \\ 
  G46 & Wholesale & 3.1 & 10.0 & 4.5 & 15.0 \\ 
  G47 & Retail & 15.5 & 10.0 & 0.6 & 14.1 \\ 
  H49 & Land transport & 2.5 & 67.0 & 0.2 & 14.9 \\ 
  H50 & Water transport & 0.6 & 67.0 & 0.7 & 15.0 \\ 
  H51 & Air transport & 1.2 & 67.0 & 0.5 & 15.0 \\ 
  H52 & Warehousing & 0.1 & 67.0 & 0.4 & 15.0 \\ 
  H53 & Postal & 0.1 & 0.0 & 0.1 & 14.8 \\ 
  I & Accommodation-Food & 7.9 & 80.0 & 0.8 & 15.0 \\ 
  J58 & Publishing & 0.5 & 0.0 & 0.6 & 14.7 \\ 
  J59\_J60 & Video-Sound-Broadcasting & 1.0 & 0.0 & 1.2 & 9.9 \\ 
  J61 & Telecommunications & 1.8 & 0.0 & 0.8 & 15.0 \\ 
  J62\_J63 & IT & 0.2 & 0.0 & 2.7 & 13.6 \\ 
  K64 & Finance & 3.0 & 0.0 & 3.1 & 14.9 \\ 
  K65 & Insurance & 6.0 & 0.0 & 1.6 & 14.9 \\ 
  K66 & Auxil. Finance-Insurance & 0.1 & 0.0 & 2.1 & 15.0 \\ 
  L68 & Real estate & 23.8 & 0.0 & 1.0 & 15.0 \\ 
  M69\_M70 & Legal & 0.1 & 0.0 & 1.4 & 14.4 \\ 
  M71 & Architecture-Engineering & 0.1 & 0.0 & 1.4 & 15.1 \\ 
  M72 & R\&D & 0.0 & 0.0 & 1.1 & 14.9 \\ 
  M73 & Advertising & 0.0 & 0.0 & 0.3 & 14.0 \\ 
  M74\_M75 & Other Science & 0.4 & 0.0 & 1.0 & 14.6 \\ 
  N & Private Administration & 1.0 & 0.0 & 2.8 & 14.1 \\ 
  O84 & Public Administration & 0.5 & 0.0 & 12.5 & 0.7 \\ 
  P85 & Education & 4.7 & 0.0 & 6.4 & 1.8 \\ 
  Q & Health & 3.7 & 0.0 & 14.8 & 0.2 \\ 
  R\_S & Other Service & 6.7 & 5.0 & 1.3 & 8.5 \\ 
  T & Household activities & 0.7 & 0.0 & 0.0 & 14.8 \\ 
   \hline
\end{tabular}
\caption{
{\bf Industry-specific demand shock details. }
Column $c$ denotes relative shares of final consumer consumption (which here aggregates column \emph{C1} and \emph{C2} in the WIOD) and $\epsilon_{i}^D$ represents final consumption shocks. 
Columns $f$ and $f$ \emph{shock} are relative shares of other final demand and the shocks applied to other final demand, respectively. 
All values are in \%.
} 
\label{tab:FO_shocks_demand}
\end{table}

\FloatBarrier

\section{Inventory data and calibration} 
\label{apx:inventory}

In the previous version of our work \citep{pichler2020production}, we used U.S. data from the BEA to calibrate the inventory target parameters $n_j$ in Eq. \eqref{eq:order_interm}. Here, we use more detailed UK data from the Annual Business Survey (ABS) . The ABS is the main structural business survey conducted by the ONS\footnote{ \url{ https://www.ons.gov.uk/businessindustryandtrade/business/businessservices/methodologies/annualbusinesssurveyabs}
}. It is sampled from all non-farm, non-financial private businesses in the UK (about two-thirds of the UK economy); data are available up to the 4-digit NACE level, but for our purposes 2-digit NACE industries are sufficient. The survey asks for information on a number of variables, including turnover and inventory stocks (at the beginning and end of each year). Data are available from 2008 to 2018. They show a general increase in turnover and inventory stocks (consistent with growth of the UK economy in the same period), with moderate year-on-year fluctuations.

\begin{figure}
    \centering
\includegraphics[width = \textwidth]{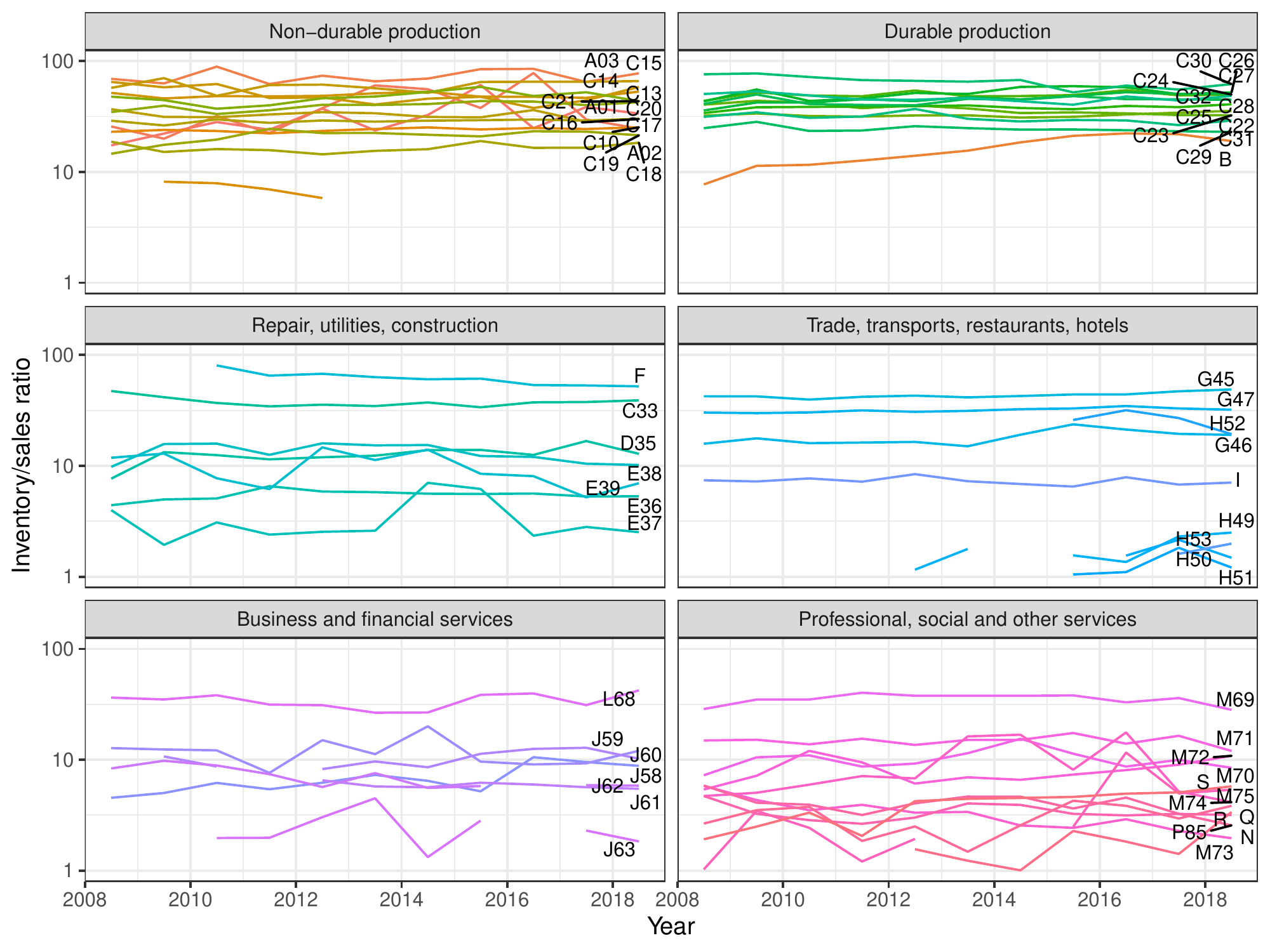}
    \caption{
    \textbf{Inventory/sales ratios over time.} 
    }
        \label{fig:inventory_levels}
\end{figure}

To proxy inventory levels available in February 2020, we proceed as follows, for each industry: (i) we take the simple average between beginning- and end-of year inventory stock levels; (ii) we calculate the ratio between this average and yearly turnover, and multiply this number by 365, because we consider a daily timescale: this inventory-to-turnover ratio is our proxy of $n_j$ (as can be seen in Figure~\ref{fig:inventory_levels}, the inventory/sales ratios are remarkably constant over time, suggesting that inventory stocks at the beginning of the Covid-19 pandemic could reliably be estimated from past data); (iii) we consider a weighted average between inventory-to-turnover ratios across all years between 2008 and 2018, giving higher weight to more recent years;\footnote{More specifically, we consider exponential weights, such that weights from a given year $X$ are proportional to $0.95^{(2018-X)}$. Some years are missing due to confidentiality problems, and data for some sectors have clear problems in some years. We deal with missing values by giving zero weights to years with missing values and renormalizing weights over the available years.} (iv) we aggregate 2-digit NACE industries to WIOD sectors (for example, we aggregate C10, C11 and C12 to C10\_C12); (v) we fill in the missing sectors (K64, K65, K66, O84, T) by imputing the average inventory-to-turnover ratio across all service sectors. 

\begin{figure}
    \centering
\includegraphics[width = \textwidth]{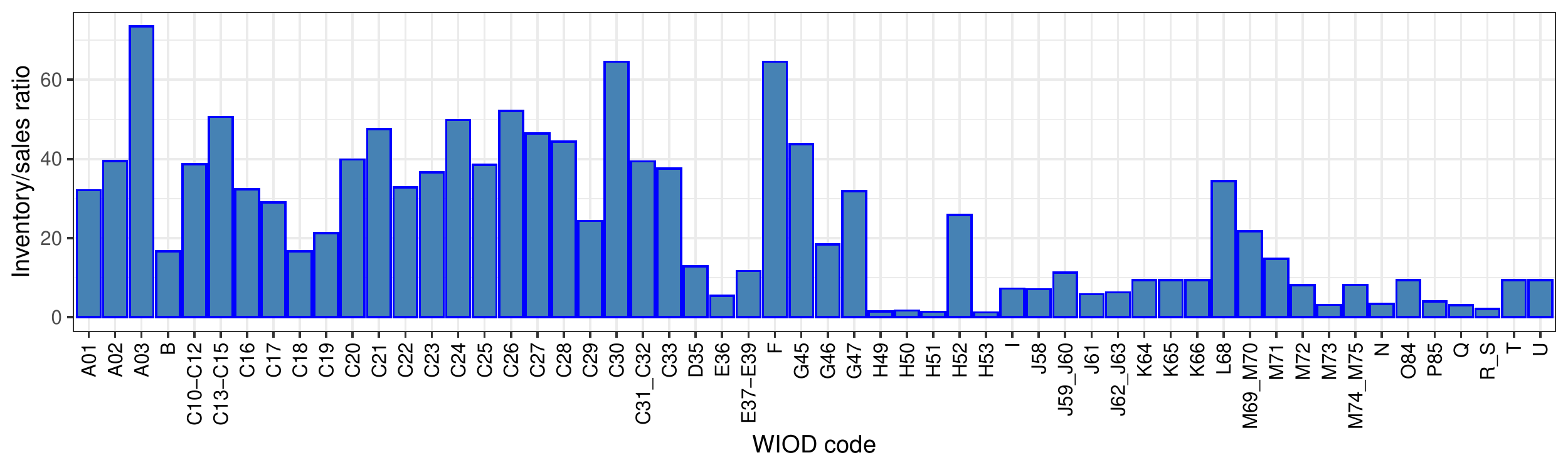}
    \caption{
    \textbf{Inventory/sales ratios for all WIOD industries.} 
    }
        \label{fig:uk_inv_sales_ratios}
\end{figure}

The final inventory to sales ratios $n_j$ are shown in Figure~\ref{fig:uk_inv_sales_ratios}. As can be seen, inventories are much larger relative to sales in production, construction and trade, while they are generally lower in services, although there is considerable heterogeneity across sectors. 

\newpage

\FloatBarrier

\section{Critical vs. non-critical inputs} 
\label{apx:ihs}

A survey was designed to address the question when production can continue during a lockdown.
For each industry, IHS Markit analysts were asked to rate every input of a given industry. The exact formulation of the question was as follows:
``For each industry in WIOD, please rate whether each of its inputs are essential. We will present you with an industry X and ask you to rate each input Y. The key question is: Can production continue in industry X if input Y is not available for two months?''
Analysts could rate each input according to the following allowed answers:
\begin{itemize}
    \item \textbf{0} -- This input is \textit{not} essential
    \item \textbf{1} -- This input is essential
    \item \textbf{0.5} -- This input is important but not essential
    \item \textbf{NA} -- I have no idea
\end{itemize}
To avoid confusion with the unrelated definition of essential industries which we used to calibrate first-order supply shocks, we refer to inputs as \textit{critical} and \textit{non-critical} instead of \textit{essential} and \textit{non-essential.}

Analysts were provided with the share of each input in the expenses of the industry. It was also made explicit that the ratings assume no inventories such that a rating captures the effect on production if the input is not available.

Every industry was rated by one analyst, except for industries Mining and Quarrying (B) and Manufacture of Basic Metals (C24) which were rated by three analysts. 
In case there are several ratings we took the average of the ratings and rounded it to 1 if the average was at least $2/3$ and 0 if the average was at most $1/3$. Average input ratings lying between these boundaries are assigned the value 0.5.

\begin{figure}[htbp]
\centering
\includegraphics[trim = {0cm 0cm 0cm 0cm}, clip,width=\textwidth]{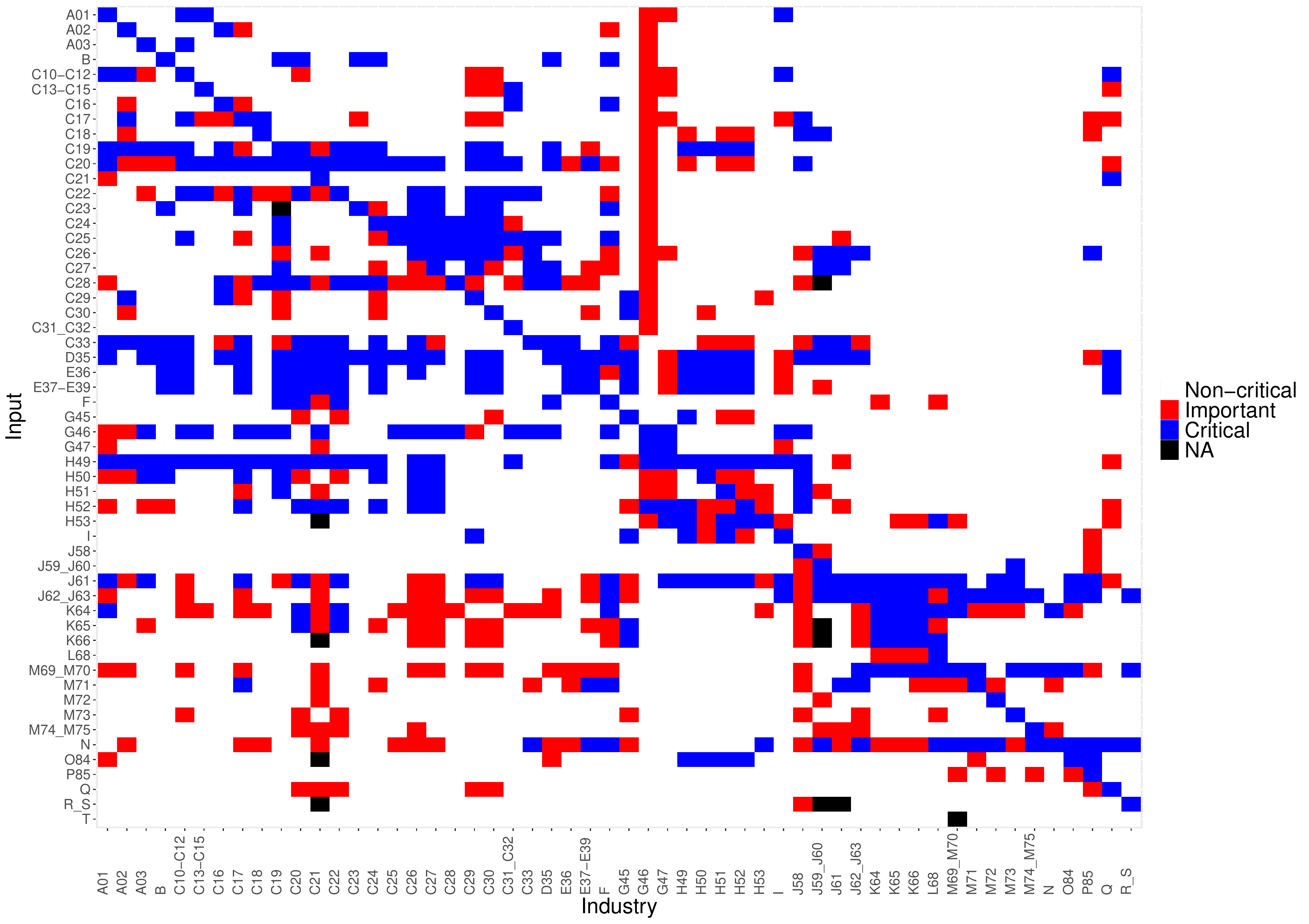}
    \caption{Criticality scores from IHS Markit analysts. 
    Rows are inputs (supply) and columns industries using these inputs (demand).
    The blue color indicates critical (score=1), red important (score=0.5) and white non-critical (score=0) inputs. Black denotes inputs which have been rated with NA. The diagonal elements are considered to be critical by definition. For industries with multiple input-ratings we took the average of all ratings and assigned a score=1 if the averaged score was at least $2/3$ and a score=0 if the average was smaller or equal to $1/3$. 
    }
    \label{fig:ihs_matrix}
\end{figure}

The ratings for each industry and input are depicted in Figure~\ref{fig:ihs_matrix}. A column denotes an industry and the corresponding rows its inputs. 
Blue colors indicate \textit{critical}, red \textit{important, but not critical} and white \textit{non-critical} inputs. Note that under the assumption of a Leontief production function every element would be considered to be critical, yielding a completely blue-colored matrix. The results shown here indicate that the majority of elements are non-critical inputs (2,338 ratings with score = 0), whereas only 477 industry-inputs are rates as critical. 365 inputs are rated as important, although not critical (score = 0.5) and \textit{NA} was assigned eleven times.

\begin{figure}[!htbp]
\centering
\includegraphics[trim = {0cm 0cm 0cm 0cm}, clip,width=1\textwidth]{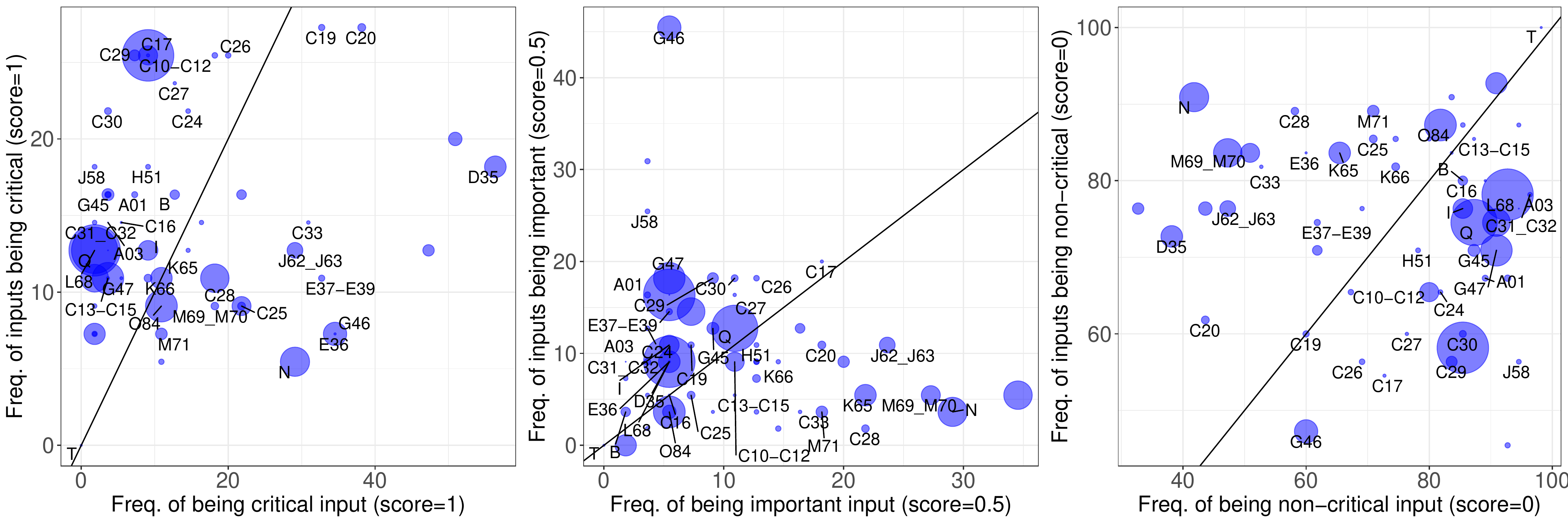}
    \caption{(Left panel) The figure shows how often an industry is rated as a critical input to other industries (x-axis) against the share of critical inputs this industry is using.
    The center and right panel are the same as the left panel, except for using half-critical and non-critical scores, respectively. In each plot the identity line is shown. Point sizes are proportional to gross output.
    }
    \label{fig:ihs_scatter}
\end{figure}

The left panel of Figure \ref{fig:ihs_scatter} shows for each industry how often it was rated as critical input to other industries (x-axis) and how many critical inputs this industry relies on in its own production (y-axis). Electricity and Gas (D35) are rated most frequently as critical inputs in the production of other industries (score=1 for almost 60\% of industries). Also frequently rated as critical are Land Transport (H49) and Telecommunications (J61). 
On the other hand, many manufacturing industries (ISIC codes starting with C) stand out as relying on a large number of critical inputs. For example, around 27\% of inputs to Manufacture of Coke and Refined Petroleum Products (C19) as well as to Manufacture of Chemicals (C20) are rated as critical.

The center panel of Figure \ref{fig:ihs_scatter} shows the equivalent plot for 0.5 ratings (important, but not critical inputs). Financial Services (K64) are most frequently rated as important inputs which do not necessarily stop the production of an industry if not available.
Conversely, the industry relying on many important, but not binding inputs is Wholesale and Retail Trade (G46) of which almost half of its inputs got rated with a score = 0.5.
This makes sense given that this industry heavily relies on all these inputs, but lacking one of these does not halt economic production. This case also illustrates that a Leontief production function could vastly overestimate input bottlenecks as Wholesale and Retail Trade would most likely still be able to realize output even if a several inputs were not available.

In the right panel of Figure \ref{fig:ihs_scatter} we show the same scatter plot but for non-critical inputs. 25 industries are rated to be non-critical inputs to other industries in 80\% of all cases, with Household Activities (T) and Manufacture of Furniture (C31-32) being rated as non-critical in at least 96\%.
Industries like Other Services (R-S), Other Professional, Scientific and Technical Activities (M74-75) and Administrative Activities (N) rely on mostly non-critical inputs ($>$90\%).

A detailed breakdown of the input- and industry-specific ratings are given in Table \ref{tab:ihs_results}.

\begin{table}[!htbp]
\tiny
\centering
\resizebox{1\textwidth}{!}{

\begin{tabular}{|cl|cccc|cccc|c|}
  \hline
  & & 
  \multicolumn{4}{c|}{ Input-based rankings  } &
  \multicolumn{4}{c|}{ Industry-based rankings} & \\
ISIC & Sector (abbreviated) & $1$ & $0.5$ & $0$ & NA & $1$ & $0.5$ & $0$ & NA & n \\ 
  \hline
A01 & Agriculture & 4 & 2 & 49 & 0 & 9 & 9 & 37 & 0 & 1 \\ 
  A02 & Foresty & 2 & 3 & 50 & 0 & 7 & 9 & 39 & 0 & 1 \\ 
  A03 & Fishing & 2 & 1 & 52 & 0 & 8 & 5 & 42 & 0 & 1 \\ 
  B & Mining & 7 & 1 & 47 & 0 & 9 & 2 & 44 & 0 & 3 \\ 
  C10-C12 & Manuf. Food-Beverages & 5 & 6 & 44 & 0 & 14 & 5 & 36 & 0 & 1 \\ 
  C13-C15 & Manuf. Textiles & 2 & 5 & 48 & 0 & 6 & 2 & 47 & 0 & 1 \\ 
  C16 & Manuf. Wood & 3 & 3 & 49 & 0 & 8 & 3 & 44 & 0 & 1 \\ 
  C17 & Manuf. Paper & 5 & 10 & 40 & 0 & 14 & 11 & 30 & 0 & 1 \\ 
  C18 & Media print & 3 & 6 & 46 & 0 & 6 & 3 & 46 & 0 & 1 \\ 
  C19 & Manuf. Coke-Petroleum & 18 & 4 & 33 & 0 & 15 & 6 & 33 & 2 & 1 \\ 
  C20 & Manuf. Chemical & 21 & 10 & 24 & 0 & 15 & 6 & 34 & 0 & 1 \\ 
  C21 & Manuf. Pharmaceutical & 2 & 2 & 51 & 0 & 9 & 17 & 25 & 7 & 1 \\ 
  C22 & Manuf. Rubber-Plastics & 11 & 7 & 37 & 0 & 14 & 5 & 36 & 0 & 1 \\ 
  C23 & Manuf. Minerals & 8 & 2 & 44 & 2 & 7 & 1 & 47 & 0 & 1 \\ 
  C24 & Manuf. Metals-basic & 8 & 2 & 45 & 0 & 12 & 7 & 36 & 0 & 3 \\ 
  C25 & Manuf. Metals-fabricated & 12 & 4 & 39 & 0 & 5 & 3 & 47 & 0 & 1 \\ 
  C26 & Manuf. Electronic & 10 & 7 & 38 & 0 & 14 & 10 & 31 & 0 & 1 \\ 
  C27 & Manuf. Electric & 7 & 6 & 42 & 0 & 13 & 9 & 33 & 0 & 1 \\ 
  C28 & Manuf. Machinery & 10 & 12 & 32 & 2 & 5 & 1 & 49 & 0 & 1 \\ 
  C29 & Manuf. Vehicles & 4 & 5 & 46 & 0 & 14 & 10 & 31 & 0 & 1 \\ 
  C30 & Manuf. Transport-other & 2 & 6 & 47 & 0 & 12 & 10 & 33 & 0 & 1 \\ 
  C31\_C32 & Manuf. Furniture & 1 & 1 & 53 & 0 & 8 & 4 & 43 & 0 & 1 \\ 
  C33 & Repair-Installation & 17 & 9 & 29 & 0 & 8 & 2 & 45 & 0 & 1 \\ 
  D35 & Electricity-Gas & 31 & 3 & 21 & 0 & 10 & 5 & 40 & 0 & 1 \\ 
  E36 & Water & 19 & 3 & 33 & 0 & 4 & 5 & 46 & 0 & 1 \\ 
  E37-E39 & Sewage & 18 & 3 & 34 & 0 & 6 & 8 & 41 & 0 & 1 \\ 
  F & Construction & 5 & 3 & 47 & 0 & 14 & 9 & 32 & 0 & 1 \\ 
  G45 & Vehicle trade & 2 & 5 & 48 & 0 & 9 & 7 & 39 & 0 & 1 \\ 
  G46 & Wholesale & 19 & 3 & 33 & 0 & 4 & 25 & 26 & 0 & 1 \\ 
  G47 & Retail & 2 & 3 & 50 & 0 & 6 & 10 & 39 & 0 & 1 \\ 
  H49 & Land transport & 28 & 3 & 24 & 0 & 11 & 2 & 42 & 0 & 1 \\ 
  H50 & Water transport & 9 & 8 & 38 & 0 & 8 & 5 & 42 & 0 & 1 \\ 
  H51 & Air transport & 5 & 7 & 43 & 0 & 10 & 6 & 39 & 0 & 1 \\ 
  H52 & Warehousing & 12 & 9 & 34 & 0 & 9 & 7 & 39 & 0 & 1 \\ 
  H53 & Postal & 6 & 7 & 41 & 2 & 3 & 5 & 47 & 0 & 1 \\ 
  I & Accommodation-Food & 5 & 3 & 47 & 0 & 7 & 6 & 42 & 0 & 1 \\ 
  J58 & Publishing & 1 & 2 & 52 & 0 & 10 & 14 & 31 & 0 & 1 \\ 
  J59\_J60 & Video-Sound-Broadcasting & 2 & 2 & 51 & 0 & 9 & 5 & 37 & 7 & 1 \\ 
  J61 & Telecommunications & 26 & 11 & 18 & 0 & 7 & 5 & 42 & 2 & 1 \\ 
  J62\_J63 & IT & 16 & 13 & 26 & 0 & 7 & 6 & 42 & 0 & 1 \\ 
  K64 & Finance & 10 & 19 & 26 & 0 & 6 & 3 & 46 & 0 & 1 \\ 
  K65 & Insurance & 6 & 12 & 36 & 2 & 6 & 3 & 46 & 0 & 1 \\ 
  K66 & Auxil. Finance-Insurance & 5 & 7 & 41 & 4 & 6 & 4 & 45 & 0 & 1 \\ 
  L68 & Real estate & 1 & 3 & 51 & 0 & 7 & 5 & 43 & 0 & 1 \\ 
  M69\_M70 & Legal & 12 & 15 & 28 & 0 & 5 & 3 & 46 & 2 & 1 \\ 
  M71 & Architecture-Engineering & 6 & 10 & 39 & 0 & 4 & 2 & 49 & 0 & 1 \\ 
  M72 & R\&D & 1 & 2 & 52 & 0 & 4 & 3 & 48 & 0 & 1 \\ 
  M73 & Advertising & 1 & 7 & 47 & 0 & 5 & 2 & 48 & 0 & 1 \\ 
  M74\_M75 & Other Science & 1 & 8 & 46 & 0 & 4 & 1 & 50 & 0 & 1 \\ 
  N & Private Administration & 16 & 16 & 23 & 0 & 3 & 2 & 50 & 0 & 1 \\ 
  O84 & Public Administration & 6 & 3 & 45 & 2 & 5 & 2 & 48 & 0 & 1 \\ 
  P85 & Education & 1 & 4 & 50 & 0 & 6 & 8 & 41 & 0 & 1 \\ 
  Q & Health & 1 & 6 & 48 & 0 & 7 & 7 & 41 & 0 & 1 \\ 
  R\_S & Other Service & 1 & 1 & 50 & 5 & 4 & 0 & 51 & 0 & 1 \\ 
  T & Household activities & 0 & 0 & 54 & 2 & 0 & 0 & 55 & 0 & 0 \\ 
   \hline
\end{tabular}
}
\caption{Summary table of critical input ratings by IHS Markit analysts.
Columns below \textit{Input-based rankings} show how often an industry has been rated as critical (score=1), half-critical (score=0.5) or non-critical (score=0) input for other industries, or how often the input was rates as NA.
Columns under \textit{Industry-based rankings} give how often an input has been rated as with 1, 0.5, 0 or NA for any given industry.
Column $n$ indicates the number of analysts who have rated the inputs of any given industry. Industry \textit{T} uses no inputs and is therefore not rated.
} 
\label{tab:ihs_results}

\end{table}

\FloatBarrier

\FloatBarrier

 \section{ Model production function and CES }
\label{apx:ces_prodfun}

Here we show that the production functions used in the main text are highly related to nested CES production functions. 
Specifically, we consider a CES production function with three nests of the form (we suppress time indices for convenience)
\begin{equation}
        x_{i}^\text{inp} = \Big( 
    a_i^\text{C} (z_i^\text{C})^\beta +
    a_i^\text{IMP} (z_i^\text{IMP})^\beta + 
    (a_i^\text{NC})^{1-\beta} (z_i^\text{NC})^\beta
    \Big)^{\frac{1}{\beta}},
\end{equation}
where $\beta$ is the substitution parameter.
Variables $a_i^\text{C} = \sum_{j \in \text{C}} A_{ji}$, 
$a_i^\text{IMP} = \sum_{j \in \text{IMP}} A_{ji} $ and
$a_i^\text{NC} = \sum_{j \in \text{NC}} A_{ji} $
are the input shares (technical coefficients) for critical, important and non-critical inputs, respectively. 
To be consistent with the specifications of the main text, we do not consider labor inputs here and only focus on $x_i^\text{inp}$. Alternatively, we could include labor inputs in the set of critical inputs and derive the full production function in an analogous manner. $z_i^\text{C}$, $z_i^\text{IMP}$ and $z_i^\text{NC}$ are CES aggregates of critical, important and non-critical inputs for which we have
\begin{align}
    z_i^\text{C} &= \left[ 
    \sum_{j \in \text{C}} A_{ji}^{1-\nu} S_{ji}^\nu
    \right]^{\frac{1}{\nu}}, \\
        z_i^\text{IMP} &= 
    \frac{1}{2}
    \left[ 
    \sum_{j \in \text{IMP}} A_{ji}^{1-\psi} S_{ji}^\psi
    \right]^{\frac{1}{\psi}} + \;
    \frac{1}{2} x_i^\text{cap}, \\
    z_i^\text{NC} &= \left[ 
    \sum_{j \in \text{NC}} A_{ji}^{1-\zeta} S_{ji}^\zeta
    \right]^{\frac{1}{\zeta}}.
\end{align}

If we assume that every input is critical (i.e. the set of important and non-critical inputs is empty: $a_i^\text{IMP} = a_i^\text{NC} = 0$), and by taking the limits $\beta, \nu \to -\infty$ we recover the Leontief production function
\begin{equation}
    x_i^\text{inp} =  
 \underset{j}{\min} \left\{ \frac{S_{ji}}{A_{ji}}
    \right\}.
\label{eq:leon_limit}
\end{equation}

If we assume that there is no critical or important inputs at all ($a_i^\text{C} = a_i^\text{IMP} = 0$), taking the limits $\beta  \to -\infty$ and $\zeta \to 1$ yields the linear production
\begin{equation}
    x_i^\text{inp} =
 \frac{ \sum_{j}  S_{ji}  }{ a_i^\text{NC} }.
\label{eq:linear_limit}
\end{equation}

The IHS1 and IHS3 production functions treat important inputs either as critical or non-critical, i.e. the set of critical inputs is again empty ($a_i^\text{IMP} = 0$). We can approximate these functions again by taking the limits $\beta, \nu \to -\infty$ and $\zeta \to 1$ which yields
\begin{equation}
    x_i^\text{inp} = \min \left\{ 
     \underset{j \in \text{C} }{\min} \left\{ \frac{S_{ji}}{A_{ji}} \right\}, 
    \frac{ \sum_{j \in \text{NC} }  S_{ji}  }{ a_i^\text{NC} }
    \right\}.
\label{eq:ihs13_limit}
\end{equation}
Note that the difference between the two production functions lies in the different definition of critical and non-critical inputs. The functional form in Eq.~\eqref{eq:ihs13_limit} applies to both cases equivalently. 

The IHS2 production function where the set of important inputs is non-empty can be similarly approximated.
In addition to the limits above, we also take the limit $\psi \rightarrow -\infty$ to obtain
\begin{equation}
    x_i^\text{inp} = \min \left\{ 
     \underset{j \in \text{C} }{\min} \left\{ \frac{S_{ji}}{A_{ji}} \right\}, 
     \frac{1}{2}
     \underset{j \in \text{IMP} }{\min} \left\{  \frac{S_{ji}}{A_{ji}} + x_i^\text{cap} \right\},
    \frac{ \sum_{j \in \text{NC} }  S_{ji}  }{ a_i^\text{NC} }
    \right\}.
\label{eq:ihs2_limit}
\end{equation}

Eqs.~\eqref{eq:ihs13_limit} and \eqref{eq:ihs2_limit} are not exactly identical to the IHS production functions used in the main text (Eqs.~\eqref{eq:xinp_ihs1} -- \eqref{eq:xinp_ihs3}), but very similar. 
The only difference is that in the IHS functions non-critical inputs do not play a role for production at all, whereas here they enter the equations as a linear term. However, simulations show that this difference is practically irrelevant. Simulating the model with the CES-derived functions instead of the IHS production functions yield exactly the same results, indicating that the linear term representing non-critical inputs is never a binding constraint.
\FloatBarrier

\section{Details on validation} 
\label{apx:validation}

In this appendix we provide further details about validation (Section~\ref{sec:econimpact} in the main paper). In Appendix~\ref{apx:validation_data} we describe the data sources that we used for validation, and we explain how we made empirical data comparable to simulated data. In Appendix \ref{apx:selectedscenario} we give more details about the selected scenario (Section~\ref{sec:selectedscenarioanalysis} in the main paper). 

\subsection{Validation data}
\label{apx:validation_data}
\begin{itemize}
    \item Index of agriculture (release: 12/08/2020): \url{https://www.ons.gov.uk/generator?format=xls&uri=/economy/grossdomesticproductgdp/timeseries/ecy3/mgdp/previous/v26}
    \item Index of production (release: 12/08/2020): \url{https://www.ons.gov.uk/file?uri=/economy/economicoutputandproductivity/output/datasets/indexofproduction/current/previous/v59/diop.xlsx}
    \item Index of construction (release: 12/08/2020): \url{https://www.ons.gov.uk/file?uri=/businessindustryandtrade/constructionindustry/datasets/outputintheconstructionindustry/current/previous/v67/bulletindataset2.xlsx}
    \item Index of services (release: 12/08/2020): \url{https://www.ons.gov.uk/file?uri=/economy/economicoutputandproductivity/output/datasets/indexofservices/current/previous/v61/ios1.xlsx}
\end{itemize}

All ONS indexes are monthly seasonally-adjusted chained volume measures, based such that the index averaged over all months in 2016 is 100. Although these indexes are used to proxy value added in UK national accounts, they are actually gross output measures, as determining input use is too burdersome for monthly indexes. 

There is not a perfect correspondence between industry aggregates as considered by ONS and in WIOD. For example, the ONS only releases data for the agricultural sector as a whole, without distinguishing between crop and animal production (A01), fishing and aquaculture (A02) and forestry (A03). In this case, when comparing simulated and empirical data we aggregate data from the simulations, using initial output shares as weights. More commonly, there is a finer disaggregation in ONS data than in WIOD. For example, ONS provides separate information on food manufacturing (C10) and on beverage and tobacco manufacturing (C11, C12), while these three sectors are aggregated into just one sector (C10\_C12) in WIOD. In this case, we aggregate empirical data using the weights provided in the indexes of production and services. These weights correspond to output shares in 2016, the base year for all time series. 

Finally, after performing aggregation we rebase all time series so that output in February 2020 takes value 100.

\subsection{Description of the selected scenario}
\label{apx:selectedscenario}

\begin{figure}[H]
    \centering
\includegraphics[width = 1.0\textwidth]{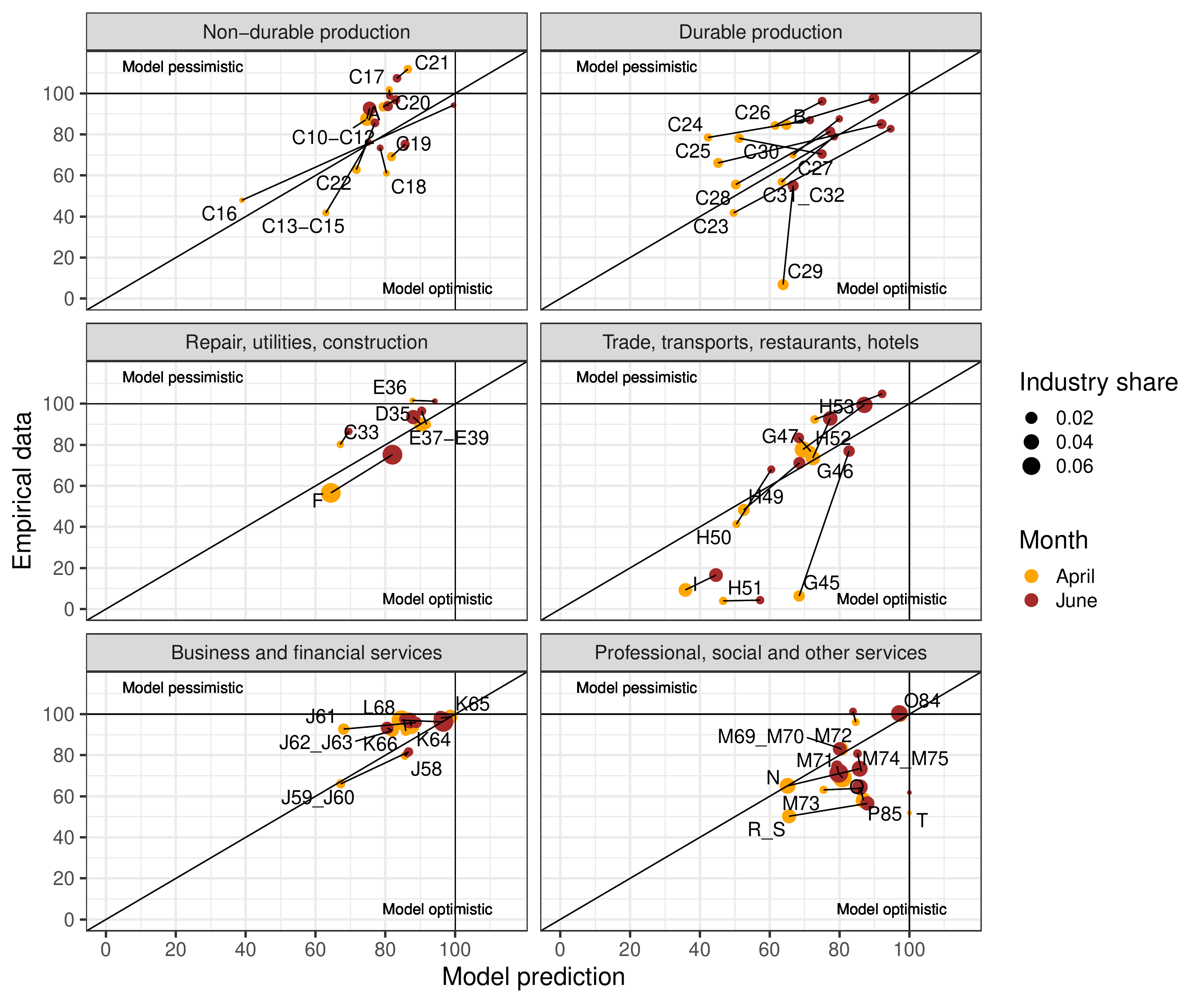}
    \caption{\textbf{Comparison between model predictions and empirical data.} We plot production (gross output) for each of 53 industries, both as predicted by our model and as obtained from the ONS' indexes. Each panel refers to a different group of industries. Different colors refer to production in April and June 2020. Black lines connect the same industry across these three months. All sectoral productions are normalized to their pre-lockdown levels, and each point size is proportional to the steady-state gross output of the corresponding sector.}
        \label{fig:new_model_april_may_june_facets}
\end{figure}

Figure \ref{fig:new_model_april_may_june_facets} shows the recovery path of all industries, both in the model and in the data. Interpretation is the same as in Figure \ref{fig:new_model_april_may_june_all} in the main text. For readability, industries are grouped into six broad categories.

\FloatBarrier

\section{Sensitivity analysis}
\label{apx:sensitivity}
To gain a better understanding of the model behavior, we conduct a series of sensitivity analyses by showing aggregate output time series under alternative parametrizations. In particular, we vary exogenous shock inputs, the production function specification and the parameters $\tau$, $\gamma_H$, $\gamma_F$ and $\Delta s$.

\begin{figure}[H]
\includegraphics[trim = {0cm 0cm 0cm 0cm}, clip, width = .95\textwidth]{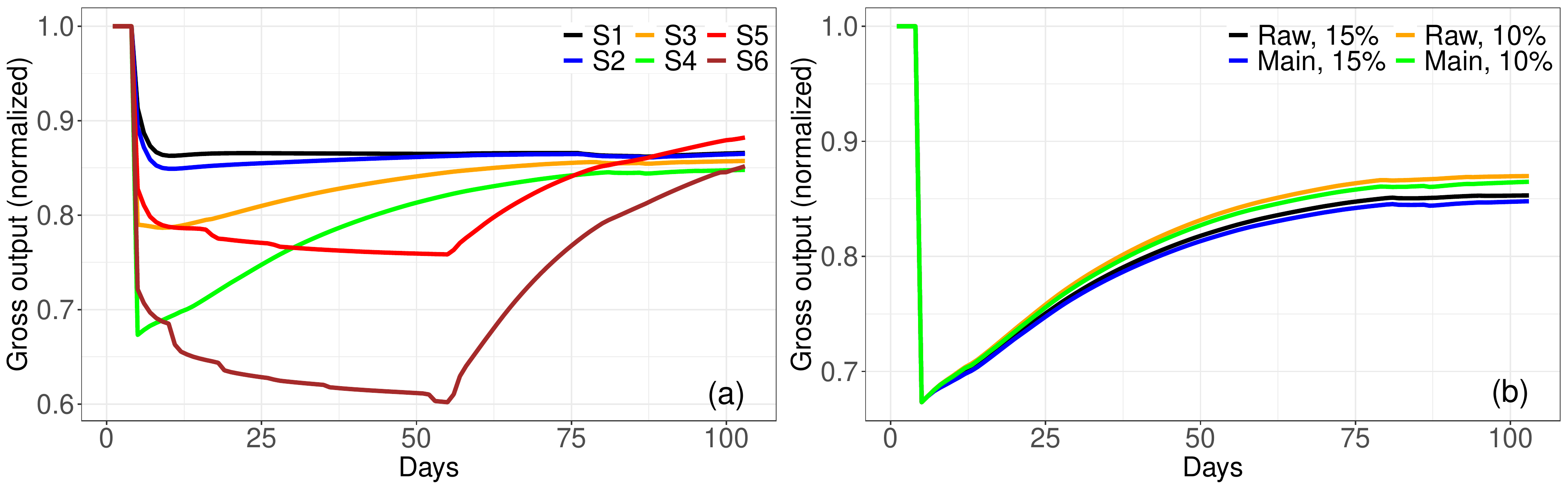}
    \caption{
    \textbf{Sensitivity analysis with respect to supply and demand shocks.}
    Shocks are applied at day one. Except for supply shocks (a) and demand shocks (b), the model is initialized as in the baseline run of the main text.
    Legend (b): Raw and Main indicate the estimates from \cite{CBO2006} and its adapted version used in the main text, respectively. 15\% and 10\% refer to the two investment/export shock scenarios.
    }
        \label{fig:suppdemsens}
\end{figure}

\paragraph{Supply shocks.}
Figure~\ref{fig:suppdemsens}(a) shows model dynamics of total output for the different supply shock scenarios considered. 
It becomes clear that alternative specifications of supply shocks can influence model results enormously. Supply shocks derived directly from the UK lockdown policy ($S_1-S_4$) yield a similar recovery pattern but nevertheless entail markedly different levels of overall impacts. When initializing the model with the shock estimates from \cite{del2020supply} ($S_5$) and \cite{fana2020covid} ($S_6$), we obtain very different dynamics.

\paragraph{Demand shocks.}
We find that our model is less sensitive with respect to changing final demand shocks within plausible ranges. Here, we compare the baseline model results with alternative specifications of consumption and ``other final demand'' (investment, export) shocks. Instead of modifying the \cite{CBO2006} consumption shocks as discussed in Section \ref{sec:consumptiondemandshockscenarios}, we now also use their raw estimates. We further consider shocks of 10\% to investment and exports which is milder than the 15\% shocks considered in the main text.
Figure~\ref{fig:suppdemsens}(a) indicates that differences are less pronounced between the alternative demand shocks compared to the supply shock scenarios, particularly during the early phase of the lockdown. Differences of several percentage points only emerge after an extended recovery period.

\begin{figure}[H]
\centering
\includegraphics[trim = {0cm 0cm 0cm 0cm}, clip, width = .95\textwidth]{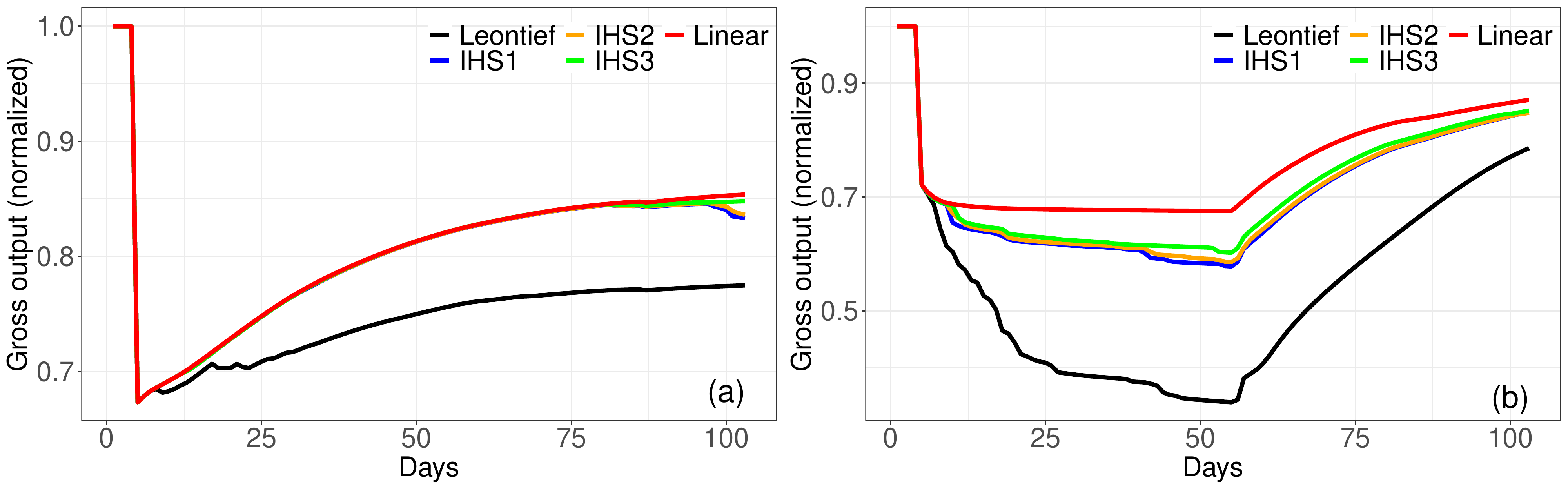}
    \caption{
    \textbf{Sensitivity analysis with respect to production functions.}
    Shocks are applied at day one. 
    (a) is based on the $S_4$ and (b) on the $S_5$ supply shock scenario.
    Otherwise, the model is initialized as in the baseline run of the main text.
    }
        \label{fig:prodfsens}
\end{figure}

\paragraph{Production function.}
In Figure~\ref{fig:prodfsens}(a) and (b) we show simulation results for alternative production function specifications, when using the baseline $S_4$ and the more severe $S_5$ supply shock scenarios, respectively. 
Regardless of the supply shock scenarios considered, Leontief production yields substantially more pessimistic predictions than the other production functions.
For the milder baseline supply shock scenarios, aggregate predictions are fairly similar across the other production functions, although we observe some differences after an extended period of simulation. 
Differences between the linear and IHS production functions are larger when considering the more severe shock scenario in Figure~\ref{fig:prodfsens}(b).

\begin{figure}[H]
\includegraphics[trim = {0cm 0cm 0cm 0cm}, clip, width = 1\textwidth]{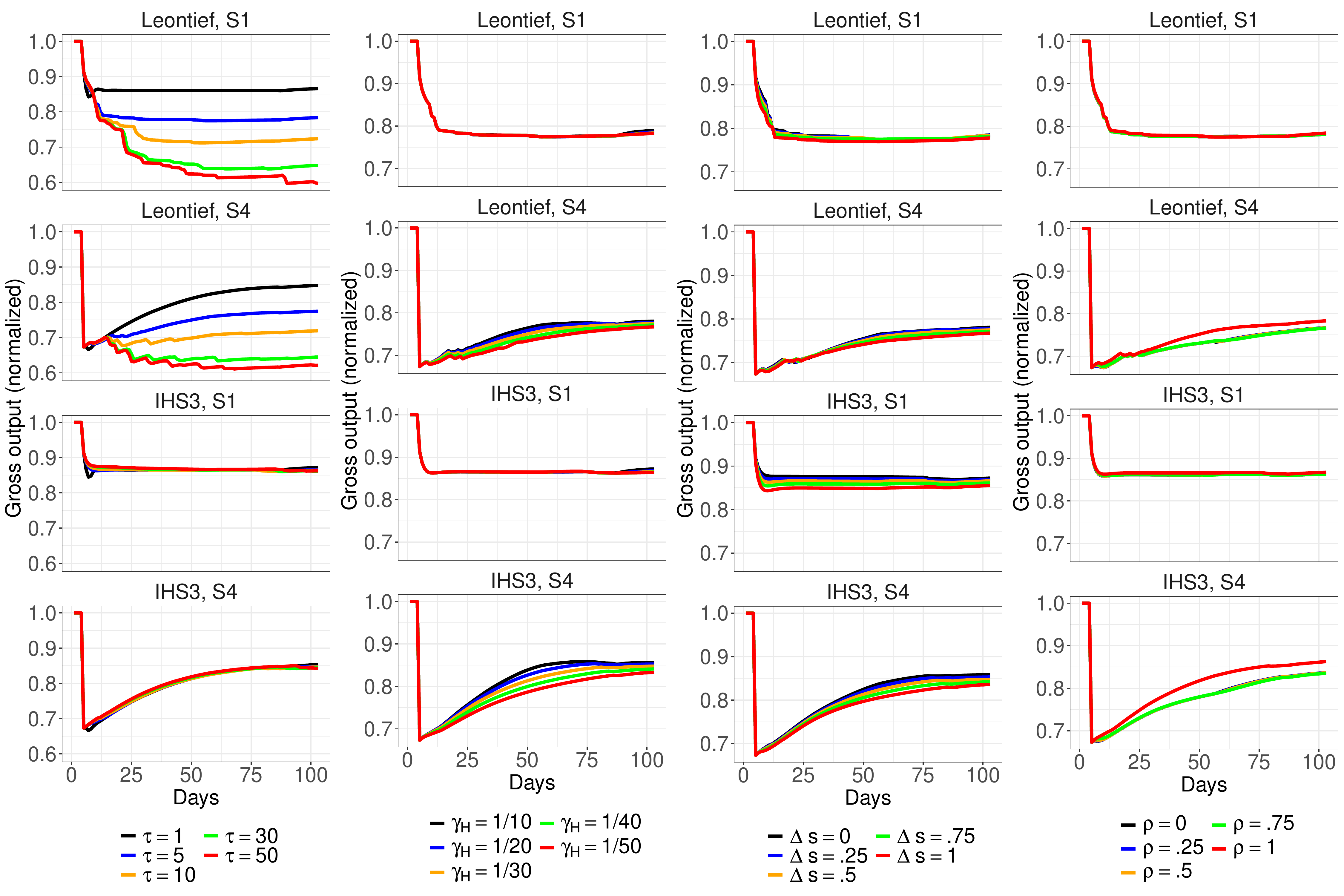}
    \caption{ 
    \textbf{Results of sensitivity analysis for parameters $\tau$, $\gamma_H$, $\Delta s$ and $\rho$.}
    All plots show normalized gross output on the y-axis and days after the shocks are applied on the x-axis. Panel rows differ with respect to the combination of production function and supply shock scenario. Panel columns differ with respect to $\tau$ (left), $\gamma_H$ (center left), $\Delta s$ (center right) and $\rho$ (right) as indicated in the legend below each column. Except for the four parameters, production function and supply shock scenario, the model is initialized as in the baseline run of the main text.
    In all simulations we used $\gamma_F = 2 \gamma_H$.
    }
        \label{fig:sensitivity_params}
\end{figure}

In Figure~\ref{fig:sensitivity_params} we explore model simulations with respect to different parametrizations of the inventory adjustment parameter $\tau$ (left column), hiring/firing parameters $\gamma_H$ and $\gamma_F$ (center left column), change in savings rate parameter $\Delta s$ (center right column) and consumption adjustment speed $\rho$ (right column) under alternative production function-supply shock combinations (panel rows).

\paragraph{Inventory adjustment time $\tau$.}
We find that the inventory adjustment time becomes a key parameter under Leontief production and much less so under the alternative IHS3 specification. The shorter the inventory adjustment time (smaller $\tau$), the more shocks are mitigated.
Contrary to inventory adjustment $\tau$, model results do not change much with respect to alternative specifications of $\gamma_H$ when going from Leontief production to the IHS3 function. Here, differences in model outcomes are rather due to shock severity. Model predictions are almost identical for all choices of $\gamma_H$ under mild $S_1$ shocks but differ somewhat for more substantial $S_4$ shocks. In the $S_4$ scenarios we observe that recovery is quickest for larger values of $\gamma_H$ (stiff labor markets), since firms would lose less productive capacity in the immediate aftermath of the shocks. However, this would come at the expense of reduced profits for firms which are able to more effectively reduce labor costs for smaller values of $\gamma_H$.

\paragraph{Change in saving rate $\Delta s$.}
We find only very little variation with respect to the whole range of $\Delta s$, regardless of the underlying production function and supply shock scenario. As expected, we find the largest adverse economic impacts in case of $\Delta s = 1$, i.e. when consumers save all the extra money which they would have spent if there was no lockdown, and the mildest impacts if $\Delta s = 0$.

\paragraph{Consumption adjustment speed $\rho$.}
Similarly, the model is not very sensitive with respect to consumption adjustment speed $\rho$. The smaller $\rho$, the quicker households adjust consumption with respect to (permanent) income shocks. Note that parameter $\rho$ is based on daily time scales. Economic impacts tend to be less adverse when consumers aim to keep original consumption levels upright (large $\rho$) and more adverse if income shocks are more relevant for consumption (small $\rho$). Overall, the effects are small, in particular for the $S_1$ shock scenarios.

\FloatBarrier

\newpage
\section{Notation}
\label{apx:notation}

\begin{table}[H]
\scriptsize
\begin{center}
    \begin{tabular}{ | p{2cm} | p{12cm} |}
    \hline
    Symbol  & Name  \\ 
    \hline
    $N$ & Number of industries  \\
    $t$ & Time index \\
    $t_\text{start\_lockdown}$ & Start date of lockdown (March 23)\\
    $t_\text{end\_lockdown}$ & End date of lockdown (May 13) \\
    $x_{i,t}$ & Total output of industry $i$   \\
    $x_{i,t}^{\text{cap}}$ & Industry production capacity based on available labor   \\
    $x_{i,t}^{\text{inp}}$ & Industry production capacity based on available inputs   \\
    $d_{i,t}$ & Total demand for industry $i$   \\
    $Z_{ij,t}$ & Intermediate consumption of good $i$ by industry $j$  \\
    $O_{ij,t}$ & Intermediate orders (demand from industry $j$ to industry $i$)   \\
    $c_{i,t}$ & Household consumption of good $i$   \\
    $c_{i,t}^d$ & Demand of household consumption of good $i$   \\
    $f_{i,t}$ & Non-household final demand of good $i$  \\
    $f_{i,t}^d$ & Demand non-household final demand of good $i$  \\
    $l_{i,t}$ & Labor compensation to workers of industry $i$  \\
    $\tilde l_t$ & Total labor compensation    \\
    $\tilde{l}_{t}^p$ & Expectations for permanent labor income   \\
    $\tilde l^*_t$ & Total labor compensation plus social benefits   \\
    $\tilde c_t$ & Total household consumption  \\
    $\tilde{c}_{t}^d$ & Aggregate consumption demand   \\
    $n_{j}$ & Number of days of targeted inventory for industry $j$   \\
    $A_{i,j}$ & Payments to $i$ per unit produced of $j$ (technical coefficients) \\
    $S_{ij,t}$ & Stock of material $i$ held in $j$'s inventory   \\
    $\tau$ & Speed of inventory adjustment   \\
    $\theta_{i,t}$ & Share of goods from industry $i$ in consumption demand   \\
    $\bar \theta_{i,t}$ & Share of goods from industry $i$ in consumption demand (unnormalized)  \\
    $\rho$ & Speed of adjustment of aggregate consumption  \\
    $m$ & Share of labor income used to consume final domestic goods   \\
    $\xi_t$ & Fraction of pre-pandemic labor income that households expect to retain in the long-run  \\
    $\epsilon_t$ & Consumption exogenous shock   \\
    $\tilde{\epsilon}_{i}^D$ & Relative changes in demand for goods of industry $i$ during lockdown   \\
    $\tilde{\epsilon}_{i,t}$ & Relative changes in demand for goods of industry $i$   \\
    $\tilde \epsilon_t$ & Aggregate consumption shock   \\
    $\Delta l_{i,t}$ &  Desired change of labor supply of industry $i$\\
    $ l_{i,t}^\text{max}$ & Maximum labor supply for industry $i$\\
    $\gamma_\text{H}$, $\gamma_\text{F}$ & Speed of upward/downward labor adjustment (hiring/firing)  \\
    $\Delta s$ & Change in saving rate   \\
    $\nu$ & Consumption shock term representing beliefs in L-shaped recovery \\
    $b$ & Share of labor income compensated as social benefit \\
    $\text{RLI}_i$ & Remote Labor Index of industry $i$\\
    $\text{ESS}_i$ & Essential score of industry $i$\\
    $\text{PPI}_i$ & Physical Proximity Index of industry $i$\\
    $\iota$ & Scaling factor for $\text{PPI}_i$ \\ 
    $\mathcal{V}_i$ & Set of critical inputs for industry $i$\\
    $\mathcal{U}_i$ & Set of important inputs for industry $i$\\
    \hline
        \end{tabular}
\end{center}
\caption{Notation summary.}
\label{tab:notation_econ}
\scriptsize
\end{table}

\FloatBarrier

\end{document}